\documentclass[twocolumn]{aastex62}

\usepackage{color}
\usepackage[normalem]{ulem}

\received{}
\accepted{}
\submitjournal{AJ}

\begin{document}

\title{ACRONYM III: Radial Velocities for 336 Candidate Young Low-Mass Stars in the Solar Neighborhood, Including 77 Newly Confirmed Young Moving Group Members}

\correspondingauthor{Adam C. Schneider}
\email{aschneid10@gmail.com}

\author{Adam C. Schneider}
\affil{School of Earth and Space Exploration, Arizona State University, Tempe, AZ, 85282, USA}

\author{Evgenya L. Shkolnik}
\affil{School of Earth and Space Exploration, Arizona State University, Tempe, AZ, 85282, USA}

\author{Katelyn N. Allers}
\affil{Department of Physics and Astronomy, Bucknell University, Lewisburg, PA 17837, USA}

\author{Adam L. Kraus}
\affil{The University of Texas at Austin, Department of Astronomy, Austin, TX 78712, USA}

\author{Michael C. Liu}
\affil{Institute for Astronomy, University of Hawaii, Honolulu, HI 96822, USA}

\author{Alycia J. Weinberger}
\affil{Department of Terrestrial Magnetism, Carnegie Institution of Washington, Washington, DC 20015, USA}

\author{Laura Flagg}
\affil{Physics \& Astronomy Department, Rice University, Houston, TX 77005, USA}

\begin{abstract}

Young, low-mass stars in the solar neighborhood are vital for completing the mass function for nearby, young coeval groups, establishing a more complete census for evolutionary studies, and providing targets for direct-imaging exoplanet and/or disk studies.  We present properties derived from high-resolution optical spectra for 336 candidate young nearby, low-mass stars.  These include measurements of radial velocities and age diagnostics such as H$\alpha$ and Li $\lambda$6707 equivalent widths.  Combining our radial velocities with astrometry from {\it Gaia} DR2, we provide full 3D kinematics for the entire sample.  We combine the measured spectroscopic youth information with additional age diagnostics (e.g., X-ray and UV fluxes, CMD positions) and kinematics to evaluate potential membership in nearby, young moving groups and associations.  We identify 77 objects in our sample as bonafide members of 10 different moving groups, 14 of which are completely new members or have had their group membership reassigned.  We also reject 46 previously proposed candidate moving group members. Furthermore, we have newly identified or confirmed the youth of numerous additional stars that do not belong to any currently known group, and find 69 co-moving systems using {\it Gaia} DR2 astrometry.  We also find evidence that the Carina association is younger than previously thought, with an age similar to the $\beta$ Pictoris moving group ($\sim$22 Myr).

\end{abstract}


\section{Introduction} 

The discovery of young stars ($<$300 Myr) in the neighborhood of the Sun ($<$100 pc) has allowed for significant advances in several areas of astrophysics.  Because of their age and proximity, such stars are optimal targets for directly detecting low-mass planetary companions (e.g., \citealt{chauv04, chauv05}, \citealt{marois08, marois10}, \citealt{lag10}, \citealt{bowler13, bowler17}, \citealt{naud14}, \citealt{chauv17}, \citealt{dup18}), in-depth studies of circumstellar disks (e.g., \citealt{su06}, \citealt{rebull08}, \citealt{zuck11}, \citealt{riv14}, \citealt{rod15}, \citealt{binks17}, \citealt{meng17}), mass function investigations (e.g., \citealt{kraus14}, \citealt{shk17}, \citealt{gagne17}), and studies of various evolutionary properties ranging from rotation and activity (e.g., \citealt{montes01} \citealt{dela04}, \citealt{malo14}, \citealt{shk14}, \citealt{schneid18}) to chemical abundances (e.g., \citealt{ment08},  \citealt{dasilva09}, \citealt{viana09}, \citealt{zerjal19}) to multiplicity (e.g, \citealt{janson14}, \citealt{shan17}) to empirical isochrones (e.g, \citealt{liu16}).

Nearby, young moving groups are coeval collections of stars that share the same age and similar space motions (e.g., \citealt{kast97}). Confirming membership to these kinematically linked groups requires knowledge of an object's distance, proper motion, radial velocity, and age.  While proper motions have typically been available from numerous all-sky surveys, accurate distances and radial velocities have been more difficult to acquire as they require dedicated observations.  With the second data release of the {\it Gaia} mission (hereafter {\it Gaia} DR2; \citealt{gaia16, gaia18}), parallaxes are now readily available for the vast majority of nearby, stellar moving group members.  Such data is being used to identify new moving group members as well as completely new moving groups (e.g., \citealt{fah18}, \citealt{gagne18d}).  In many instances, the only pieces of information now missing to confirm a potential moving group member are its radial velocity and youth.  The {\it Gaia} mission will eventually include radial velocities for $\sim$150 million stars \citep{soub18}, and the DR2 release includes radial velocities for over 7 million bright stars, though with increasing uncertainties at the faint end ($\gtrsim$2 km s$^{-1}$; \citealt{gaia18}). In this third installment of the All-sky Co-moving Recovery Of Nearby Young Members (ACRONYM) series (\citealt{kraus14}, \citealt{shk17}), we present radial velocities and youth characteristics for 336 nearby, low-mass stars that are suspected to be young in order to identify new, and confirm previously proposed, moving group members.    

\section{Observations}

Nearby young candidates were observed with the goal of identifying young, nearby, low-mass stars.  These include the list of X-ray active M dwarfs from the NStars catalog \citep{reid03, reid04} and from \cite{riaz06}, kinematic moving group candidates from \cite{malo13}, and UV-active M dwarfs using the methods outlined in \cite{shk11}.  For X-ray active M dwarfs, we required the log of the X-ray flux ($F_{\rm X}$) to 2MASS $J$-band flux ($F_J$) to be greater than --2.5, a value chosen to be comparable to members of the Pleiades \citep{shk09}.  For UV-active M dwarf candidates, we required 3$\sigma$ detections in both the {\it GALEX} FUV and NUV bands, and required the FUV and NUV to 2MASS $J$-band flux ratios to be greater than 10$^{-5}$ and 10$^{-4}$, respectively \citep{shk11}.  High resolution optical spectra were acquired for all targets to measure radial velocities and youth diagnostics.  Radial velocity standards were observed each night in order to determine radial velocities of our target stars.  Care was taken to observe radial velocity standards with similar spectral types as the target sample. All observations are summarized in Table 1. 

\subsection{Keck/HIRES}
Fifty-two targets were observed with the High Resolution Echelle Spectrometer (HIRES; \citealt{vogt94}) on the Keck I telescope.  The observing strategy and data reduction routine are outlined in \cite{shk09, shk17}, which we briefly summarize here.  All observations were taken with the 0\farcs861 slit, resulting in a spectral resolution of $\lambda$/$\Delta$$\lambda$ $\approx$ 58,000 covering a wavelength range of 4900--9300 \AA.  The GG475 filter was utilized with the red cross-disperser in order to maximize the throughput at the longest wavelengths (where M dwarfs emit their peak emission).  All data reduction was accomplished using the facility pipeline MAKEE.  

\subsection{CFHT/ESPaDOnS}
We observed 15 targets with the Echelle SpectroPolarimatric Device for the Observation of Stars (ESPaDOnS; \citealt{don06}) on the Canada-France-Hawaii Telescope (CFHT).  The observing strategy and data reduction routine are outlined in \cite{shk09}, which we briefly summarize here.  ESPaDOnS covers a wavelength range of 3700--10400 \AA, with a spectral resolution of $\lambda$/$\Delta$$\lambda$ $\approx$ 68,000.  Data reduction was achieved using the fully automated reduction package {\it Libre ESpRIT} \citep{don07}.  

\subsection{Magellan/MIKE}
The Magellan Inamori Kyocera Echelle (MIKE) optical spectrograph located on the Clay telescope at the Magellan Observatory was used to characterize 122 of our targets.   The observing strategy and data reduction routine are outlined in \cite{shk11, shk17}, which we briefly summarize here.  We used the 0\farcs5 slit with for all MIKE observations, which gives a spectral resolution of $\lambda$/$\Delta$$\lambda$ $\approx$ 35,000 over the wavelength range  4900--10000 \AA.  All MIKE data were reduced using the CarPy pipeline \citep{kels00,kels03}.

\subsection{Du Pont/Echelle}
Two-hundred and twenty-nine targets were observed with the Echelle Spectrograph on the 2.5 m Irenee Du Pont Telescope. All observations were taken with a 1\farcs0 slit, resulting in a resolution of $\lambda$/$\Delta$$\lambda$ $\approx$ 32,000 covering the 4000--9000 \AA\ wavelength range.  Standard IRAF echelle routines were used to reduce the data.

\startlongtable
\begin{deluxetable*}{llccrc}
\tablecaption{Observations Summary}
\tablehead{
\colhead{2MASS} & \colhead{Other} & \colhead{Telescope/} & \colhead{Obs.\ Date} & \colhead{RV} & \colhead{SB?}  \\
\colhead{Designation} & \colhead{Name} & \colhead {Instrument} & \colhead{(UT)} & \colhead{(km s$^{-1}$)} &  }
\startdata
00153670$-$2946003 & GJ 3017 & Magellan/MIKE & 2009 Dec 31 & 0.61$\pm$1.31 & \dots \\ 
00213729$-$4605331 & GJ 3029 & Magellan/MIKE & 2009 Jun 06 & -19.32$\pm$0.12 & \dots \\  
00254902$+$4501315 & [FS2003] 0016 & Keck/HIRES & 2012 Dec 28 & 6.30$\pm$0.32 & \dots \\ 
00354412$-$0541102 & LP 645-53 & Magellan/MIKE & 2009 Jun 07 & -15.63$\pm$0.17 & \dots \\ 
00501079$-$0337532 & \dots & Magellan/MIKE & 2010 Jan 01 & 0.27$\pm$1.24 & \dots \\ 
00503319$+$2449009 & GJ 3061 & DuPont/Echelle & 2009 Aug 21 & 5.42$\pm$0.38 & \dots \\ 
00503319$+$2449009 & GJ 3061 & Magellan/MIKE & 2009 Dec 31 & 7.44$\pm$1.24 & \dots \\ 
00503319$+$2449009 & GJ 3061 & Magellan/MIKE & 2009 Dec 31 & 7.53$\pm$1.58 & \dots \\ 
00503319$+$2449009 & GJ 3060 & Magellan/MIKE & 2009 Dec 31 & 5.60$\pm$0.79 & \dots \\ 
00503319$+$2449009 & GJ 3060 & Magellan/MIKE & 2009 Dec 31 & 5.64$\pm$0.87 & \dots \\ 
00560596$+$4153282 & \dots & Keck/HIRES & 2013 Oct 24 & -33.12$\pm$0.17 & \dots \\ 
01031408$+$2005523 & GJ 1026 A & DuPont/Echelle & 2009 Aug 21 & 21.51$\pm$0.38 & \dots \\ 
01031408$+$2005523 & GJ 1026 B & DuPont/Echelle & 2009 Aug 21 & 21.87$\pm$0.41 & \sout{SB1} \\
01031408$+$2005523 & GJ 1026 B & Magellan/MIKE & 2009 Dec 31 & 20.11$\pm$0.92 & \sout{SB1} \\
01031408$+$2005523 & GJ 1026 B & Magellan/MIKE & 2009 Dec 31 & 21.53$\pm$0.39 & \sout{SB1} \\
01050677$+$2815049 & 1RXS J010506.0+281505 & Keck/HIRES & 2013 Oct 24 & 3.85$\pm$0.39 & \dots \\
01090150$-$5100494 & DG Phe & Magellan/MIKE & 2009 Jun 08 & -1.15$\pm$0.81 & \dots \\
01112542$+$1526214 & GJ 3076 & Keck/HIRES & 2012 Dec 28 & 2.84$\pm$0.12 & \dots \\ 
01132401$-$2254077 & GJ 1033 & DuPont/Echelle & 2013 Dec 19 & -0.13$\pm$0.80 & \dots \\
01132958$-$0738088 & StKM 1-124 & DuPont/Echelle & 2013 Dec 18 & 115.97$\pm$2.37 & SB1 \\
01205998$-$2408520 & BPS CS 29514-0033 & Magellan/MIKE & 2009 Jun 07 & 3.48$\pm$5.98 & \dots \\
01210504$-$0402082 & G 271-42 & Magellan/MIKE & 2010 Jan 01 & 9.30$\pm$1.83 & \dots \\ 
01244246$-$1540454 & G 272-13 & DuPont/Echelle & 2013 Dec 18 & 18.09$\pm$0.60 & \dots \\
01245068$-$3844389 & SERC 296A & Magellan/MIKE & 2011 Jun 14 & 16.20$\pm$0.51 & \dots \\ 
01302034$-$2557105 & GSC 06426-01758 & Magellan/MIKE & 2009 Jun 07 & 13.56$\pm$0.24 & \dots \\
01335800$-$1738235 & LP 768-113 & Magellan/MIKE & 2010 Jan 01 & 5.87$\pm$0.28 & \dots \\
01345037$-$0254397 & \dots & Magellan/MIKE & 2010 Jan 01 & -11.27$\pm$1.46 & \dots \\
01452133$-$3957204 & CD-40 436 & DuPont/Echelle & 2013 Dec 18 & 21.06$\pm$0.59 & \dots \\ 
01484087$-$4830519 & GSC 08044-00859 & Magellan/MIKE & 2010 Jan 01 & 21.32$\pm$0.31 & \dots \\
01511997$+$1324525 & [ZEH2003] RX J0151.3+1324 3 & Magellan/MIKE & 2010 Jan 01 & 25.96$\pm$0.42 & \dots \\
01564714$-$0021127 & GSC 04686-00596 & Magellan/MIKE & 2010 Jan 01 & 3.43$\pm$1.33 & \dots \\
02130073$+$1803460 & LP 409-52 & DuPont/Echelle & 2009 Aug 21 & 2.68$\pm$0.43 & \sout{SB1} \\
02130073$+$1803460 & LP 409-52 & Magellan/MIKE & 2009 Dec 31 & 2.97$\pm$0.61 & \sout{SB1} \\
02133021$-$4654505 & WISE J021330.24-465450.3 & Magellan/MIKE & 2010 Jan 01 & 10.33$\pm$1.02 & \dots \\
02135155$-$4129304 & CD-42 759 & DuPont/Echelle & 2013 Dec 18 & 8.33$\pm$1.09 & \dots \\
02135155$-$4129304 & CD-42 759 & DuPont/Echelle & 2013 Dec 18 & 10.28$\pm$1.93 & \dots \\  
02165488$-$2322133 & 1RXS J021655.0-232216 & Magellan/MIKE & 2010 Jan 01 & 3.81$\pm$2.53 & \dots \\
02202235$-$0808253 & LP 650-215 & Keck/HIRES & 2013 Oct 24 & 3.37$\pm$1.07 & \dots \\ 
02291365$-$1009419 & \dots & Magellan/MIKE & 2010 Jan 01 & 12.94$\pm$0.40 & \dots \\ 
02344773$-$2222451 & \dots & Magellan/MIKE & 2010 Jan 01 & 8.36$\pm$1.70 & \dots \\ 
02371574$-$2222590 & \dots & Magellan/MIKE & 2010 Jan 01 & 3.54$\pm$0.44 & \dots \\ 
02374615$-$0705480 & LHS 1427 & Keck/HIRES & 2013 Oct 24 & -8.01$\pm$0.08 & \dots \\
02390078$-$1937040 & \dots & Magellan/MIKE & 2010 Jan 01 & 23.41$\pm$0.69 & \dots \\ 
02411909$-$5725185 & GSC 08494-00369 & Magellan/MIKE & 2010 Jan 01 & 7.03$\pm$0.43 & \dots \\ 
02411909$-$5725185 & GSC 08494-00369 & Magellan/MIKE & 2010 Jan 01 & 8.68$\pm$0.60 & \dots \\ 
02442137$+$1057411 & MCC 401 & DuPont/Echelle & 2013 Dec 20 & 5.60$\pm$1.09 & \dots \\
02445715$-$4407313 & \dots & Magellan/MIKE & 2010 Jan 01 & 16.79$\pm$0.67 & \dots \\  
02461477$-$0459182 & GJ 3180 & Magellan/MIKE & 2009 Jun 07 & 35.00$\pm$0.22 & \dots \\
02490228$-$1029220 & UCAC4 398-003401 & Magellan/MIKE & 2010 Jan 01 & 16.85$\pm$0.43 & \dots \\
02492136$-$4416063 & \dots & Magellan/MIKE & 2010 Jan 01 & 29.06$\pm$0.31 & \dots \\ 
02511150$-$4753077 & GSC 08054-00859 & DuPont/Echelle & 2013 Dec 19 & -15.11$\pm$0.88 & \dots \\
02523096$-$1548357 & \dots & Keck/HIRES & 2012 Dec 28 & 73.68$\pm$0.59 & \dots \\ 
02535980$+$3206373 & \dots & Keck/HIRES & 2006 Aug 12 & -36.26$\pm$0.81 & \sout{SB1} \\
02535980$+$3206373 & \dots & Keck/HIRES & 2012 Dec 28 & -34.64$\pm$0.17 & \sout{SB1} \\
02545247$-$0709255 & \dots & Keck/HIRES & 2012 Dec 28 & 25.11$\pm$0.18 & SB2 \\ 
02560388$-$0036332 & LP 591-156 & DuPont/Echelle & 2009 Aug 21 & 46.29$\pm$0.38 & \sout{SB1} \\
02560388$-$0036332 & LP 591-156 & Magellan/MIKE & 2009 Dec 31 & 46.04$\pm$0.40 & \sout{SB1} \\
03033668$-$2535329 & CD-26 1122 & Magellan/MIKE & 2012 Jan 16 & 16.94$\pm$0.71 & \dots \\
03051118$-$3405239 & LP 942-107 & DuPont/Echelle & 2013 Dec 18 & 24.32$\pm$0.64 & \dots \\  
\enddata
\tablecomments{Table 1 is available in its entirety in a machine-readable form in the online journal. A portion is shown here for guidance regarding its form and content.}
\end{deluxetable*}

\section{Analysis}

\subsection{Spectral Types}

Spectral types are taken from the literature, with references given in Table 2.  For those objects without an available spectral type, we measure the type using the TiO5 index defined in \cite{reid95}. The spectral type distribution of the entire sample is shown in Figure 1.  

\begin{figure}
\plotone{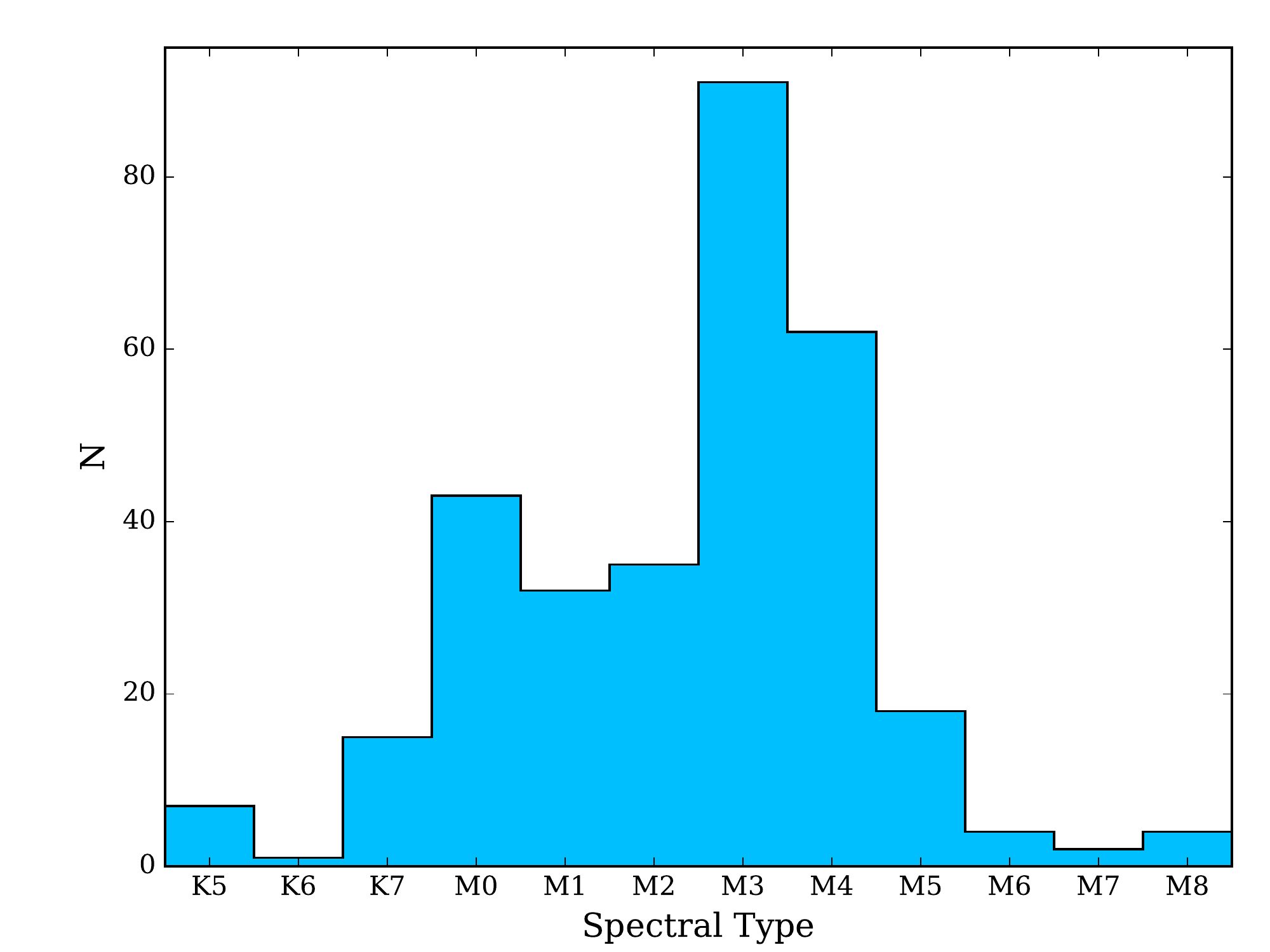}
\caption{The spectral type distribution of our sample.}
\end{figure}

\subsection{Radial Velocities}
One of the main objectives of this work is to measure radial velocities for potentially young, nearby, low mass stars. For each of our high-resolution spectra, we isolate the reddest orders ($\gtrsim$7000 \AA) for radial velocity measurements as that is where the signal to noise ratio (S/N) is highest for low mass stars.  Each order was cross-correlated with radial velocity standards of similar spectral type using the {\it fxcor} routine from IRAF.  Radial velocities were measured by finding the Gaussian peak of the cross-correlation function in each order.  We average the radial velocities of all orders to determine a final value, and use the standard deviation of the radial velocity from each order as the measurement uncertainty.  All measured radial velocities are given in Table 1.  The median uncertainty for our radial velocity measurements is $\sim$0.7 km s$^{-1}$.  Objects found to be likely spectroscopic binaries via multiple resolved peaks in the resulting cross-correlations functions are noted in Table 1 and will be discussed further in a future publication.

In Table 2, we combine our radial velocity measurements with astrometry from {\it Gaia} DR2 \citep{gaia18} to calculate full XYZUVW kinematics for each object in our sample.  For objects with more than one radial velocity measurement, we take the weighted average.  For some objects, astrometry from {\it Gaia} DR2 was unavailable.  In these instances, we filled in astrometry from other sources, as described in the Table 2 astrometry notes.

\startlongtable
\begin{deluxetable}{llccccccc}
\tablecaption{Sample Properties }
\tablehead{
\colhead{Column Label} & \colhead{Description}  \\}
\startdata
2MASS & 2MASS Designation \\
SpT & Spectral Type \\
SpT\_ref & Reference for SpT \\
Dist & Distance from Gaia DR2 \\
e\_Dist & Error in distance \\
$\mu$$_{\alpha}$ & Proper motion in right ascension \\
e\_$\mu$$_{\alpha}$ & Error in $\mu$$_{\alpha}$ \\
$\mu$$_{\delta}$ & Proper motion in declination \\ 
e\_$\mu$$_{\delta}$ & Error in $\mu$$_{\delta}$ \\
ast\_note & Astrometry note \\
RV & Radial velocity \\
e\_RV & Error in radial velocity \\
X & X position \\ 
e\_X & Error in X \\ 
Y & Y position \\ 
e\_Y & Error in Y \\ 
Z & Z position \\
e\_Z & Error in Z \\
U & U velocity \\ 
e\_U & Error in U \\ 
V & V velocity \\ 
e\_V & Error in V \\ 
W & W velocity \\
e\_W & Error in W \\
Gmag & {\it Gaia} G magnitude \\
e\_Gmag & Error in Gmag \\
RPmag & {\it Gaia} G$_{\rm RP}$ magnitude \\
e\_RPmag & Error in RPmag \\
BPmag & {\it Gaia} G$_{\rm BP}$ magnitude \\
e\_BPmag & Error in BPmag \\
Jmag & 2MASS J magnitude \\
e\_Jmag & Error in Jmag \\
Hmag & 2MASS H magnitude \\
e\_Hmag & Error in Hmag \\
Kmag & 2MASS K$_{\rm S}$ magnitude \\
e\_Kmag & Error in Kmag \\
Li & Lithium $\lambda$6707 equivalent width \\
H$\alpha$ & H$\alpha$ equivalent width \\ 
log $L_{\rm X}$ & X-ray luminosity \\
e\_log $L_{\rm X}$ & Error in log $L_{\rm X}$ \\
FUV & GALEX FUV magnitude \\
e\_FUV & Error in FUV \\
$f_{\rm FUV}$/$f_{\rm J}$ & FUV to J flux ratio \\
NUV & GALEX NUV magnitude \\
e\_NUV & Error in NUV \\
$f_{\rm NUV}$/$f_{\rm J}$ & NUV to J flux ratio \\
\enddata
\tablecomments{This table is available in its entirety in a machine-readable form in the online journal.}
\end{deluxetable}

We searched the literature for previously measured radial velocities for our targets.  We found 540 individual radial velocity measurements for 241 of the targets in our sample.  Thus, 91 of our targets have radial velocities presented for the first time.  In many cases where a previously measured radial velocity is available, our measurement is of significantly higher precision.  All previously measured radial velocities are given in Table 3. Excluding spectroscopic binaries noted in Table 1, the median $\sigma$ difference (difference divided by combined uncertainty) for all previously measured radial velocities compared to our measurements is 2.45$\sigma$.  Note that cases with large radial velocity differences may not necessarily indicate an erroneous measurement, as they could instead be evidence of spectroscopic binarity. 

\begin{figure*}
\plotone{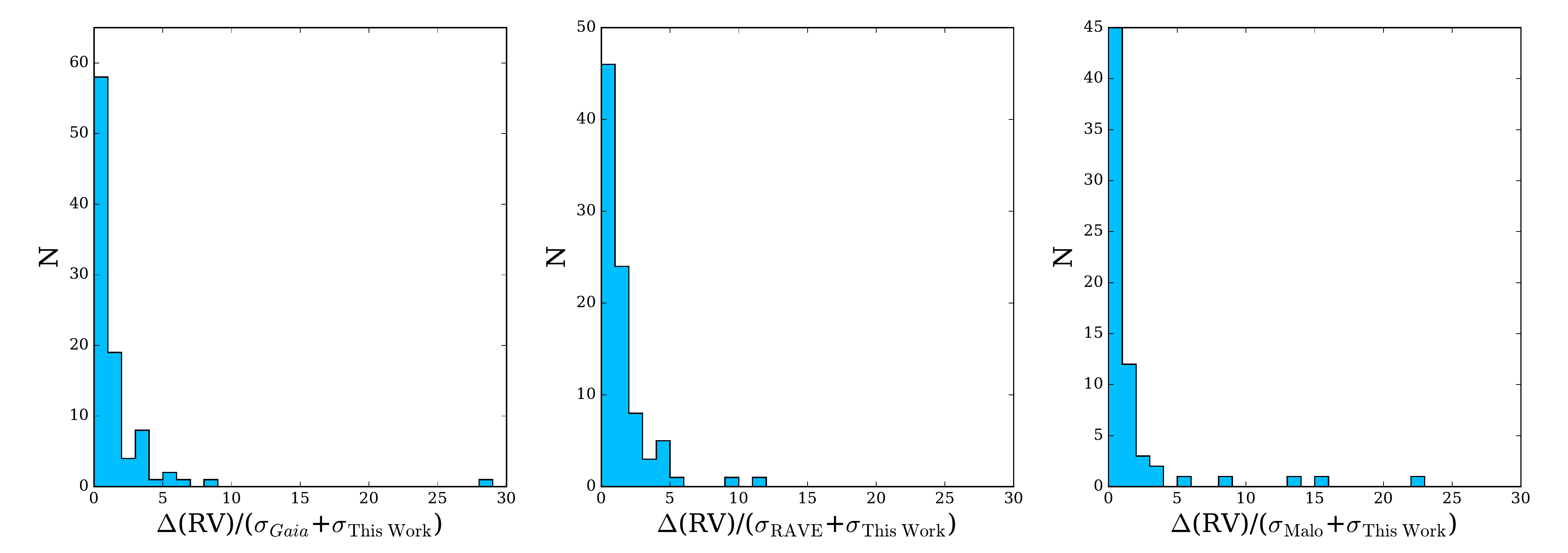}
\caption{Comparison of radial velocities from this work with those determined for the same stars from {\it Gaia} DR2 (left), RAVE (center) and \cite{malo14} (right).  The abscissa corresponds to the absolute difference between radial velocity measurements divided by the combined uncertainties for those measurements.  Objects noted as spectroscopic binaries in Table 1 are not included in this figure. }  
\end{figure*}

\begin{deluxetable}{crrrrrrrrr}
\tablecaption{RV Comparisons}
\tablehead{
\colhead{2MASS} & \colhead{RV$_{\rm This\ Work}$} & \colhead{RV$_{\rm Lit}$} & \colhead{Ref} \\
\colhead{Name} & \colhead{(km s$^{-1}$)} & \colhead{(km s$^{-1}$)}  &   }
\startdata
00153670$-$2946003 & 0.61$\pm$1.31 & 63.4 & 7 \\
\dots & 0.61$\pm$1.31 & 0.02$\pm$4.84 & 61 \\
00213729$-$4605331 & -19.32$\pm$0.12 & 49.2 & 7 \\
\dots & -19.32$\pm$0.12 & -30.70$\pm$0.32 & 24 \\
00501079$-$0337532 & 0.27$\pm$1.24 & 5.16$\pm$2.98 & 61 \\
\dots & 0.27$\pm$1.24 & 5.95$\pm$4.46 & 64 \\
00503319$+$2449009 & 5.69$\pm$0.35 & 9.2 & 6 \\
\dots & 5.69$\pm$0.35 & 6.2 & 15 \\
\dots & 5.69$\pm$0.35 & 6.00$\pm$1.10 & 23 \\
\dots & 5.69$\pm$0.35 & 6.40$\pm$11.50 & 22 \\
\dots & 5.69$\pm$0.35 & 5.60$\pm$1.00 & 16 \\
\enddata
\tablerefs{(1) \cite{wilson67}; (2) \cite{wool70}; (3) \cite{och80}; (4) \cite{and85}; (5) \cite{gliese91}; (6) \cite{reid95}; (7) \cite{hawley96}; (8) \cite{up96}; (9) \cite{cov97}; (10) \cite{thack97}; (11) \cite{del98}; (12) \cite{strass00}; (13) \cite{torres00}; (14) \cite{montes01}; (15) \cite{gizis02}; (16) \cite{moch02}; (17) \cite{song02}; (18) \cite{massey03}; (19) \cite{nord04}; (20) \cite{boch05}; (21) \cite{scholz05}; (22) \cite{boby06}; (23) \cite{gont06}; (24) \cite{torres06}; (25) \cite{blake07}; (26) \cite{gue07}; (27) \cite{kha07}; (28) \cite{fern08}; (29) \cite{west08}; (30) \cite{hue09}; (31) \cite{lep09}; (32) \cite{rein09}; (33) \cite{blake10}; (34) \cite{mal10}; (35) \cite{seif10}; (36) \cite{kiss11}; (37) \cite{rod11}; (38) \cite{west11}; (39) \cite{chub12}; (40) \cite{desh12}; (41) \cite{sacco12}; (42) \cite{malo13}; (43) \cite{moor13}; (44) \cite{binks14}; (45) \cite{ell14}; (46) \cite{malo14}; (47) \cite{newt14}; (48) \cite{burg15}; (49) \cite{desi15}; (50) \cite{guo15}; (51) \cite{luo15}; (52) \cite{ter15a}; (53) \cite{ter15b}; (54) \cite{west15}; (55) \cite{zhong15}; (56) \cite{binks16}; (57) \cite{brew16}; (58) \cite{fah16}; (59) \cite{luo16}; (60) \cite{sper16}; (61) \cite{kunder17}; (62) \cite{riedel17}; (63) \cite{frasca18}; (64) \cite{gaia18}; (65) \cite{rein18}  }
\tablecomments{Table 3 is available in its entirety in a machine-readable form in the online journal. A portion is shown here for guidance regarding its form and content.}
\end{deluxetable}

Three catalogs constituted more than half of the previously reported radial velocities of our targets; \cite{malo14}, the 5th data release of the Radial Velocity Experiment (RAVE; \citealt{kunder17}), and {\it Gaia} DR2 \citep{gaia18}. The number of objects in common with {\it Gaia} DR2 are 99, with a median difference of 2.06$\sigma$.  The number of objects that also have radial velocities in RAVE is 93, with a median difference of 3.91$\sigma$.  The number of objects in common with \cite{malo14} is 74, with a median difference of 0.58$\sigma$.  A comparison of our measured radial velocities with those from {\it Gaia} DR2, RAVE, and \cite{malo14} is shown in Figure 2.  The median radial velocity uncertainties for {\it Gaia} DR2, RAVE, and \cite{malo14} measurements are 1.8 km s$^{-1}$, 3.9 km s$^{-1}$, and 0.5 km s$^{-1}$, respectively.  

\subsection{Age Diagnostics}
For each object in our sample, we measure H$\alpha$ and Li $\lambda$6707 equivalent widths and report the values in Table 2.  The detection limit for Li equivalent widths is taken to be 0.05 \AA\ following \cite{shk09}.  Because we aimed for the S/N for each observed spectrum, and the resolutions are similar for each spectrograph, this limit applies to all observations. We also gathered X-ray data from ROSAT \citep{trump82} using the Second ROSAT all-sky survey (2RXS; \citealt{boller16}).  Each detection was converted to an X-ray flux using equations (5) and (6) from \cite{riaz06}, which was then converted to an X-ray luminosity with the following equation:

\begin{equation}
L_{\rm X} = 1.2^{\rm 38} f_{\rm X} d^{\rm 2}
\end{equation}

\noindent which comes from NASA's High Energy Astrophysics Science Archive Research Center (HEASARC)\footnote{\url https://heasarc.gsfc.nasa.gov/W3Browse/all/rassdsstar.html}.

We also include FUV and NUV photometry from the {\it Galaxy Evolution Explorer (GALEX)}, as young ($<$300 Myr) moving group members have been shown to have higher UV flux levels than field objects in the same mass range (e.g., \citealt{rod11}, \citealt{shk11}, \citealt{schneid18}).  

We also provide {\it Gaia} DR2 photometry for the entire sample in Table 2, which is used to construct color-magnitude diagrams in subsequent sections.  2MASS JHK photometry is also provided for convenience.  

\section{Discussion}

\subsection{Objects with Ongoing Accretion}
Strong H$\alpha$ emission can signify ongoing accretion on low-mass stars. We identify objects that are potentially accreting using the empirical accretion boundary for H$\alpha$ equivalent widths presented in \cite{barr03}.  Note that the empirical relation in \cite{barr03} is only applicable for objects with spectral types earlier than M5.5.  Figure 3 shows the H$\alpha$ measurements for our entire sample sorted by spectral type.  Only one object in our sample with a spectral type later than M5.5 shows strong H$\alpha$ absorption ($\geq$ -20 \AA), the well known flare star DG Phe.  Five objects are noted to be above the accretion boundary in the spectral type range K5 to M5.5.  However, previous work has shown that the boundary defined in \cite{barr03} does not exclusively identify accretors, as several well known non-accreting flare stars lie above the curve \citep{shk09}.  Therefore, for any object lying above the \cite{barr03} boundary, we also evaluate the 10\%-width defined in \cite{white03}, which has been shown to better identify accreting objects.  \cite{white03} define an accretion boundary for the 10\%-width of $>$200 km s$^{-1}$ for non-optically veiled stars. 

Of the five objects above the \cite{barr03} curve, one object is the known accretor and $\beta$ Pic member V4046 Sgr (2MASS J18141047$-$3247344), with a spectral type of K5+K7 for which we measure a 10\%-width of 447 km s$^{-1}$.  2MASS 05082729$-$2101444 is a second $\beta$ Pic member, confirmed in Section 4.2.3, with a 10\%-width of 197 km s$^{-1}$, right at the \cite{white03} accretion boundary.  2MASS 05425587$-$0718382 (aka V* V909 Ori) is another well-known flare star just beyond the accretion boundary in Figure 3, and with a 10\%-width of 241 km s$^{-1}$, is likely accreting.  This object is rejected as a potential Columba member in Section 4.2.6., though is likely young based on its position in color-magnitude diagrams and its H$\alpha$ emission.  2MASS 08440914-7833457 is a member of the $\sim$11 Myr old $\eta$ Cha cluster in Section 4.2.8, and is accreting, with a 10\%-width of 212 km s$^{-1}$.  2MASS 06002304$-$4401217 is an object reclassified as a $\sim$45 Myr old member of the Tucana-Horologium Association (see Section 4.2.10), though has a 10\%-width of only 121 km s$^{-1}$, and is thus not likely to be accreting.  

\begin{figure}
\plotone{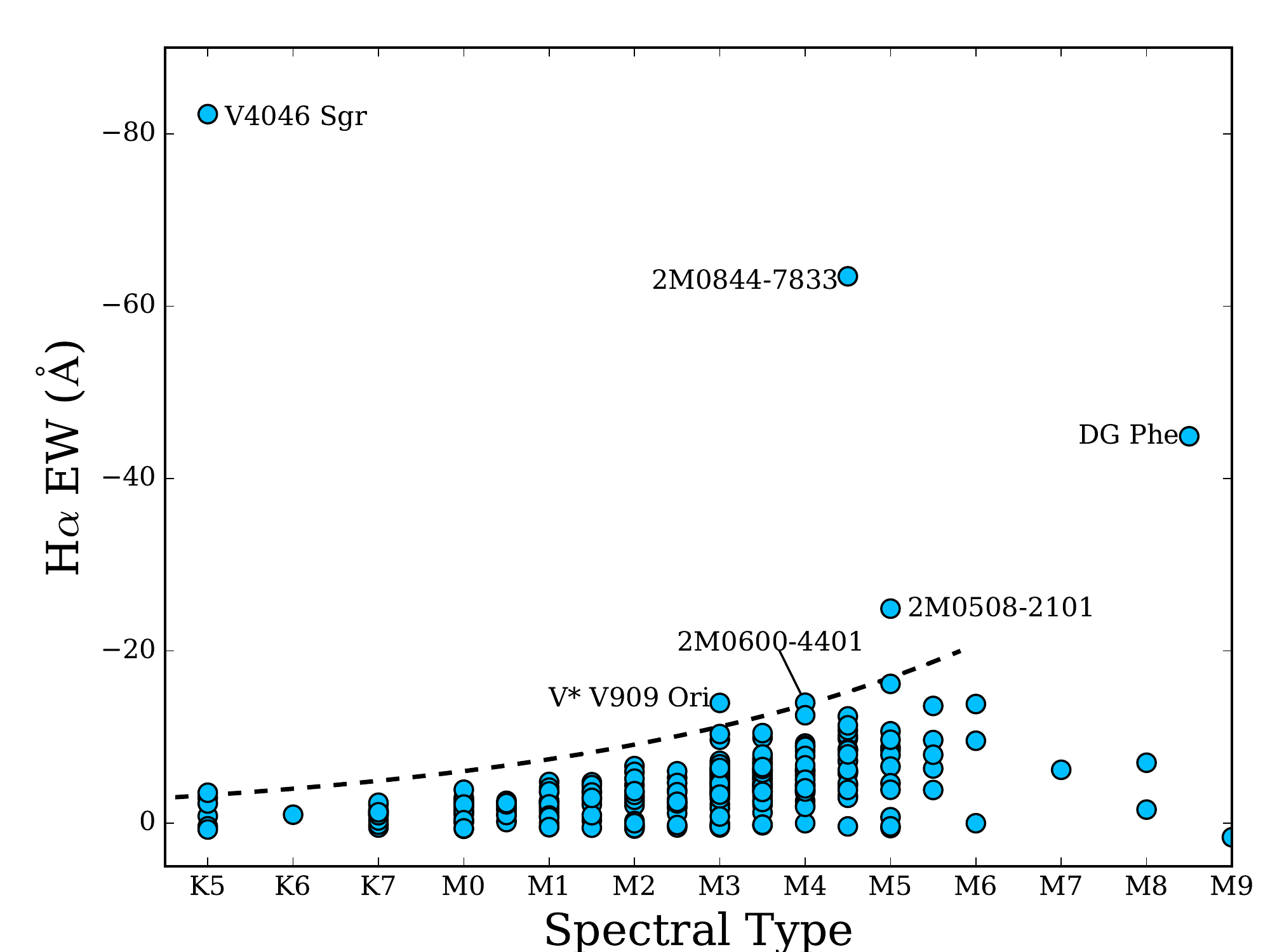}
\caption{H$\alpha$ equivalent width versus spectral type for our entire sample.  The dashed line represents the empirical accretor/non-accretor boundary defined in \cite{barr03}.  All objects on the accretor side of the boundary are labeled and discussed further in the text.  }
\end{figure}

\subsection{Co-Moving Companions}
Co-moving systems are valuable benchmarks for testing formation and evolutionary models because the objects in such systems share many of the same physical parameters, such as age, distance, and composition.  The astrometric data from {\it Gaia DR2} allows for a search for co-moving companions to each object in our sample.  For each object, we chose a search radius corresponding to 0.1 pc separation at the object's distance.  This separation corresponds to a $>$50\% survivability rate according to \cite{weinberg87}.  However, we note that co-moving systems with much larger separations are known to exist (e.g., \citealt{oh17}) and thus our search radius can be thought of as conservative.  The average distance for objects in our sample is $\sim$34 pc, which would correspond to a search radius of $\sim$10\arcmin.  Using the appropriate search radii, we investigated for objects with parallaxes within 0.1 pc of each object in our sample.    

To ensure objects are co-moving, we first verify that any potential companions have similar parallaxes by requiring $\Delta$$D$ - 3$\sigma$$_{D}$ $<$ 1 pc (e.g., \citealt{hollands18}), where $\Delta$$D$ is the difference between measured distances, and $\sigma$$_{D}$ is the combined distance uncertainty for the target and potential companion.  We then ensure similar proper motions by requiring $\Delta$$\mu$/$\mu$ $<$ 0.21, where $\Delta$$\mu$ $\equiv$ ($\Delta$$\mu_{\alpha}$ + $\Delta$$\mu_{\delta}$)$^{\onehalf}$ (e.g., \citealt{dup12}).  Note that this proper motion condition for companionship is valid for proper motions greater than $\sim$150 mas yr$^{-1}$.  We relax this condition to $\Delta$$\mu$/$\mu$ $<$ 0.31 for smaller proper motions.  Only two systems required this lower proper motion threshold (systems 9 and 29 in Table 4). 

A total of 69 co-moving systems were found, and are listed in Table 4.  Systems containing multiple objects in our target list are only listed once.  Sixty-three of the sixty-nine systems contain a single co-moving companion, while the remaining 6 systems contain 3 or more components.  Figure 4 shows color-magnitude diagrams (CMDs) of all companions using {\it Gaia} DR2 parallaxes and photometry.  The panels show G, G$_{\rm BP}$, and G$_{\rm RP}$ photometry($\lambda$$_{0}$ = 673, 532, and 797 nm, respectively; \citealt{jordi10}).  The figure shows that the spectral types of the companions found to objects in our sample are a mix of earlier and later spectral types, along with a handful of white dwarfs.   

\begin{figure*}
\plotone{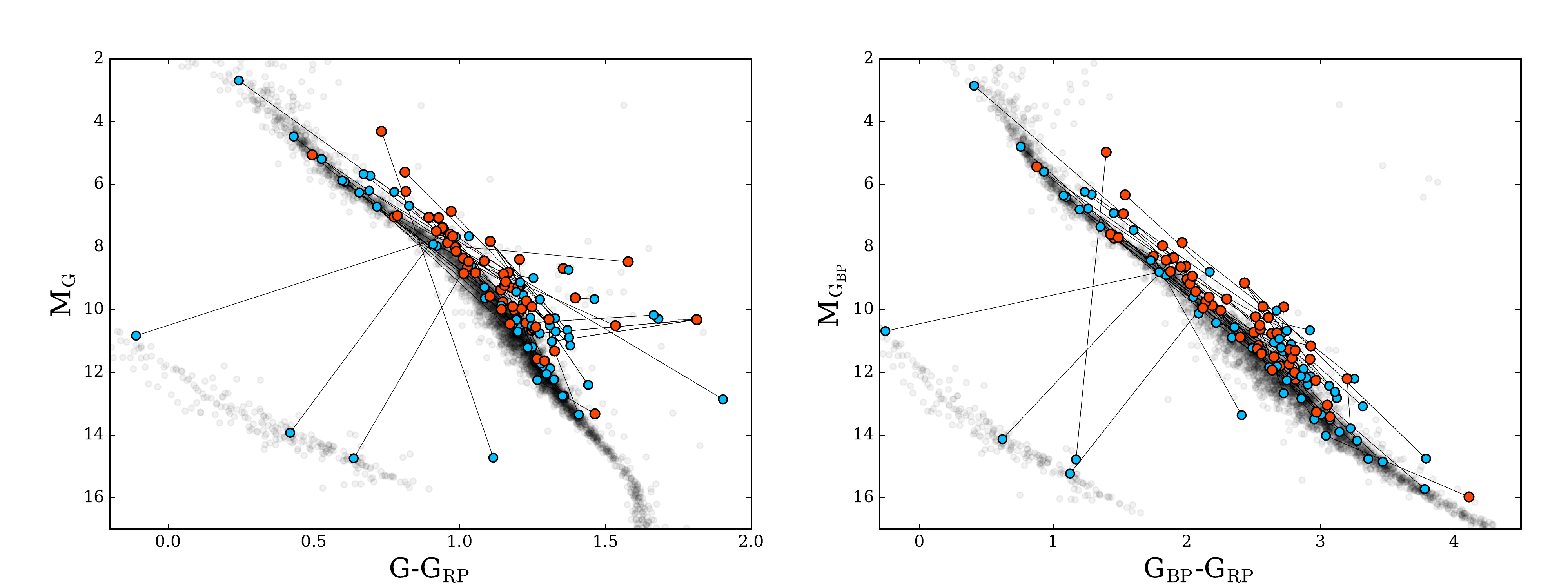}
\caption{{\it Gaia} CMDs of all co-moving systems found in this work.  Objects in our sample are displayed in red, while companions are light blue.    For reference, we show all objects from {\it Gaia} DR2 within 25 pc as background gray symbols. }
\end{figure*}

\startlongtable
\begin{deluxetable*}{lrrrrrrrr}
\tablecaption{Co-Moving Systems}
\tablehead{
\colhead{Sys.} & \colhead{Comp.} & \colhead{Gaia R.A.} & \colhead{Gaia Dec.} & \colhead{plx} & \colhead{$\mu$$_{\delta}$} & \colhead{$\mu$$_{\alpha}$} & \colhead{Separation} & \colhead{Separation}  \\
\colhead{No.} & \colhead{No.} & \colhead{($\degr$)} & \colhead{($\degr$)} & \colhead{(mas)} & \colhead{(mas yr$^{-1}$)} & \colhead{(mas yr$^{-1}$)} & \colhead{(\arcsec)} & \colhead{(AU)}  }
\startdata
1 & 1.1 & 6.45495504961 & 45.025376915 & 15.8615$\pm$0.0801 & 102.891$\pm$0.08 & -9.526$\pm$0.036 & \dots & \dots \\
\dots & 1.2 & 6.45743133884 & 45.0269135029 & 16.0416$\pm$0.1055 & 102.809$\pm$0.111 & -8.653$\pm$0.05 & 8.4 & 523 \\
2 & 2.1 & 14.024420588 & 41.891185381 & 16.331$\pm$0.0607 & -78.979$\pm$0.103 & 4.156$\pm$0.105 & \dots & \dots \\
\dots & 2.2 & 14.019747768 & 41.8900123923 & 15.9529$\pm$0.0489 & -83.873$\pm$0.066 & 3.93$\pm$0.081 & 13.2 & 828 \\
3 & 3.1 & 15.8126716282 & 20.0984581507 & 62.1$\pm$0.2033 & 677.548$\pm$0.436 & 26.765$\pm$0.296 & \dots & \dots \\
\dots & 3.2 & 15.812059101 & 20.0980757415 & 62.627$\pm$0.0425 & 672.249$\pm$0.079 & 50.851$\pm$0.067 & 2.5 & 40 \\
4 & 4.1 & 18.3735917791 & -7.63608776673 & 15.3648$\pm$0.0516 & 74.651$\pm$0.075 & -68.292$\pm$0.06 & \dots & \dots \\
\dots & 4.2 & 18.3729635322 & -7.63663544495 & 15.6057$\pm$0.1904 & 70.301$\pm$0.259 & -66.398$\pm$0.207 & 3.0 & 191 \\
5 & 5.1 & 20.2718431581 & -4.03570030568 & 11.1213$\pm$0.0475 & 169.306$\pm$0.085 & -8.59$\pm$0.053 & \dots & \dots \\
\dots & 5.2 & 20.2727145599 & -4.03661690337 & 10.9178$\pm$0.0791 & 171.761$\pm$0.148 & -8.094$\pm$0.074 & 4.5 & 417 \\
6 & 6.1 & 27.1710820305 & -48.5146518446 & 25.5403$\pm$0.0235 & 110.899$\pm$0.037 & -53.968$\pm$0.036 & \dots & \dots \\
\dots & 6.2 & 27.1995848064 & -48.5212175903 & 25.6167$\pm$0.0416 & 110.54$\pm$0.058 & -53.667$\pm$0.061 & 72.0 & 2809 \\
7 & 7.1 & 33.4650594831 & -41.491991267 & 11.9977$\pm$0.0246 & 56.572$\pm$0.03 & -18.784$\pm$0.035 & \dots & \dots \\
\dots & 7.2 & 33.4655848276 & -41.4911563705 & 11.9931$\pm$0.0188 & 54.584$\pm$0.023 & -19.257$\pm$0.027 & 3.3 & 277 \\
8 & 8.1 & 34.2288032428 & -23.3703726789 & 13.7216$\pm$0.0778 & 26.234$\pm$0.146 & 1.122$\pm$0.14 & \dots & \dots \\
\dots & 8.2 & 34.2278548085 & -23.369533735 & 13.6387$\pm$0.0468 & 21.64$\pm$0.09 & -0.189$\pm$0.083 & 4.4 & 319 \\
9 & 9.1 & 37.306737751 & -10.1615812986 & 16.3434$\pm$0.076 & -34.882$\pm$0.106 & -18.49$\pm$0.105 & \dots & \dots \\
\dots & 9.2 & 37.3065913493 & -10.1620032446 & 16.7055$\pm$0.0835 & -25.918$\pm$0.131 & -10.12$\pm$0.117 & 1.6 & 96 \\
10 & 10.1 & 40.3298322728 & -57.4217128511 & 11.3641$\pm$0.0451 & 14.729$\pm$0.072 & 29.64$\pm$0.081 & \dots & \dots \\
\dots & 10.2 & 40.3290993909 & -57.4215856312 & 11.6443$\pm$0.0783 & 17.917$\pm$0.103 & 30.519$\pm$0.132 & 1.5 & 128 \\
11 & 11.1 & 41.0893497836 & 10.9611708044 & 20.5066$\pm$0.1372 & 67.385$\pm$0.285 & -56.036$\pm$0.242 & \dots & \dots \\
\dots & 11.2 & 41.0950166414 & 10.9594993429 & 21.3227$\pm$0.0865 & 73.847$\pm$0.178 & -58.413$\pm$0.146 & 20.9 & 981 \\
12 & 12.1 & 43.718796959 & -7.15734233716 & 21.5234$\pm$0.0666 & 38.938$\pm$0.149 & -55.285$\pm$0.13 & \dots & \dots \\
\dots & 12.2 & 43.7176761993 & -7.15704206394 & 21.3807$\pm$0.0896 & 34.261$\pm$0.194 & -64.692$\pm$0.165 & 4.1 & 194 \\
13 & 13.1 & 51.0280110852 & 23.7845173999 & 48.2954$\pm$0.0494 & 215.05$\pm$0.1 & -120.236$\pm$0.07 & \dots & \dots \\
\dots & 13.2 & 51.0277404632 & 23.7851922099 & 48.0921$\pm$0.0617 & 200.851$\pm$0.127 & -114.325$\pm$0.089 & 2.6 & 54 \\
\dots & 13.3 & 51.0545524391 & 23.7714861393 & 48.4516$\pm$0.0677 & 205.518$\pm$0.104 & -110.346$\pm$0.08 & 99.2 & 2048 \\
14 & 14.1 & 53.0148285759 & -51.6652419697 & 13.0328$\pm$0.0228 & 50.184$\pm$0.039 & 7.261$\pm$0.049 & \dots & \dots \\
\dots & 14.2 & 53.016112676 & -51.6656193486 & 13.0197$\pm$0.0449 & 50.644$\pm$0.079 & 6.542$\pm$0.097 & 3.2 & 244 \\
15 & 15.1 & 54.9491321525 & 33.4751835323 & 25.465$\pm$0.119 & -38.631$\pm$0.135 & -3.727$\pm$0.109 & \dots & \dots \\
\dots & 15.2 & 54.9538270286 & 33.4734135346 & 25.4853$\pm$0.0849 & -35.974$\pm$0.111 & -2.512$\pm$0.083 & 15.5 & 607 \\
16 & 16.1 & 64.4876758851 & -38.4513485042 & 25.5609$\pm$0.5218 & -95.676$\pm$1.166 & -64.997$\pm$1.007 & \dots & \dots \\
\dots & 16.2 & 64.4923466117 & -38.4293196842 & 26.5906$\pm$0.0688 & -100.751$\pm$0.111 & -67.663$\pm$0.142 & 80.4 & 3023 \\
17 & 17.1 & 71.0441624381 & -70.323716455 & 47.5062$\pm$0.0253 & -120.466$\pm$0.046 & -65.845$\pm$0.055 & \dots & \dots \\
\dots & 17.2 & 71.0450943035 & -70.3243619533 & 47.4314$\pm$0.0646 & -102.042$\pm$0.149 & -46.054$\pm$0.195 & 2.6 & 54 \\
18 & 18.1 & 77.5207049404 & -23.6709966491 & 17.3479$\pm$0.0408 & 38.099$\pm$0.06 & -9.074$\pm$0.062 & \dots & \dots \\
\dots & 18.2 & 77.5202711962 & -23.6706992996 & 17.275$\pm$0.0487 & 36.988$\pm$0.073 & -13.443$\pm$0.071 & 1.8 & 103 \\
19 & 19.1 & 78.2567403265 & -70.4607095853 & 30.9077$\pm$0.0738 & 107.157$\pm$0.157 & 234.762$\pm$0.191 & \dots & \dots \\
\dots & 19.2 & 78.257702683 & -70.4603692812 & 31.0644$\pm$0.0557 & 119.442$\pm$0.125 & 223.431$\pm$0.113 & 1.7 & 54 \\
20 & 20.1 & 78.6142599655 & -15.247715066 & 13.145$\pm$0.0414 & 34.552$\pm$0.06 & -15.176$\pm$0.07 & \dots & \dots \\
\dots & 20.2 & 78.6134888585 & -15.2477399307 & 13.1076$\pm$0.0653 & 35.722$\pm$0.096 & -13.416$\pm$0.109 & 2.7 & 204 \\
\dots & 20.3 & 78.6201038834 & -15.2485743739 & 13.1976$\pm$0.0467 & 33.49$\pm$0.068 & -15.008$\pm$0.079 & 20.5 & 1556 \\
21 & 21.1 & 79.9874389463 & -11.4123560204 & 14.1972$\pm$0.0438 & 27.299$\pm$0.07 & -21.627$\pm$0.069 & \dots & \dots \\
\dots & 21.2 & 79.982721039 & -11.4159132296 & 14.4424$\pm$0.0627 & 26.531$\pm$0.097 & -22.415$\pm$0.093 & 21.0 & 1454 \\
22 & 22.1 & 87.5469822497 & 9.66803304279 & 39.7713$\pm$0.182 & 257.061$\pm$0.275 & 239.913$\pm$0.21 & \dots & \dots \\
\dots & 22.2 & 87.55008623 & 9.66473068132 & 39.6816$\pm$0.1052 & 263.405$\pm$0.142 & 239.802$\pm$0.116 & 16.2 & 408 \\
23 & 23.1 & 90.0958524601 & -44.0225800557 & 14.4695$\pm$0.0525 & 25.378$\pm$0.086 & 15.806$\pm$0.097 & \dots & \dots \\
\dots & 23.2 & 90.0965597837 & -44.0227244432 & 14.4132$\pm$0.0538 & 23.151$\pm$0.087 & 15.075$\pm$0.113 & 1.9 & 132 \\
24 & 24.1 & 91.5725680245 & -27.9012654117 & 27.9688$\pm$0.0376 & -10.53$\pm$0.062 & 32.007$\pm$0.078 & \dots & \dots \\
\dots & 24.2 & 91.5691507809 & -27.9056805417 & 27.9831$\pm$0.0294 & -16.297$\pm$0.039 & 34.359$\pm$0.045 & 19.3 & 688 \\
25 & 25.1 & 93.4239587636 & -28.2545771594 & 24.2785$\pm$0.31 & 27.198$\pm$0.576 & 86.668$\pm$0.516 & \dots & \dots \\
\dots & 25.2 & 93.4239732732 & -28.2542821755 & 23.3908$\pm$0.2541 & 31.168$\pm$0.488 & 90.478$\pm$0.491 & 1.1 & 45 \\
\dots & 25.3 & 93.506342479 & -28.2611592687 & 23.606$\pm$0.0791 & 27.352$\pm$0.11 & 90.76$\pm$0.128 & 262.3 & 11112 \\
\dots & 25.4 & 93.5068390109 & -28.2606529165 & 23.8732$\pm$0.0426 & 26.88$\pm$0.059 & 92.051$\pm$0.069 & 263.7 & 11047 \\
26 & 26.1 & 93.4385067821 & -23.8614582851 & 59.7617$\pm$0.0243 & -47.733$\pm$0.026 & 110.9$\pm$0.042 & \dots & \dots \\
\dots & 26.2 & 93.4464230422 & -23.9064161051 & 60.0064$\pm$0.0432 & -31.043$\pm$0.05 & 110.803$\pm$0.07 & 163.9 & 2732 \\
27 & 27.1 & 98.8201005228 & -19.1919537555 & 8.2974$\pm$0.0336 & -10.615$\pm$0.05 & -28.548$\pm$0.064 & \dots & \dots \\
\dots & 27.2 & 98.8200891154 & -19.1923472134 & 8.2239$\pm$0.0418 & -8.326$\pm$0.067 & -26.52$\pm$0.066 & 1.4 & 172 \\
28 & 28.1 & 99.5012976895 & -40.9331925133 & 25.9455$\pm$0.0339 & -0.901$\pm$0.072 & 100.528$\pm$0.064 & \dots & \dots \\
\dots & 28.2 & 99.4934539826 & -40.9322052962 & 25.9428$\pm$0.0551 & 1.569$\pm$0.106 & 102.576$\pm$0.101 & 21.6 & 834 \\
29 & 29.1 & 107.99652442 & -35.1711642829 & 27.0922$\pm$0.0767 & -18.251$\pm$0.197 & -59.482$\pm$0.158 & \dots & \dots \\
\dots & 29.2 & 107.996305062 & -35.1713921997 & 27.3922$\pm$0.5966 & -34.898$\pm$0.937 & -50.518$\pm$1.01 & 1.0 & 38 \\
30 & 30.1 & 122.46684401 & 3.01931596993 & 10.2325$\pm$0.0356 & -15.159$\pm$0.064 & -39.245$\pm$0.04 & \dots & \dots \\
\dots & 30.2 & 122.467210115 & 3.01985424758 & 10.2347$\pm$0.0379 & -17.262$\pm$0.065 & -39.799$\pm$0.042 & 2.3 & 229 \\
\dots & 30.3 & 122.474103517 & 3.01885247148 & 10.3936$\pm$0.0544 & -16.952$\pm$0.099 & -37.131$\pm$0.063 & 26.2 & 2516 \\
31 & 31.1 & 124.747394623 & -72.6653660338 & 12.3669$\pm$0.0194 & -22.13$\pm$0.034 & 53.681$\pm$0.037 & \dots & \dots \\
\dots & 31.2 & 124.558006465 & -72.6262036884 & 12.4117$\pm$0.0409 & -21.677$\pm$0.073 & 53.863$\pm$0.083 & 247.5 & 19937 \\
32 & 32.1 & 125.694579057 & -57.4459684881 & 77.4877$\pm$0.0716 & -372.239$\pm$0.157 & 481.326$\pm$0.157 & \dots & \dots \\
\dots & 32.2 & 125.696502898 & -57.443984652 & 77.4495$\pm$0.0706 & -371.155$\pm$0.165 & 437.05$\pm$0.158 & 8.1 & 104 \\
33 & 33.1 & 127.026873626 & -30.6754567721 & 6.9508$\pm$0.7706 & -31.647$\pm$1.355 & 10.77$\pm$1.414 & \dots & \dots \\
\dots & 33.2 & 127.029755867 & -30.6949199474 & 7.5778$\pm$0.0336 & -31.421$\pm$0.042 & 7.343$\pm$0.048 & 70.6 & 9321 \\
34 & 34.1 & 130.35521572 & -57.6006901021 & 13.8219$\pm$0.0586 & -22.963$\pm$0.13 & 22.945$\pm$0.116 & \dots & \dots \\
\dots & 34.2 & 130.355051694 & -57.6002910335 & 13.776$\pm$0.071 & -29.457$\pm$0.139 & 22.859$\pm$0.136 & 1.5 & 107 \\
35 & 35.1 & 133.075291641 & -19.2211391051 & 21.6057$\pm$0.0511 & -60.649$\pm$0.071 & 50.531$\pm$0.058 & \dots & \dots \\
\dots & 35.2 & 133.062595689 & -19.2473920359 & 21.3902$\pm$0.2603 & -59.27$\pm$0.256 & 49.141$\pm$0.299 & 103.9 & 4857 \\
36 & 36.1 & 133.186640696 & 22.5149166382 & 30.533$\pm$0.1048 & -40.493$\pm$0.158 & -153.386$\pm$0.1 & \dots & \dots \\
\dots & 36.2 & 133.185905577 & 22.5138295121 & 30.6652$\pm$0.1024 & -36.737$\pm$0.155 & -149.973$\pm$0.096 & 4.6 & 150 \\
37 & 37.1 & 134.972160736 & -20.9080635285 & 16.9812$\pm$0.0342 & 35.261$\pm$0.055 & -72.393$\pm$0.049 & \dots & \dots \\
\dots & 37.2 & 134.977646101 & -20.9081030456 & 16.7952$\pm$0.0895 & 33.204$\pm$0.155 & -72.638$\pm$0.139 & 18.4 & 1098 \\
38 & 38.1 & 136.022677642 & -15.9219506654 & 36.6103$\pm$0.0483 & -109.034$\pm$0.072 & -32.638$\pm$0.076 & \dots & \dots \\
\dots & 38.2 & 136.085742643 & -15.9143826651 & 36.52$\pm$0.0517 & -107.828$\pm$0.083 & -30.936$\pm$0.079 & 220.0 & 6025 \\
39 & 39.1 & 138.148246805 & -15.2844999181 & 22.0586$\pm$0.0493 & -51.328$\pm$0.068 & -42.01$\pm$0.093 & \dots & \dots \\
\dots & 39.2 & 138.140600035 & -15.2900770922 & 21.9064$\pm$0.0545 & -51.798$\pm$0.088 & -43.857$\pm$0.111 & 33.3 & 1520 \\
40 & 40.1 & 139.691670049 & 26.7515403197 & 40.3161$\pm$0.0716 & -196.741$\pm$0.102 & -353.518$\pm$0.072 & \dots & \dots \\
\dots & 40.2 & 139.671646501 & 26.7630015702 & 40.2542$\pm$0.1079 & -194.238$\pm$0.15 & -353.18$\pm$0.108 & 76.5 & 1899 \\
41 & 41.1 & 145.658280011 & -62.4837073856 & 23.5547$\pm$0.0847 & -149.056$\pm$0.173 & 83.372$\pm$0.168 & \dots & \dots \\
\dots & 41.2 & 145.657482461 & -62.4837573699 & 23.9316$\pm$0.0425 & -158.669$\pm$0.08 & 73.762$\pm$0.114 & 1.3 & 56 \\
\dots & 41.3 & 145.645472032 & -62.4758856653 & 23.9434$\pm$0.0238 & -157.239$\pm$0.045 & 79.715$\pm$0.044 & 35.3 & 1475 \\
42 & 42.1 & 153.089944124 & -1.47119090641 & 25.1246$\pm$0.0727 & -117.944$\pm$0.147 & -10.212$\pm$0.207 & \dots & \dots \\
\dots & 42.2 & 153.08918359 & -1.47086683642 & 25.1079$\pm$0.0777 & -112.706$\pm$0.189 & 1.903$\pm$0.268 & 3.0 & 118 \\
43 & 43.1 & 154.555865125 & -20.4775953826 & 39.4836$\pm$0.087 & -396.396$\pm$0.135 & 114.239$\pm$0.122 & \dots & \dots \\
\dots & 43.2 & 154.548627208 & -20.4720560377 & 39.3106$\pm$0.0789 & -390.304$\pm$0.132 & 116.753$\pm$0.112 & 31.5 & 802 \\
44 & 44.1 & 155.518050963 & -32.5575877603 & 8.2821$\pm$0.031 & -120.028$\pm$0.053 & -23.291$\pm$0.056 & \dots & \dots \\
\dots & 44.2 & 155.5142724 & -32.5585139454 & 7.2181$\pm$0.7932 & -120.84$\pm$1.348 & -23.821$\pm$1.464 & 11.9 & 1654 \\
45 & 45.1 & 156.355439692 & -49.3105816883 & 39.1202$\pm$0.0508 & -205.315$\pm$0.083 & 61.405$\pm$0.07 & \dots & \dots \\
\dots & 45.2 & 156.359926909 & -49.3041217036 & 39.1902$\pm$0.0416 & -201.765$\pm$0.068 & 68.964$\pm$0.057 & 25.5 & 651 \\
46 & 46.1 & 179.956109675 & -42.7405704848 & 16.1778$\pm$0.2008 & -72.299$\pm$0.253 & -16.576$\pm$0.201 & \dots & \dots \\
\dots & 46.2 & 179.970249869 & -42.733752644 & 16.1329$\pm$0.0866 & -73.037$\pm$0.105 & -18.987$\pm$0.072 & 44.7 & 2772 \\
47 & 47.1 & 182.970867544 & 12.8202838136 & 16.3878$\pm$0.0793 & -71.618$\pm$0.13 & -57.639$\pm$0.074 & \dots & \dots \\
\dots & 47.2 & 182.970853152 & 12.8199660561 & 16.1273$\pm$0.0976 & -74.267$\pm$0.18 & -63.983$\pm$0.092 & 1.1 & 71 \\
\dots & 47.3 & 183.035094024 & 12.8011015213 & 16.4079$\pm$0.0451 & -73.195$\pm$0.071 & -60.992$\pm$0.044 & 235.8 & 14371 \\
48 & 48.1 & 186.683308106 & -12.4884892808 & 33.3095$\pm$0.078 & -161.876$\pm$0.148 & -73.467$\pm$0.099 & \dots & \dots \\
\dots & 48.2 & 186.682789758 & -12.4886743198 & 32.6332$\pm$0.079 & -170.942$\pm$0.171 & -89.395$\pm$0.101 & 1.9 & 59 \\
49 & 49.1 & 203.881093361 & -62.1871496784 & 4.6833$\pm$0.2095 & -22.922$\pm$0.226 & -12.044$\pm$0.277 & \dots & \dots \\
\dots & 49.2 & 203.860170132 & -62.1851102716 & 4.0776$\pm$1.5383 & -18.962$\pm$2.782 & -11.943$\pm$3.049 & 35.9 & 8805 \\
50 & 50.1 & 207.68317614 & -21.6924529707 & 47.4932$\pm$0.0596 & -35.895$\pm$0.113 & -377.302$\pm$0.102 & \dots & \dots \\
\dots & 50.2 & 207.598875406 & -21.6238049614 & 47.5095$\pm$0.0663 & -33.504$\pm$0.111 & -376.208$\pm$0.11 & 375.0 & 7893 \\
51 & 51.1 & 220.495473154 & -16.8880611197 & 46.683$\pm$0.0522 & -170.941$\pm$0.093 & -247.728$\pm$0.081 & \dots & \dots \\
\dots & 51.2 & 220.494457166 & -16.8179334119 & 46.6443$\pm$0.0757 & -171.652$\pm$0.125 & -247.082$\pm$0.11 & 252.5 & 5413 \\
52 & 52.1 & 220.908135379 & -4.24352280153 & 19.3726$\pm$0.1992 & -101.846$\pm$0.294 & -69.609$\pm$0.275 & \dots & \dots \\
\dots & 52.2 & 220.907855485 & -4.2434345354 & 19.4511$\pm$0.1454 & -112.583$\pm$0.229 & -66.234$\pm$0.203 & 1.1 & 54 \\
53 & 53.1 & 237.528545136 & -45.4227023254 & 21.3324$\pm$0.0492 & 32.526$\pm$0.109 & -33.392$\pm$0.079 & \dots & \dots \\
\dots & 53.2 & 237.568220468 & -45.4018597943 & 21.3601$\pm$0.0533 & 39.755$\pm$0.129 & -33.474$\pm$0.068 & 125.2 & 5863 \\
54 & 54.1 & 254.334352878 & -53.7258072971 & 19.7563$\pm$0.12 & -10.849$\pm$0.187 & -84.087$\pm$0.143 & \dots & \dots \\
\dots & 54.2 & 254.339199055 & -53.7247727621 & 19.8171$\pm$0.1371 & -16.204$\pm$0.156 & -85.696$\pm$0.138 & 11.0 & 554 \\
55 & 55.1 & 261.990267808 & -40.2737800294 & 24.9806$\pm$0.0702 & -22.095$\pm$0.134 & -61.126$\pm$0.094 & \dots & \dots \\
\dots & 55.2 & 261.989622771 & -40.2736500924 & 25.025$\pm$0.065 & -16.192$\pm$0.123 & -58.788$\pm$0.087 & 1.8 & 73 \\
56 & 56.1 & 272.81636071 & -78.9886459065 & 85.2009$\pm$0.3832 & 97.719$\pm$0.717 & 275.387$\pm$0.759 & \dots & \dots \\
\dots & 56.2 & 272.814906041 & -78.9883766915 & 85.8373$\pm$0.3909 & 60.499$\pm$0.728 & 304.053$\pm$0.745 & 1.4 & 16 \\
57 & 57.1 & 273.543692455 & -32.7931482752 & 13.8111$\pm$0.0642 & 3.491$\pm$0.111 & -52.754$\pm$0.087 & \dots & \dots \\
\dots & 57.2 & 273.591990323 & -32.7697082458 & 13.9995$\pm$0.0521 & 3.081$\pm$0.101 & -52.639$\pm$0.08 & 168.8 & 12057 \\
58 & 58.1 & 279.909851791 & 16.3866196343 & 14.3647$\pm$0.0404 & -20.777$\pm$0.06 & -106.458$\pm$0.072 & \dots & \dots \\
\dots & 58.2 & 279.899837715 & 16.3820506469 & 14.515$\pm$0.1599 & -19.859$\pm$0.267 & -107.222$\pm$0.33 & 38.3 & 2639 \\
59 & 59.1 & 294.287243177 & -51.5670290492 & 20.4834$\pm$0.0421 & 91.125$\pm$0.068 & -26.99$\pm$0.048 & \dots & \dots \\
\dots & 59.2 & 294.284544043 & -51.5687325295 & 20.4788$\pm$0.0579 & 87.338$\pm$0.085 & -23.971$\pm$0.06 & 8.6 & 420 \\
60 & 60.1 & 295.853742242 & -37.3705104884 & 41.264$\pm$0.0648 & 162.885$\pm$0.098 & -183.23$\pm$0.062 & \dots & \dots \\
\dots & 60.2 & 295.853258439 & -37.3702613909 & 44.321$\pm$0.3694 & 155.208$\pm$0.539 & -182.811$\pm$0.387 & 1.6 & 37 \\
61 & 61.1 & 320.370809461 & -66.9188363806 & 31.7121$\pm$0.3965 & 105.13$\pm$0.671 & -85.324$\pm$0.782 & \dots & \dots \\
\dots & 61.2 & 320.353108276 & -66.9163705624 & 31.2692$\pm$0.0434 & 95.668$\pm$0.062 & -100.258$\pm$0.081 & 26.5 & 848 \\
62 & 62.1 & 326.832200245 & -48.0550277575 & 15.7704$\pm$0.1229 & 54.193$\pm$0.151 & -89.707$\pm$0.215 & \dots & \dots \\
\dots & 62.2 & 326.816594573 & -48.0445477226 & 15.9461$\pm$0.0552 & 49.848$\pm$0.07 & -84.906$\pm$0.096 & 53.2 & 3338 \\
63 & 63.1 & 332.922938405 & -20.7366560482 & 24.1987$\pm$0.0867 & 147.272$\pm$0.138 & -61.669$\pm$0.122 & \dots & \dots \\
\dots & 63.2 & 332.926084587 & -20.7386890403 & 24.2517$\pm$0.0877 & 146.754$\pm$0.144 & -64.44$\pm$0.126 & 12.9 & 531 \\
64 & 64.1 & 342.48416723 & 17.744598513 & 13.0626$\pm$0.2053 & -28.967$\pm$0.378 & -64.631$\pm$0.29 & \dots & \dots \\
\dots & 64.2 & 342.480450677 & 17.7442276081 & 13.6598$\pm$0.126 & -27.113$\pm$0.257 & -55.2$\pm$0.178 & 12.8 & 938 \\
65 & 65.1 & 347.973950223 & -45.1338600039 & 21.1338$\pm$0.1211 & 74.896$\pm$0.105 & -94.183$\pm$0.172 & \dots & \dots \\
\dots & 65.2 & 347.967423785 & -45.1366897831 & 20.8274$\pm$0.0683 & 87.469$\pm$0.05 & -93.499$\pm$0.091 & 19.5 & 934 \\
66 & 66.1 & 350.19685687 & -67.3896567818 & 24.3611$\pm$0.0738 & 78.28$\pm$0.12 & -81.043$\pm$0.119 & \dots & \dots \\
\dots & 66.2 & 350.197395285 & -67.3888825742 & 24.4343$\pm$0.0656 & 84.043$\pm$0.104 & -78.354$\pm$0.101 & 2.9 & 118 \\
67 & 67.1 & 351.888430293 & -22.0418804446 & 14.5199$\pm$0.1203 & 61.6$\pm$0.233 & -50.315$\pm$0.173 & \dots & \dots \\
\dots & 67.2 & 351.885024487 & -22.0388739146 & 14.577$\pm$0.0509 & 60.982$\pm$0.103 & -50.158$\pm$0.074 & 15.7 & 1077 \\
68 & 68.1 & 353.342398779 & -12.6685927216 & 27.809$\pm$0.4086 & 189.131$\pm$0.549 & 24.159$\pm$0.49 & \dots & \dots \\
\dots & 68.2 & 353.350093919 & -12.664513128 & 29.4073$\pm$0.064 & 185.523$\pm$0.091 & 24.355$\pm$0.078 & 30.8 & 1046 \\
69 & 69.1 & 358.423396079 & -65.9480499764 & 39.5909$\pm$0.0328 & -43.768$\pm$0.047 & 87.533$\pm$0.047 & \dots & \dots \\
\dots & 69.2 & 358.415716065 & -65.9467703123 & 39.5453$\pm$0.0268 & -42.373$\pm$0.039 & 84.84$\pm$0.044 & 12.2 & 308 \\
\enddata
\end{deluxetable*}

\subsection{Moving Group Membership}
We evaluate each object for potential moving group membership using the BANYAN $\Sigma$ tool \citep{gagne18a}, which evaluates the membership probabilities by comparing positions to known members of different groups using a Bayesian classifier.  BANYAN $\Sigma$ considers all kinematic information (position, proper motion, parallax, and radial velocity) to compare 6-dimensional XYZUVW coordinates to known groups using multivariate Gaussians.  While we evaluate membership in all groups included in BANYAN $\Sigma$, potential matches from our sample were found for 10 groups listed in Table 5.  Any object with a BANYAN $\Sigma$ probability of $>$1\% for a particular group was investigated for potential membership.  The TW Hya and Scorpius-Centaurus Associations are part of a larger, targeted search for new members by our research group, and will be discussed in a forthcoming separate study. 

\begin{deluxetable*}{lcccccccc}
\tablecaption{Moving Group Evaluation Summary}
\tablehead{
\colhead{Group} & \colhead{Age} & \colhead{Age} & \colhead{New\tablenotemark{a}} & \colhead{Confirmed} & \colhead{Rejected}  \\
\colhead{Name} & \colhead{(Myr)} & \colhead{Ref.} & \colhead{Members} & \colhead{Members} & \colhead{Members}  }
\startdata
AB Dor & 149$^{+51}_{-19}$ & 1 & 0 & 13 & 5 \\
Argus & 40-50 & 2 & 0 & 7 & 21 \\
$\beta$ Pictoris & 22$\pm$6 & 3 & 1 & 8 & 10 \\
Carina & 45$^{+11}_{-7}$ & 1 & 3 & 8 & 1 \\
Carina-Near & $\sim$200  & 4 & 6 & 0 & 0 \\
Columba & 42$^{+6}_{-4}$ & 1 & 1 & 10 & 5 \\
$\epsilon$ Cha & 3.7$^{+4.6}_{-1.4}$ & 5 & 0 & 4 & 0 \\
$\eta$ Cha & 11$\pm$3 & 1 & 0 & 8 & 0 \\
Octans & 35$\pm$5 & 6 & 0 & 1 & 0 \\
Tuc-Hor & 45$\pm$4 & 1 & 3 & 4 & 2 \\
\colrule
Total & & & 14 & 63 & 46\\
\enddata
\tablenotetext{a}{This number includes those objects that have been previously suggested to be a member of a different group, but have been reassigned in this work to this group.}
\tablerefs{(1) \cite{bell15}; (2) \cite{zuck18}; (3) \cite{shk17}; (4) \cite{zuck06}; (5) \cite{murph13}; (6) \cite{murph15b} }
\end{deluxetable*}

Note that a kinematic match to a particular group using BANYAN $\Sigma$ does not guarantee membership because kinematic interlopers with different ages are possible.  For all kinematic matches, we also evaluate membership using all available age information gathered in Table 4 to ensure consistency with the age of any matching group.  Note that any objects in our list that are already included in the bonafide members of BANYAN $\Sigma$ are not discussed further.  Table 5 summarizes the number of new, confirmed, and rejected members found for all groups.  Table 6 provides age and kinematic information for new, confirmed, and reassigned moving group and association members, while Table 7 provides information for all rejected moving group and association members.  Notes for each group including comments on individual sources are presented in the following sections.

\begin{longrotatetable}
\begin{deluxetable}{lrrrrrrrrrrrrr}
\tabletypesize{\footnotesize}
\tablecaption{New and Confirmed Moving Group Members}
\tablehead{
\colhead{2MASS} & \colhead{Spec.} & \colhead{Li EW\tablenotemark{a}} & \colhead{H$\alpha$ EW} & \colhead{log $L_{\rm X}$} & \colhead{$f_{\rm FUV}$/$f_{\rm J}$} & \colhead{$f_{\rm NUV}$/$f_{\rm J}$} & \colhead{XYZ} & \colhead{UVW}  & \colhead{BANYAN $\Sigma$} & \colhead{New?}\tablenotemark{a}  \\
\colhead{Name} & \colhead{Type} & \colhead{(\AA)} & \colhead{(\AA)} & \colhead{(erg s$^{-1}$)} & & &  \colhead{(pc)} & \colhead{(km s$^{-1}$)} & \colhead{(\%)}  }
\startdata
\colrule
\sidehead{AB Dor}
\colrule
01484087$-$4830519 & M1.5 & \dots & -2.68 & 29.36$\pm$0.07 & 3.41e-05 & 2.04e-04 & (2.58,-15.78,-35.74) & (-7.51,-28.13,-11.48) & 99.8 & \dots\\
02130073$+$1803460 & M4.5 & \dots & -5.49 & \dots & 6.33e-05 & 2.05e-04 & (-17.55,10.62,-17.63) & (-5.33,-27.42,-15.48) & 99.9 & \dots\\
03100305$-$2341308 & M3.5 & \dots & -6.20 & \dots & 7.96e-05 & 2.09e-04 & (-11.31,-7.77,-22.49) & (-4.75,-27.53,-16.92) & 99.3 & \dots\\
04084031$-$2705473 & M4.0 & \dots & -3.60 & 28.92$\pm$0.10 & 8.77e-05 & 2.13e-04 & (-15.80,-15.64,-23.35) & (-5.47,-28.22,-16.53) & 99.6 & \dots\\
04353618$-$2527347 & M3.5 & \dots & -5.80 & 28.71$\pm$0.08 & 5.37e-05 & 2.28e-04 & (-9.15,-8.98,-10.82) & (-5.10,-28.05,-18.16) & 98.4 & \dots\\
04514615$-$2400087 & M3.0 & \dots & -6.23 & 29.17$\pm$0.09 & \dots & 2.24e-04 & (-29.39,-28.34,-29.91) & (-4.47,-26.77,-8.52) & 65.2 & \dots\\
05531299$-$4505119 & M0.5 & 0.132 & -1.57 & 29.23$\pm$0.04 & 1.86e-05 & \dots & (-8.58,-25.99,-15.01) & (-8.05,-28.67,-12.02) & 99.9 & \dots\\
06373215$-$2823125 & M2.5 & \dots & -1.79 & 29.14$\pm$0.10 & \dots & \dots & (-22.44,-35.02,-11.40) & (-8.13,-27.50,-15.07) & 99.9 & \dots\\
12383713$-$2703348 & M2.5 & \dots & -2.31 & 28.80$\pm$0.10 & 2.10e-05 & 1.22e-04 & (9.92,-17.60,14.53) & (-7.43,-28.51,-11.92) & 99.7 & \dots\\
14415908$-$1653133 & M3.0 & \dots & 0.00 & \dots & 1.88e-04 & 2.11e-04 & (15.49,-6.43,13.32) & (-11.76,-25.47,-16.49) & 60.2 & \dots\\
21464282$-$8543046 & M3.5 & \dots & -10.46 & 28.82$\pm$0.05 & 6.23e-05 & \dots & (8.00,-10.82,-7.78) & (-7.52,-27.45,-13.94) & 100.0 & \dots\\
21471964$-$4803166 & M4.0 & \dots & -6.72 & 29.32$\pm$0.11 & 4.94e-05 & 1.76e-04 & (41.13,-7.39,-47.70) & (-5.68,-29.52,-14.00) & 16.5 & \dots\\
23115362$-$4508004 & M3.0 & \dots & -6.41 & 29.81$\pm$0.07 & 6.51e-05 & \dots & (20.51,-5.83,-42.24) & (-4.67,-26.66,-10.46) & 84.1 & \dots\\
\colrule
\sidehead{Argus}
\colrule
05471788$-$2856130 & M3.5 & \dots & -5.54 & 29.64$\pm$0.07 & \dots & \dots & (-33.46,-45.69,-27.42) & (-20.98,-15.20,-4.12) & 99.7 & \dots\\
09445422$-$1220544 & M5.0 & \dots & -16.17 & 28.71$\pm$0.04 & 3.94e-05 & 1.25e-04 & (-4.28,-10.54,6.56) & (-22.36,-13.61,-3.75) & 99.5 & \dots\\
10252563$-$4918389 & M4.0 & \dots & -4.52 & 28.86$\pm$0.09 & \dots & \dots & (4.38,-25.00,3.08) & (-24.26,-6.60,-6.75) & 5.9 & \dots\\
12233860$-$4606203 & M4.0 & \dots & -5.08 & 28.71$\pm$0.09 & \dots & 1.17e-04 & (10.29,-19.42,6.51) & (-24.03,-8.03,-5.63) & 89.4 & \dots\\
17275761$-$4016243 & M4.0 & \dots & -7.81 & 28.91$\pm$0.14 & \dots & \dots & (39.17,-7.98,-2.12) & (-20.07,-8.15,-1.98) & 81.4 & \dots\\
20072376$-$5147272 & K6.0 & \dots & -0.99 & 29.79$\pm$0.04 & 4.58e-05 & 3.75e-04 & (27.96,-6.54,-18.30) & (-25.74,-16.72,-5.09) & 95.3 & \dots\\
22274882$-$0113527 & M3.5 & \dots & -3.64 & 28.98$\pm$0.10 & 9.73e-05 & \dots & (8.95,18.16,-21.66) & (-24.17,-16.59,-0.16) & 84.4 & \dots\\
\colrule
\sidehead{$\beta$ Pic}
\colrule
02442137$+$1057411 & M1.0 & 0.739 & -2.63 & \dots & \dots & \dots & (-33.95,10.76,-33.31) & (-8.96,-18.36,-5.00) & 3.9 & \dots\\
02490228$-$1029220 & M2.0 & 0.266 & -2.42 & 29.34$\pm$0.20 & 3.97e-05 & 1.78e-04 & (-20.65,-2.75,-32.97) & (-12.38,-10.28,-11.32) & 1.3 & \dots \\
05082729$-$2101444 & M5.0 & \dots & -24.90 & 28.89$\pm$0.13 & 2.57e-05 & 1.53e-04 & (-30.78,-27.73,-25.52) & (-14.65,-20.39,-7.73) & 2.5 & \dots\\
05294468$-$3239141 & M4.5 & \dots & -5.88 & 29.03$\pm$0.08 & 2.27e-05 & 1.03e-04 & (-14.16,-21.49,-15.16) & (-11.90,-16.36,-9.04) & 99.9 & \dots\\
05320450$-$0305291 & M2.0 & 0.073 & -3.02 & 29.80$\pm$0.04 & \dots & 1.93e-04 & (-29.25,-14.56,-11.27) & (-19.08,-15.64,-8.73) & 51.3 & \dots\\
09462782$-$4457408 & M5.5 & 0.542 & -13.61 & \dots & \dots & \dots & (1.60,-46.39,5.29) & (-13.99,-17.17,-7.78) & 40.9 & Y\\
13545390$-$7121476 & M2.5 & \dots & -2.84 & 28.86$\pm$0.06 & \dots & 2.32e-04 & (13.89,-17.78,-3.62) & (-12.02,-15.03,-10.41) & 99.4 & \dots\\
16572029$-$5343316 & M3.0 & \dots & -0.78 & 29.14$\pm$0.16 & \dots & \dots & (45.37,-21.65,-5.88) & (-7.12,-15.99,-10.52) & 98.7 & \dots\\
19233820$-$4606316 & M0.0 & 0.447 & -1.01 & 29.75$\pm$0.11 & 3.04e-05 & 3.32e-04 & (64.02,-9.10,-29.50) & (-7.02,-16.34,-9.64) & 97.0 & \dots\\
\colrule
\sidehead{Carina}
\colrule
06112997$-$7213388 & M4.0 & \dots & -6.16 & 29.40$\pm$0.06 & 4.56e-05 & 1.42e-04 & (11.12,-48.65,-27.40) & (-11.88,-24.50,-4.88) & 97.7 & \dots\\
06234024$-$7504327 & M3.5 & \dots & -5.65 & 29.52$\pm$0.33 & \dots & \dots & (11.77,-40.75,-22.48) & (-5.30,-20.45,-6.12) & 98.6 & \dots\\
07013884$-$6236059 & M0.5 & \dots & -1.54 & 29.70$\pm$0.05 & 9.52e-05 & 2.66e-04 & (4.33,-81.70,-34.29) & (-11.40,-23.04,-4.93) & 94.6 & Y\\
07065772$-$5353463 & M0.0 & 0.380 & -1.36 & 29.66$\pm$0.06 & 4.46e-05 & 1.83e-04 & (-4.32,-43.79,-15.54) & (-10.75,-20.78,-6.24) & 96.2 & R\tablenotemark{c} \\ 
08040534$-$6316396 & M2.0 & \dots & -5.17 & 29.81$\pm$0.09 & \dots & \dots & (8.46,-74.44,-22.07) & (-11.13,-22.50,-6.22) & 100.0 & \dots\\
08194309$-$7401232 & M4.5 & \dots & -9.80 & 29.49$\pm$0.13 & \dots & 1.48e-04 & (17.96,-58.19,-22.48) & (-9.29,-28.84,-8.53) & 82.7 & \dots\\
08412528$-$5736021 & M3.0 & \dots & -7.24 & \dots & \dots & \dots & (5.13,-71.17,-11.93) & (-9.49,-21.70,-5.11) & 96.7 & \dots \\
09032434$-$6348330 & M0.5 & 0.380 & -2.14 & 29.92$\pm$0.08 & \dots & \dots & (14.38,-75.70,-15.34) & (-10.58,-22.68,-6.44) & 100.0 & \dots\\
09180165$-$5452332 & M4.0 & 0.285 & -8.96 & 29.25$\pm$0.13 & \dots & \dots & (4.95,-52.65,-3.57) & (-11.61,-26.26,-6.41) & 96.9 & \dots\\
09315840$-$6209258 & M3.5 & \dots & -6.15 & 29.94$\pm$0.09 & \dots & \dots & (15.90,-76.08,-10.61) & (-10.24,-21.08,-6.19) & 99.9 & \dots\\
14284804$-$7430205 & M1.0 & 0.099 & -2.19 & 29.66$\pm$0.08 & \dots & \dots & (36.41,-44.30,-13.11) & (-9.25,-20.94,-5.72) & 4.2 & R\tablenotemark{d} \\
\colrule
\sidehead{Carina-Near}
\colrule
06134171$-$2815173 & M3.5 & \dots & -5.93 & 29.16$\pm$0.07 & 6.09e-05 & 3.04e-04 & (-22.08,-31.75,-14.18) & (-28.32,-16.28,0.85) & 9.5 & Y\\
06380031$-$4056011 & M3.5 & \dots & -6.52 & 28.98$\pm$0.09 & \dots & \dots & (-12.60,-34.01,-13.03) & (-25.58,-18.93,-3.29) & 99.7 & R\tablenotemark{d} \\
07170438$-$6311123 & M2.0 & \dots & -4.43 & 29.41$\pm$0.07 & \dots & \dots & (4.43,-60.49,-23.56) & (-13.61,-13.78,-4.57) & 1.6 & R\tablenotemark{f} \\
11462310$-$5238519 & M4.5 & \dots & -8.46 & \dots & \dots & 3.55e-04 & (22.38,-52.65,9.03) & (-19.04,-16.72,-1.61) & 45.9 & R\tablenotemark{b} \\
19435432$-$0546363 & M4.0 & \dots & -5.65 & 28.94$\pm$0.11 & 6.12e-05 & 1.71e-04 & (23.53,15.70,-7.21) & (-31.38,-15.13,1.45) & 65.3 & R\tablenotemark{d} \\ 
20284361$-$1128307 & M3.5 & \dots & -6.30 & 28.85$\pm$0.07 & 4.88e-05 & 1.83e-04 & (13.65,8.99,-8.23) & (-23.85,-17.91,-3.00) & 70.0 & R\tablenotemark{e} \\ 
\colrule
\sidehead{Columba}
\colrule
03241504$-$5901125 & K7.0 & 0.224 & -0.94 & 29.75$\pm$0.10 & 4.62e-05 & 3.35e-04 & (4.10,-57.06,-65.05) & (-11.03,-22.83,-5.04) & 8.2 & \dots\\
03320347$-$5139550 & M2.0 & \dots & -2.06 & 29.75$\pm$0.07 & 2.33e-05 & 1.43e-04 & (-5.66,-47.72,-59.82) & (-12.39,-21.79,-3.71) & 97.5 & \dots\\
04091413$-$4008019 & M3.5 & \dots & -6.04 & 29.59$\pm$0.12 & 3.26e-05 & 1.90e-04 & (-22.59,-45.88,-55.64) & (-12.70,-21.84,-4.39) & 99.9 & \dots\\
05100427$-$2340407 & M3.0 & \dots & -2.20 & 29.89$\pm$0.05 & 3.69e-05 & 1.47e-04 & (-34.45,-34.58,-30.66) & (-14.04,-22.04,-5.26) & 99.9 & \dots\\
05100488$-$2340148 & M2.0 & \dots & -6.63 & \dots & 5.94e-05 & 2.25e-04 & (-34.60,-34.73,-30.79) & (-13.01,-22.52,-5.76) & 100.0 & \dots\\
05175793$-$5433503 & M2.0 & \dots & -5.96 & 29.72$\pm$0.13 & \dots & 1.09e-04 & (-8.11,-60.00,-42.68) & (-12.16,-22.91,-5.10) & 83.4 & Y\\
05195695$-$1124440 & M3.5 & \dots & -9.91 & 29.65$\pm$0.09 & \dots & 1.74e-04 & (-53.30,-34.72,-30.24) & (-14.30,-21.71,-5.51) & 99.2 & \dots\\
05241317$-$2104427 & M4.0 & \dots & -4.84 & 29.25$\pm$0.11 & \dots & \dots & (-37.13,-35.27,-27.45) & (-13.62,-21.95,-6.02) & 100.0 & \dots\\
05395494$-$1307598 & M3.0 & \dots & -3.65 & 29.51$\pm$0.13 & \dots & \dots & (-58.54,-44.11,-29.17) & (-15.50,-22.55,-6.76) & 42.7 & \dots\\
05432676$-$3025129 & M0.5 & \dots & -2.60 & 29.97$\pm$0.07 & \dots & \dots & (-44.52,-63.78,-39.77) & (-14.45,-23.33,-8.36) & 1.3 & \dots\\
06262932$-$0739540 & M1.0 & 0.228 & -2.43 & \dots & \dots & \dots & (-64.68,-48.69,-12.84) & (-12.56,-20.72,-6.11) & 71.5 & \dots\\
\colrule
\sidehead{$\epsilon$ Cha}
\colrule
11493184$-$7851011 & M0.0 & 0.579 & -2.96 & 30.07$\pm$0.06 & 8.10e-05 & 3.25e-04 & (48.15,-83.91,-28.38) & (-9.77,-20.95,-11.13) & 99.9 & \dots \\
12194369$-$7403572 & M1.0 & 0.575 & -4.13 & \dots & \dots & 2.03e-04 & (50.77,-85.45,-19.92) & (-9.25,-21.01,-9.65) & 99.8 & \dots\\
12202177$-$7407393 & M1.0 & 0.663 & -3.69 & \dots & 1.29e-04 & 1.96e-04 & (74.70,-125.45,-29.40) & (-18.45,-24.39,-8.85) & 20.9 & \dots\\
12210499$-$7116493 & K7.0 & 0.549 & -0.93 & 30.32$\pm$0.09 & \dots & \dots & (49.88,-84.76,-14.79) & (-9.19,-20.65,-8.84) & 95.4 & \dots\\
\colrule
\sidehead{$\eta$ Cha}
\colrule
08361072$-$7908183 & M5.5 & 0.621 & -9.65 & \dots & \dots & \dots & (34.85,-84.45,-36.82) & (-10.89,-20.99,-11.38) & 92.2 & \dots\\
08385150$-$7916136 & M5.0 & 0.652 & -10.67 & \dots & \dots & \dots & (35.83,-86.00,-37.47) & (-10.72,-20.74,-11.49) & 38.7 & \dots\\
08413030$-$7853064 & M5.0 & 0.685 & -8.46 & \dots & \dots & \dots & (34.91,-84.97,-36.38) & (-11.89,-20.35,-11.09) & 41.6 & \dots\\
08413703$-$7903304 & M3.5 & 0.529 & -2.63 & \dots & \dots & \dots & (35.13,-84.83,-36.52) & (-10.95,-20.58,-10.84) & 57.6 & \dots\\
08422710$-$7857479 & M4.0 & 0.680 & -4.51 & \dots & \dots & \dots & (35.04,-84.88,-36.37) & (-11.25,-20.86,-11.57) & 38.9 & \dots\\
08423879$-$7854427 & M3.0 & 0.487 & -4.25 & 29.00$\pm$0.29 & \dots & \dots & (34.76,-84.37,-36.07) & (-10.68,-20.98,-11.22) & 84.9 & \dots\\
08440914$-$7833457 & M4.5 & 0.536 & -63.46 & \dots & \dots & \dots & (34.54,-84.99,-35.81) & (-10.90,-20.35,-11.14) & 79.1 & \dots\\
08441637$-$7859080 & M4.5 & 0.643 & -10.14 & \dots & \dots & \dots & (34.63,-83.59,-35.72) & (-11.35,-20.65,-12.92) & 99.6 & \dots\\
\colrule
\sidehead{Octans}
\colrule
02411909$-$5725185 & M3.0 & \dots & -4.63 & 29.57$\pm$0.14 & 2.20e-05 & 1.71e-04 & (7.11,-51.18,-71.23) & (-12.82,-2.56,-8.81) & 100.0 & \dots\\
\colrule
\sidehead{Tuc-Hor}
\colrule
00153670$-$2946003 & M4.0 & \dots & -6.32 & 28.76$\pm$0.15 & 2.32e-05 & 1.33e-04 & (5.15,1.12,-36.09) & (-9.60,-20.98,-2.64) & 99.9 & \dots\\
03454058$-$7509121 & M4.0 & \dots & -8.36 & 29.34$\pm$0.07 & \dots & 1.61e-04 & (15.69,-42.86,-34.78) & (-9.61,-21.19,-1.00) & 100.0 & \dots\\
05142736$-$1514514 & M3.5 & \dots & -5.42 & 29.74$\pm$0.08 & \dots & \dots & (-53.96,-39.79,-35.95) & (-13.54,-22.30,-2.33) & 20.4 & R\tablenotemark{c}\\
05142878$-$1514546 & M3.5 & \dots & -4.53 & \dots & \dots & \dots & (-53.75,-39.64,-35.80) & (-13.16,-21.71,-2.39) & 24.1 & R\tablenotemark{c}\\
06002304$-$4401217 & M4.0 & \dots & -13.98 & 29.75$\pm$0.05 & \dots & \dots & (-20.16,-58.02,-31.67) & (-11.70,-23.70,-3.12) & 59.9 & R\tablenotemark{c}\\
19225071$-$6310581 & M3.0 & \dots & -6.21 & 29.65$\pm$0.12 & 5.47e-05 & 1.88e-04 & (41.60,-21.03,-24.27) & (-8.68,-19.27,-0.01) & 97.9 & \dots\\
23204705$-$6723209 & M5.0 & \dots & -9.69 & 29.25$\pm$0.08 & 8.26e-05 & 2.34e-04 & (19.83,-19.43,-30.24) & (-8.74,-21.03,-0.55) & 99.9 & \dots\\
\enddata
\tablenotetext{a}{`Y' refers to completely new members, while `R' refers to reassigned members (i.e., those objects previously proposed to belong to a different group).}
\tablenotetext{b}{Previously suggested to be a member of the Scorpius-Centaurus OB association in \cite{rod11}.}
\tablenotetext{c}{Previously suggested to be a member of the Columba association in \cite{malo13, malo14}.}
\tablenotetext{d}{Previously suggested to be a member of the Argus association in \cite{malo13}.}
\tablenotetext{e}{Previously suggested to be a member of the Argus association in \cite{ried14}.}
\tablenotetext{f}{Previously suggested as an ambiguous $\beta$ Pic/Carina/Columba/AB Dor member in \cite{malo13}}
\end{deluxetable}
\end{longrotatetable}

\begin{longrotatetable}
\begin{deluxetable}{lrrrrrrrrrrrrr}
\tablecaption{Rejected Moving Group Members}
\tablehead{
\colhead{2MASS} & \colhead{Spec.} & \colhead{Li EW\tablenotemark{a}} & \colhead{H$\alpha$ EW} & \colhead{log $L_{\rm X}$} & \colhead{$f_{\rm FUV}$/$f_{\rm J}$} & \colhead{$f_{\rm NUV}$/$f_{\rm J}$} & \colhead{XYZ} & \colhead{UVW}  & \colhead{Reassigned?}   \\
\colhead{Name} & \colhead{Type} & \colhead{(\AA)} & \colhead{(\AA)} & \colhead{(erg s$^{-1}$)} & & &  \colhead{(pc)} & \colhead{(km s$^{-1}$)} & \colhead{}  }
\startdata
\colrule
\sidehead{AB Dor}
\colrule
02523096$-$1548357 & M2.5 & \dots & -2.35 & 29.72$\pm$0.17 & 3.43e-05 & 1.71e-04 & (-54.46,-16.88,-97.76) & (-30.87,-79.45,-54.38) & \dots \\
04424932$-$1452268 & M4.0 & \dots & -6.00 & 29.00$\pm$0.10 & \dots & 1.75e-04 & (-24.41,-15.55,-20.33) & (-1.67,-30.52,-13.76) & \dots \\
05130132$-$7027418 & M3.5 & \dots & -5.11 & 29.16$\pm$0.04 & \dots & 5.45e-05 & (5.38,-26.40,-17.91) & (-37.95,-8.56,8.43) & \dots \\
05240991$-$4223054 & M0.5 & \dots & -2.41 & 29.80$\pm$0.09 & \dots & 1.84e-04 & (-31.89,-77.48,-55.42) & (-11.00,-31.02,-11.84) & \dots \\
07115917$-$3510157 & M3.0 & \dots & -4.86 & 29.11$\pm$0.09 & \dots & \dots & (-14.27,-33.25,-7.29) & (10.52,4.10,-5.89) & \dots \\
\colrule
\sidehead{Argus}
\colrule
00503319$+$2449009A & M3.5 & \dots & -4.18 & 28.97$\pm$0.06 & 6.59e-05 & 1.57e-04 & (-6.36,9.92,-9.23) & (-13.79,-5.23,-5.34) & \dots \\
00503319$+$2449009B & M3.5 & \dots & -4.53 & \dots & 6.59e-05 & 1.57e-04 & (-6.36,9.92,-9.23) & (-13.76,-5.28,-5.30) & \dots \\
03282609$-$0537361 & M4.0 & \dots & -4.32 & 28.26$\pm$0.18 & 2.06e-05 & 9.95e-05 & (-18.84,-3.38,-20.48) & (-25.43,-16.13,-12.27) & \dots \\
03415581$-$5542287 & M4.5 & \dots & -7.28 & \dots & 1.51e-04 & 1.83e-04 & (-0.63,-18.14,-20.31) & (-43.27,-26.49,-9.91) & \dots \\
03415608$-$5542408 & M4.0 & \dots & -3.84 & 28.47$\pm$0.10 & \dots & 9.50e-05 & (-0.63,-18.14,-20.31) & (-43.04,-19.94,-2.58) & \dots \\
04464970$-$6034109 & M1.5 & \dots & -4.42 & 28.99$\pm$0.05 & \dots & \dots & (0.13,-19.20,-15.31) & (-14.20,-7.41,-1.88) & \dots \\
04595855$-$0333123 & M4.0 & \dots & -5.50 & 29.46$\pm$0.10 & 8.52e-05 & 1.63e-04 & (-50.26,-21.13,-26.97) & (-44.90,-19.46,1.19) & \dots \\
06380031$-$4056011 & M3.5 & \dots & -6.52 & 28.98$\pm$0.09 & \dots & \dots & (-12.60,-34.01,-13.03) & (-25.58,-18.93,-3.29) & Y \\    
07343426$-$2401353 & M3.5 & \dots & -6.19 & 29.06$\pm$0.13 & \dots & \dots & (-21.54,-35.95,-1.41) & (-16.88,-18.69,-3.72) & \dots \\
09423823$-$6229028 & M3.5 & \dots & -1.24 & 29.19$\pm$0.10 & \dots & \dots & (9.44,-41.05,-5.31) & (-30.35,-18.15,-8.55) & \dots \\
09455843$-$3253299 & M4.5 & \dots & -8.60 & 28.41$\pm$0.06 & 4.01e-05 & 1.06e-04 & (-1.25,-11.51,3.22) & (-21.09,-24.78,2.03) & \dots \\
11200609$-$1029468 & M2.0 & \dots & -1.42 & 29.01$\pm$0.05 & 1.02e-05 & 1.08e-04 & (-0.14,-13.05,13.68) & (-16.40,-13.75,3.95) & \dots \\
13283294$-$3654233 & M3.5 & \dots & -2.47 & 28.88$\pm$0.13 & \dots & 1.20e-04 & (21.27,-24.35,15.34) & (-39.22,-8.01,-5.41) & \dots \\
13382562$-$2516466 & M3.5 & \dots & -3.73 & 29.25$\pm$0.07 & 7.79e-05 & 2.95e-04 & (18.11,-17.40,18.51) & (-59.23,-10.66,-20.64) & \dots \\
13591045$-$1950034 & M4.5 & \dots & -4.63 & 28.33$\pm$0.07 & 1.66e-05 & 6.18e-05 & (6.68,-4.86,6.99) & (-28.11,-16.99,-9.81) & \dots \\
14284804$-$7430205 & M1.0 & 0.099 & -2.19 & 29.66$\pm$0.08 & \dots & \dots & (36.41,-44.30,-13.11) & (-9.25,-20.94,-5.72) & Y \\     
19432464$-$3722108 & M3.5 & \dots & -7.41 & 28.60$\pm$0.13 & \dots & \dots & (21.79,0.87,-10.56) & (-45.72,-17.12,-3.86) & \dots \\
19435432$-$0546363 & M4.0 & \dots & -5.65 & 28.94$\pm$0.11 & 6.12e-05 & 1.71e-04 & (23.53,15.70,-7.21) & (-31.38,-15.13,1.45) & Y \\    
20163382$-$0711456 & M0.0 & \dots & -0.35 & 29.06$\pm$0.09 & 1.40e-05 & 1.30e-04 & (25.59,18.73,-12.93) & (-23.01,-5.84,5.44) & \dots \\
20284361$-$1128307 & M3.5 & \dots & -6.30 & 28.85$\pm$0.07 & 4.88e-05 & 1.83e-04 & (13.65,8.99,-8.23) & (-23.85,-17.91,-3.00) & Y \\    
23332198$-$1240072 & K5.0 & \dots & -3.54 & 28.99$\pm$0.11 & 5.38e-05 & \dots & (5.46,13.14,-33.03) & (-27.22,-1.99,-25.93) & \dots \\
\colrule
\sidehead{$\beta$ Pic}
\colrule
05332802$-$4257205 & M4.5 & \dots & -2.95 & 28.43$\pm$0.04 & 2.77e-05 & 9.80e-05 & (-3.21,-8.18,-5.46) & (-0.30,4.85,2.09) & \dots \\
05422387$-$2758031 & M4.5 & \dots & -8.50 & 29.57$\pm$0.08 & 9.63e-05 & 2.28e-04 & (-36.00,-46.73,-29.48) & (-4.12,-9.55,-5.49) & \dots \\
06135773$-$2723550 & G5.0 & 0.076 & -5.55 & 31.16$\pm$0.07 & 7.67e-05 & 7.72e-04 & (-204.69,-285.40,-126.33) & (1.59,13.84,9.77) & \dots \\
08224744$-$5726530 & M4.5 & \dots & -6.09 & 28.34$\pm$0.05 & \dots & \dots & (0.56,-12.64,-2.56) & (-36.48,-15.61,-5.83) & \dots \\
10571139$+$0544547 & M1.0 & \dots & -1.33 & 30.09$\pm$0.09 & 5.67e-05 & 2.81e-04 & (-24.25,-54.39,86.41) & (-18.97,-29.34,-10.74) & \dots \\
11493184$-$7851011 & M0.0 & 0.579 & -2.96 & 30.07$\pm$0.06 & 8.10e-05 & 3.25e-04 & (48.15,-83.91,-28.38) & (-9.77,-20.95,-11.13) & Y \\
12115308$+$1249135 & M0.0 & \dots & -0.14 & \dots & 1.73e-05 & 1.37e-04 & (-0.60,-17.87,58.34) & (-9.78,-23.56,-7.58) & \dots \\
14255593$+$1412101 & M0.0 & 0.150 & -3.89 & 31.36$\pm$0.03 & 9.94e-05 & 4.48e-04 & (68.11,8.69,143.77) & (-42.49,-49.88,-44.76) & \dots \\
17150219$-$3333398 & M0.0 & \dots & -1.99 & 29.45$\pm$0.05 & \dots & \dots & (25.66,-3.36,1.31) & (-29.97,-13.39,-15.20) & \dots \\
23301341$-$2023271 & M3.0 & \dots & -3.60 & 29.08$\pm$0.04 & \dots & 1.07e-04 & (3.57,4.04,-15.02) & (-15.82,-24.87,-1.75) & \dots \\
\colrule
\sidehead{Carina}
\colrule
08185942$-$7239561 & M0.0 & \dots & -0.68 & 29.64$\pm$0.10 & \dots & 1.77e-04 & (20.81,-73.25,-27.20) & (-14.42,-30.25,-6.38) & \dots \\
\colrule
\sidehead{Columba}
\colrule
04322548$-$3903153 & M3.5 & \dots & -5.20 & 28.92$\pm$0.10 & \dots & 1.42e-04 & (-11.00,-20.94,-22.01) & (-10.04,-3.91,-7.96) & \dots \\
05425587$-$0718382 & M3.0 & \dots & -13.96 & 29.73$\pm$0.09 & \dots & 2.86e-04 & (-66.33,-41.01,-26.15) & (-26.72,-26.73,-12.38) & \dots \\
06511418$-$4037510 & non-M & \dots & 1.49 & \dots & \dots & 6.46e-05 & (-454.65,-1268.70,-419.99) & (-48.55,-0.76,7.54) & \dots \\
07065772$-$5353463 & M0.0 & 0.380 & -1.36 & 29.66$\pm$0.06 & 4.46e-05 & 1.83e-04 & (-4.32,-43.79,-15.54) & (-10.75,-20.78,-6.24) & Y \\    
23301341$-$2023271 & M3.0 & \dots & -3.60 & 29.08$\pm$0.04 & \dots & 1.07e-04 & (3.57,4.04,-15.02) & (-15.82,-24.87,-1.75) & \dots \\
\colrule
\sidehead{Tuc-Hor}
\colrule
02511150$-$4753077 & K5.0 & \dots & -0.83 & 31.15$\pm$0.08 & 6.83e-05 & 6.50e-04 & (-25.06,-194.89,-324.57) & (2.01,-2.29,18.89) & \dots \\
06434532$-$6424396 & M3.0 & \dots & -5.38 & 28.77$\pm$0.18 & 1.07e-04 & 5.11e-04 & (1.42,-18.25,-8.57) & (-2.75,-19.14,-7.98) & \dots \\
\enddata
\end{deluxetable}
\end{longrotatetable}

\subsubsection{AB Doradus}

In total, we confirm AB Dor membership for 13 previously suggested candidates, and find that 5 previously suggested candidates are not members, as summarized in Table 6.  Figure 5 shows the 3D XYZUVW distributions of bona fide members from \cite{gagne18a} and newly confirmed members.  Figure 6 shows a CMD of previously known members and newly confirmed AB Dor members using {\it Gaia} DR2 parallaxes and G and G$_{\rm RP}$ photometry compared to a volume limited (25 pc) sample of nearby objects. Note that for this CMD (and all CMDs that follow), we plot G-G$_{\rm RP}$ colors ranging from 0.5 to 2.0 mag to focus on low-mass members of the group.  Thus the number of known members plotted in the XYZUVW figure may not match the number of known members plotted in the CMD.

2MASS 01484087$-$4830519 was posited as a possible Horologium member in \cite{torres00}.  \cite{malo13} considered this object to have ambiguous membership, which was revised to possible membership in AB Dor in \cite{malo14}.  2MASS 05531299$-$4505119 was found to be a young star in \cite{torres06}, and was suggested to be an AB Dor candidate in \cite{malo14}.  2MASS 12383713$-$2703348 was suggested to be a potential AB Dor member in \cite{malo13, malo14}, though without a parallax.  \cite{binks16} also find 2MASS 12383713$-$2703348 to be a very likely AB Dor member, again without a parallax.  \cite{malo13} classified 2MASS 21464282$-$8543046 as an ambiguous $\beta$ Pic/Columba/AB Dor candidate, which was then reclassified to an AB Dor candidate in \cite{malo14}.  2MASS 23115362$-$4508004 was first found to be a young star in \cite{torres06} and was presented as a potential AB Dor member in \cite{torres08} and \cite{dasilva09}, though lacking a parallax measurement.  Our measured radial velocities, which are consistent with previous measurements, and {\it Gaia} DR2 parallaxes confirm AB Dor kinematics for all of these objects.  Note that 2MASS 01484087$-$4830519 and 2MASS 23115362$-$4508004 have co-moving companions in Section 4.2.

2MASS 02130073$+$1803460 was suggested as a possible AB Dor member in \cite{gagne18b} using {\it Gaia} DR2 kinematics, but lacking a radial velocity measurement.  2MASS 14415908$-$1653133 was listed as a weak AB Dor candidate in \cite{gagne15a} without a parallax or radial velocity, and is included in \cite{gagne18b} as an AB Dor candidate using its {\it Gaia} DR2 parallax.  2MASS 04353618$-$2527347 was suggested to be a possible AB Dor member in \cite{malo13} without a radial velocity or parallax measurement.  \cite{bart17} also suggest AB Dor membership using their parallax measurement of 57.64$\pm$1.78 mas, though without a radial velocity measurement.  Our measured radial velocities and age diagnostics for these objects confirm AB Dor membership. A co-moving companion to 2MASS 14415908$-$1653133 was also found in Section 4.2.  

2MASS 03100305$-$2341308 was suggested as a modest probability AB Dor candidate in \cite{riedel17} using a radial velocity measurement of 25.50$\pm$6.12 km s$^{-1}$, but lacking a distance measurement.  \cite{gagne18b} consider this object an AB Dor candidate using kinematic information from {\it Gaia} DR2.  Using our more precise radial velocity (24.6$\pm$0.34 km s$^{-1}$), we confirm the AB Dor kinematics of this star.  All youth information is consistent with AB Dor membership, and we consider this a bona fide AB Dor member.

2MASS 04084031$-$2705473 and 2MASS 04514615$-$2400087 were considered AB Dor candidates in \cite{malo13}.  Our radial velocity measurements, combined with {\it Gaia} DR2 parallaxes, show kinematics consistent with AB Dor membership.  Age diagnostics agree with AB Dor membership, so we consider both of these objects bona fide members of AB Dor.

2MASS 06373215$-$2823125 was presented as an AB Dor candidate member in \cite{malo13}, though the radial velocity measurement in \cite{malo14} led the authors to reject AB Dor membership.  Our measured radial velocity (30.55$\pm$0.64 km s$^{-1}$) disagrees with the measurement from \cite{malo14} (17.3$\pm$0.2 km s$^{-1}$).  We find, however, that when our radial velocity is employed, 2MASS 06373215$-$2823125 shows remarkable agreement with known AB Dor members.  We deem 2MASS 06373215$-$2823125 a bona fide AB Dor member, though a confirming radial velocity measurement is likely warranted.  

\cite{malo13, malo14} classified 2MASS 21471964$-$4803166 as an AB Dor candidate without a parallax measurement.  Our radial velocity combined with the parallax from {\it Gaia} DR2 confirms AB Dor membership.  Note that BANYAN $\Sigma$ returns a relatively small probability of AB Dor membership for 2MASS 21471964$-$4803166 (16.5\%).  This is due to a slight distance discrepancy, as the optimal predicted distance from BANYAN $\Sigma$ is 58.1 $\pm$ 1.8 pc when the {\it Gaia} DR2 parallax is not included in the calculation.  The actual distance from {\it Gaia} DR2 is 63.4 $\pm$ 0.5 pc, a difference of $\sim$2.3$\sigma$.  Note also that 2MASS 21471964$-$4803166 has a co-moving companion found in Section 4.2. 

2MASS 02523096$-$1548357, 2MASS 02545247$-$0709255, 2MASS 04424932$-$1452268, 2MASS 05130132$-$7027418, and 2MASS 07115917$-$3510157 were considered an AB Dor candidates in \cite{malo13}, while 2MASS 05240991$-$4223054 was suggested as a AB Dor candidate member in \cite{malo13, malo14} without a parallax measurement.  Our radial velocity measurements, combined with {\it Gaia} DR2 parallaxes, rule out AB Dor membership.  Note, however, that we consider 2MASS 04424932$-$1452268 and 2MASS 05240991$-$4223054 possible fringe AB Dor members in Section 4.4.

\begin{figure*}
\plotone{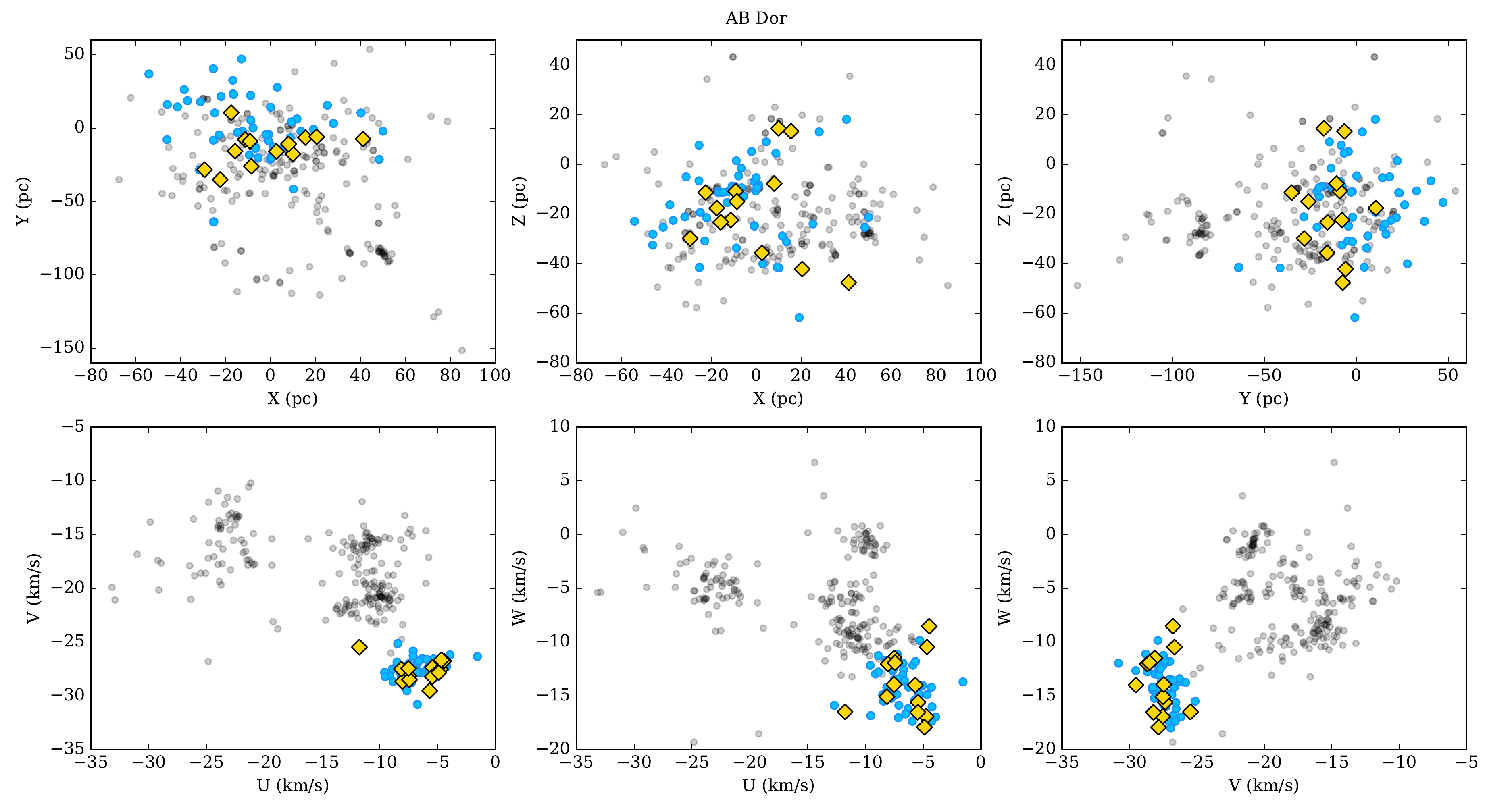}
\caption{A comparison of XYZUVW distributions of previously known AB Dor members from \cite{gagne18a} (blue circles) with newly confirmed members (yellow diamonds).  Black symbols are bonafide moving group and association members from AB Dor, Argus, $\beta$ Pic, Carina, Carina-Near, Columba, $\epsilon$ Cha, $\eta$ Cha, and Tuc-Hor from \cite{gagne18a} for reference.  }  
\end{figure*}

\begin{figure}
\plotone{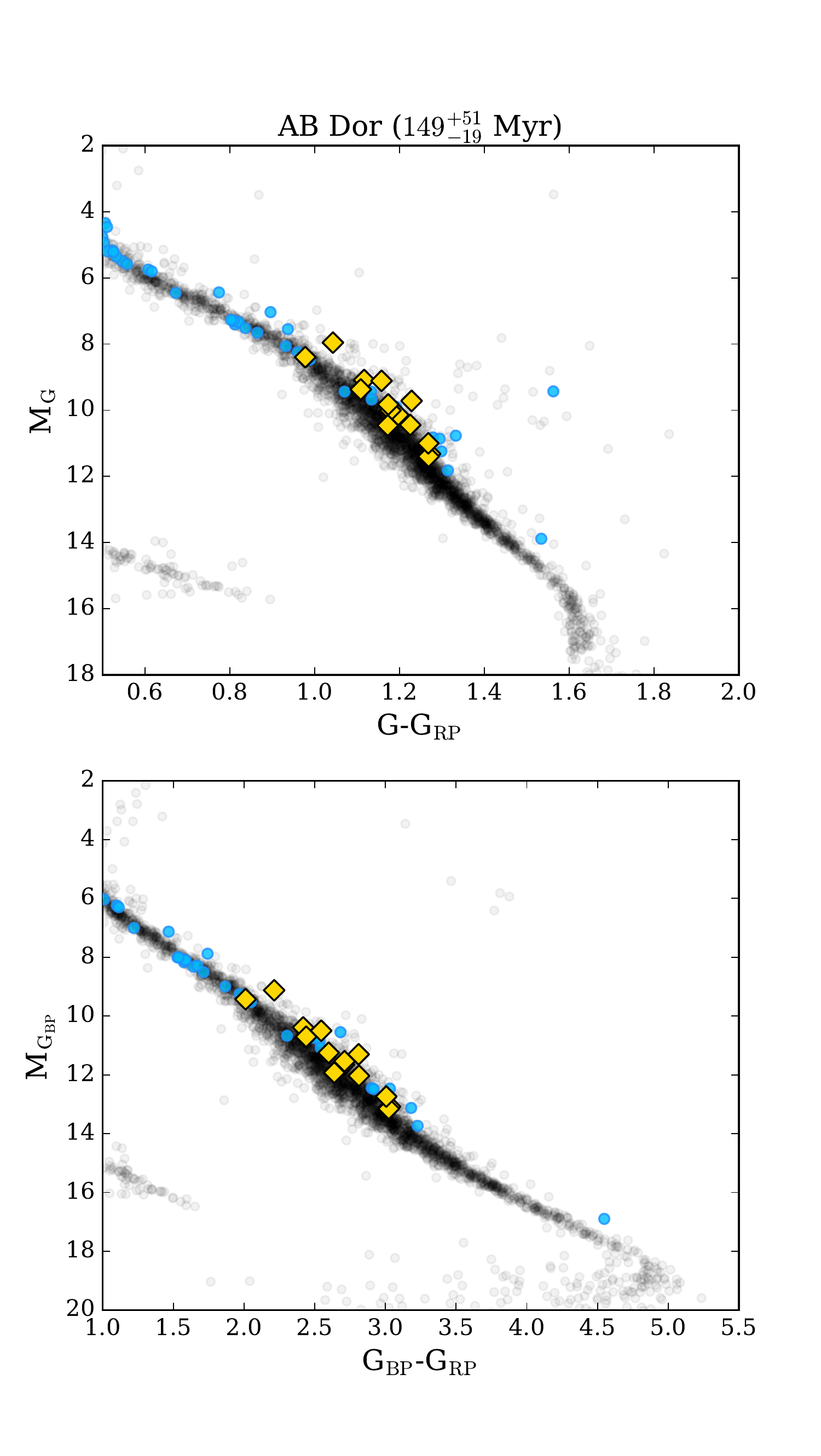}
\caption{CMDs of previously known AB Dor members from \cite{gagne18a} (blue circles) with newly confirmed members (yellow diamonds).  For reference, we show all objects from {\it Gaia} DR2 within 25 pc as background gray symbols.  The outlier in the upper panel is known AB Dor member Wolf 1225, which likely has an inaccurate {\it Gaia} G-band magnitude, as this object is not an outlier in the bottom panel which shows $G_{\rm RP}$ and $G_{\rm BP}$ magnitudes.}  
\end{figure}

\subsubsection{Argus}
Through our kinematic and age analysis, we have confirmed the membership of 7 Argus candidates.  We have also rejected 21 previously suggested candidate members.  Note that the Argus association has recently been redefined \citep{zuck18}, which is the likely reason for so many rejected candidates.  Many of these rejected candidates are still young based on their age information. Figure 7 shows the 3D XYZUVW distributions of previously known Argus members from \cite{gagne18a} and newly confirmed members.  We note here that while the newly defined Argus association in \cite{zuck18} occupies a well-defined area of UVW space, the XYZ distribution of this group is more dispersed than that of most other groups.  This may support the assertion that Argus is an unbound association, possibly connected to the open cluster IC 2391 \citep{desilva13}. Figure 8 shows a color-magnitude diagram of previously known members and newly confirmed Argus members using {\it Gaia} DR2 parallaxes and photometry. 

2MASS 05471788$-$2856130 and 2MASS 17275761$-$4016243 were put forth as ambiguous members of either $\beta$ Pic or Argus in \cite{malo13}, without a parallaxes or radial velocity measurements.  2MASS 12233860$-$4606203 and 2MASS 22274882$-$0113527 were originally suggested as Argus candidates in \cite{malo13} without radial velocities or parallaxes.  Our measured radial velocities along with the parallaxes from Gaia DR2, and the age information for these objects confirms Argus membership. Note that 2MASS 17275761$-$4016243 has a co-moving companion in Section 4.2. 

2MASS 09445422$-$1220544 was a candidate Horologium Association member in \cite{torres00}.  \cite{malo13,malo14} suggested Argus membership for this object, but lacked a parallax measurement.  \cite{bart17} reject 2MASS 09445422$-$1220544 as an Argus member using their parallax, but do not associate it with any other group. Our analysis using our radial velocity and the parallax from {\it Gaia} DR2 instead confirms 2MASS 09445422$-$1220544 as an Argus member. 

2MASS 10252563$-$4918389 was suggested as an Argus candidate in \cite{malo13, malo14}, but lacked a parallax measurement. 2MASS 20072376$-$5147272 was first considered an Argus candidate in \cite{torres08}.  It was furthered considered an Argus candidate in \cite{dasilva09} and \cite{malo13}, though lacked a parallax.   We confirm 2MASS 10252563$-$4918389 and 2MASS 20072376$-$5147272 as Argus members using their {\it Gaia} DR2 parallaxes and age information from Table 2.  Note that a co-moving companion to 2MASS 10252563$-$4918389 was found in Section 4.2.   

\cite{ried14} classified 2MASS 20284361$-$1128307 as a potential Argus member using their measured parallax, but without a radial velocity.  \cite{bowler15} also considered 2MASS 20284361$-$1128307 an Argus candidate (referencing Shkonik et al.\ in prep).  We find that instead that the kinematics and age information for 2MASS 20284361$-$1128307 better match Carina-Near than Argus.

2MASS 03282609$-$0537361, 2MASS 03415581$-$5542287, 2MASS 03415608$-$5542408, 2MASS 04595855$-$0333123, 2MASS 06380031$-$4056011, 2MASS 07343426$-$2401353, 2MASS 11200609$-$1029468, 2MASS 13283294$-$3654233, 2MASS 13382562$-$2516466, 2MASS 13591045$-$1950034, 2MASS 19432464$-$3722108, 2MASS 19435432$-$0546363, 2MASS 20163382$-$0711456, and 2MASS 23332198$-$1240072 were suggested as Argus candidates in \cite{malo13} without radial velocity or parallax measurements.  2MASS 00503319$+$2449009AB, 2MASS 04464970$-$6034109, and 2MASS 09423823$-$6229028 were suggested as Argus candidates in \cite{malo13, malo14} without parallax measurements.  2MASS 09455843$-$3253299 was suggested as an Argus candidate in \cite{gagne15a} without a parallax or radial velocity measurement.  For all of these objects, we find kinematics inconsistent with the Argus association using our measured radial velocities and {\it Gaia} DR2 parallaxes. 

\begin{figure*}
\plotone{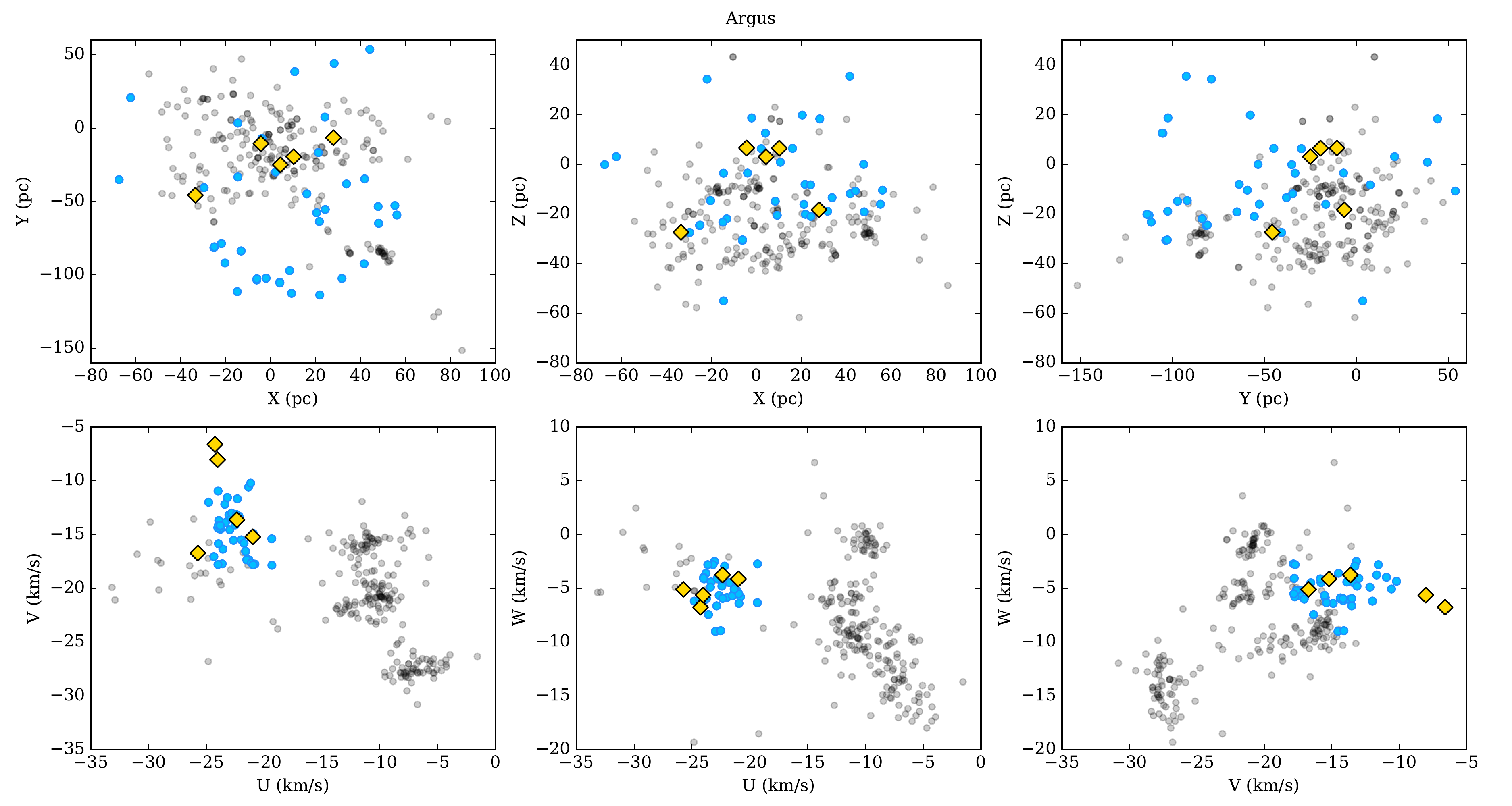}
\caption{A comparison of XYZUVW distributions of previously known Argus members from (blue circles) with newly confirmed members (yellow diamonds), and new discoveries (red squares).  Black symbols are the same as in Figure 5.}  
\end{figure*}

\begin{figure}
\plotone{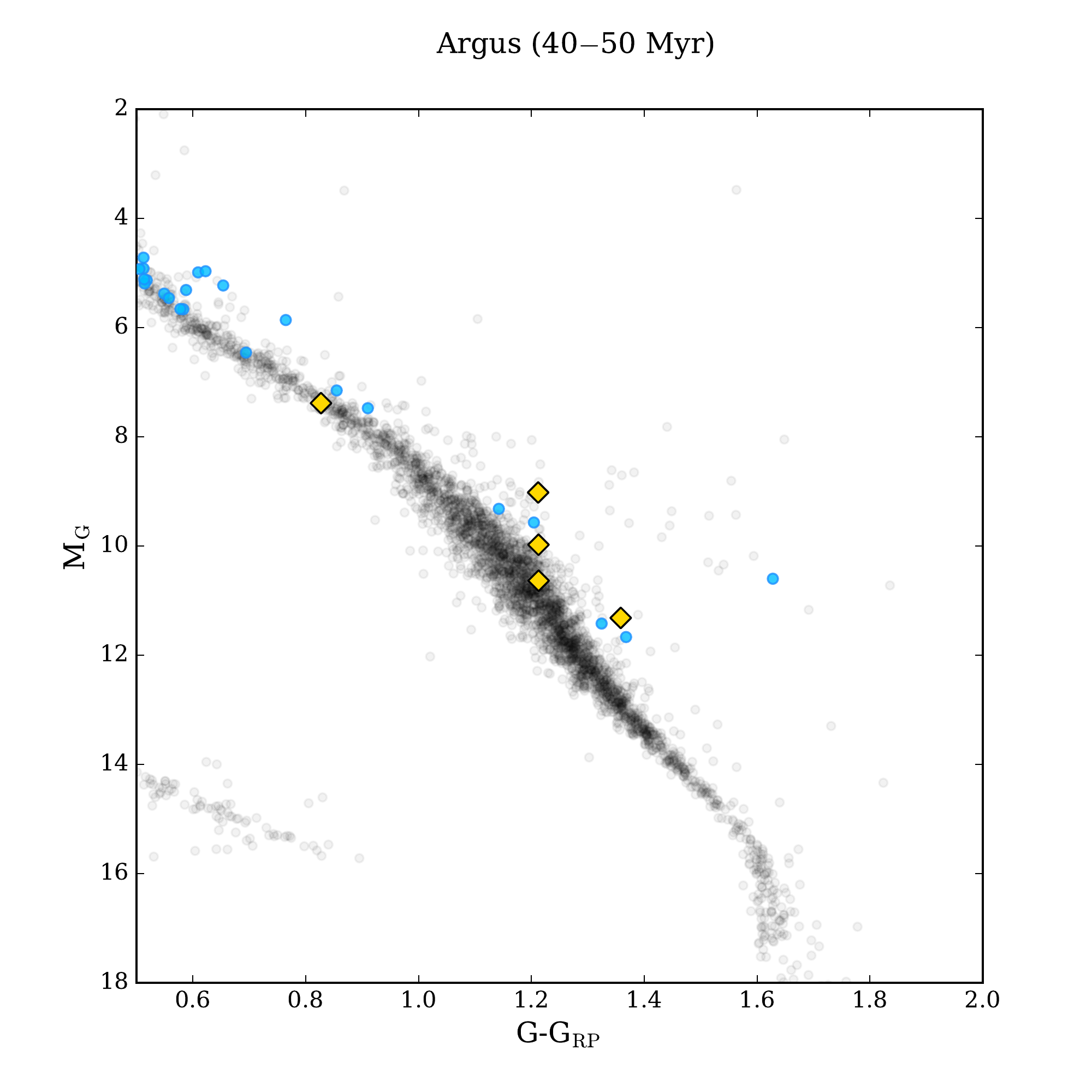}
\caption{CMD of previously known Argus members from \cite{zuck18} (blue circles) with newly confirmed members (yellow diamonds), and new discoveries (red squares).  For reference, we show all objects from {\it Gaia} DR2 within 25 pc as background gray symbols. The outlier is known Argus member CPD-54 1295, which has poor {\it Gaia} photometry.}
\end{figure}

\subsubsection{$\beta$ Pictoris}
We find 2 new $\beta$ Pic members, confirm $\beta$ Pic membership for 7 previously suggested candidates, and reject $\beta$ Pic membership for 7 previously suggested candidates, as summarized in Table 6.  Figure 9 shows the 3D XYZUVW distributions of previously known members from \cite{gagne18a} and newly confirmed members.  Figure 10 shows a color-magnitude diagram of previously known members and newly confirmed $\beta$ Pic members using {\it Gaia} DR2 parallaxes and photometry.   

2MASS 02490228$-$1029220 was suggested as a $\beta$ Pic member in \cite{bowler19}.  While BANYAN $\Sigma$ gives a small probability of $\beta$ Pic membership for this object (2.2\%), note that there is no kinematic information for this object from {\it Gaia} DR2.  The photometric distance to this object from \cite{janson12} (39.0$\pm$7.8 pc) is consistent with $\beta$ pic membership (57.9$\pm$3.9) considering the large uncertainties inherent in photometric distances, especially for young objects.  This star is definitely young considering its clear lithium detection, consistent with $\beta$ Pic membership.  We therefore consider this object a highly likely member of $\beta$ Pic, where a more accurate distance measurement could be used to further confirm its membership.  Note also that this is a close triple system \citep{berg10}, unresolved in {\it Gaia} DR2.

2MASS 09462782$-$4457408 is a new $\beta$ Pic member with a clear lithium detection, ensuring its youth, and kinematics consistent with known $\beta$ Pic members.  Note that BANYAN also finds a small probability of TW Hya membership for this object (2.5\%).

2MASS 02442137$+$1057411 was suggested as a possible $\beta$ Pic member in \cite{lep09}, though was assigned ambiguous membership in \cite{malo13}, with significant probabilities of belonging to $\beta$ Pic, Tuc-Hor, and Columba.  We present the first radial velocity measurement for this object (5.60$\pm$1.09 km s$^{-1}$), which gives this objects similar kinematics with bona fide $\beta$ Pic members, though it returns a small probability of membership from BANYAN $\Sigma$ (6.7\%).  Both the measured radial velocity (5.6$\pm$1.1 km s$^{-1}$) and distance (48.8$\pm$0.3 pc) are slightly discrepant from the optimum values from BANYAN $\Sigma$ (11.9$\pm$1.2 km s$^{-1}$ and 43.0$\pm$2.6 pc, respectively), corresponding to differences of $\sim$2.0$\sigma$ and $\sim$2.7$\sigma$. Considering that this object is certainly young, with a well-measured lithium detection, we consider it a bona fide $\beta$ Pic member. Note that this object has a co-moving companion in Section 4.2.

2MASS 05082729$-$2101444 is a candidate $\beta$ Pic member discussed in \cite{malo13, malo14}, \cite{binks14}, \cite{shk17}, and \cite{mess17}.  Our measured radial velocity (24.94$\pm$0.92 km s$^{-1}$) is more precise than and consistent with those previously measured in \cite{malo14} (23.5$\pm$1.8 km s$^{-1}$) and \cite{binks14} (22.8$\pm$3.8 km s$^{-1}$).  We investigate the kinematics of this object for the first time using a measured parallax, and find good agreement with known $\beta$ Pic members, although with a low probability from BANYAN $\Sigma$ (4.4\%).  The small probability is due to a mismatch between the predicted distance from BANYAN $\Sigma$ (29.6$\pm$5.5 pc) and the measured distance from {\it Gaia} DR2 (48.6$\pm$0.2 pc), a difference of $\sim$3.3$\sigma$.  We consider 2MASS 05082729$-$2101444 a bona fide $\beta$ Pic member.

\cite{malo13} found 2MASS 05294468$-$3239141 to be an ambiguous moving group member, then revised that determination in \cite{malo14} with a radial velocity measurement (but no parallax), and suggested AB Dor membership.  2MASS 05294468$-$3239141 was later suggested as a possible $\beta$ Pic member in \cite{ried14} with a parallax but no radial velocity, and was listed as a candidate $\beta$ Pic member in \cite{shk17} and \cite{mess17}.  Our radial velocity and the parallax from {\it Gaia} DR2 confirms $\beta$ Pic kinematics.  We deem 2MASS 05294468$-$3239141 a bona fide $\beta$ Pic member.

2MASS 05320450$-$0305291 was presented as a $\beta$ Pic candidate first in \cite{torres08}, though lacking radial velocity and parallax measurements. \cite{malo13} also suggest $\beta$ Pic membership without radial velocity or distance information.  \cite{ell14} use a radial velocity measurement of 23.84$\pm$0.55 km s$^{-1}$ and conclude 2MASS 05320450$-$0305291 is a $\beta$ Pic member, while \cite{binks16} measure a radial velocity of 26.2$\pm$1.6 km s$^{-1}$ and reject 2MASS 05320450$-$0305291 as a potential $\beta$ Pic member, though neither study had a parallax measurement.  2MASS 16572029$-$5343316 was listed as a $\beta$ Pic candidate in \cite{malo13} and again in \cite{malo14} with a measured radial velocity, but without a parallax.  \cite{malo13} listed 2MASS 19233820$-$4606316 as a potential $\beta$ Pic member, although without a parallax or radial velocity.  Both \cite{moor13} and \cite{malo14} measured radial velocities for 2MASS 19233820$-$4606316 consistent with $\beta$ Pic membership.  Using our measured radial velocities and parallaxes from {\it Gaia} DR2, we find good agreement with known $\beta$ Pic members for all three objects.  We conclude 2MASS 05320450$-$0305291, 2MASS 16572029$-$5343316, and 2MASS 19233820$-$4606316 are bona fide $\beta$ Pic members.  Note that 2MASS 05320450$-$0305291 has two co-moving companions and 2MASS 16572029$-$5343316 has a single co-moving companion in Section 4.2. 
 
\cite{malo13} first considered 2MASS 13545390$-$7121476 an AB Dor candidate, then suggested $\beta$ Pic membership in \cite{malo14} after obtaining a radial velocity.  \cite{shk17} echoes $\beta$ Pic membership, while \cite{gagne18b} also consider 2MASS 13545390$-$7121476 a $\beta$ Pic candidate using its {\it Gaia} DR2 parallax.  We also find strong evidence for $\beta$ Pic membership, and suggest 2MASS 13545390$-$7121476 be considered a bona fide $\beta$ Pic member. 

2MASS 05332802$-$4257205,  2MASS 05422387$-$2758031, and 2MASS 06135773$-$2723550 were considered $\beta$ Pic candidates in \cite{malo13}, while 2MASS 10571139$+$0544547 was presented as a $\beta$ Pic candidate in \cite{sch12}.  2MASS 12115308$+$1249135 and 2MASS 14255593$+$1412101 were suggested as a $\beta$ Pic members in \cite{sch12}.  2MASS 12115308$+$1249135 was rejected as a possible member in \cite{binks16} based on its radial velocity, while 2MASS 14255593$+$1412101 was rejected as a possible member in \cite{mess17} based on its kinematics.  2MASS 17150219$-$3333398 was given as a $\beta$ Pic candidate in \cite{malo13, malo14}, and is considered a likely $\beta$ Pic member in \cite{shk17}, though without a parallax.  Our radial velocity measurements, combined with {\it Gaia} DR2 parallaxes, rule out membership in $\beta$ Pic. 

2MASS 08224744$-$5726530 was considered and ambiguous candidate in \cite{malo13}, possibly belonging to $\beta$ Pic or AB Dor.  The radial velocity measurement of this star in \cite{malo14} prompted a reclassification to $\beta$ Pic candidate, which was echoed in \cite{shk17}.  However, using the {\it Gaia} DR2 parallax for this object, we find that its kinematics do not agree with known $\beta$ Pic members.  

2MASS 23301341$-$2023271 was initially listed as a $\beta$ Pic candidate in \cite{malo13}, while \cite{malo14} revised its potential membership status to a Columba candidate based on their radial velocity measurement.  \cite{shk17} consider 2MASS 23301341$-$2023271 a likely member of $\beta$ Pic.  Using our measured radial velocity, which is consistent with the measurement from \cite{malo14}, and a {\it Gaia} DR2 parallax, we find that 2MASS 23301341$-$2023271 does not match well with either group.  Note, however that 2MASS 23301341$-$2023271 is an SB2, and thus likely warrants a reevaluation once a systemic velocity is measured.

\begin{figure*}
\plotone{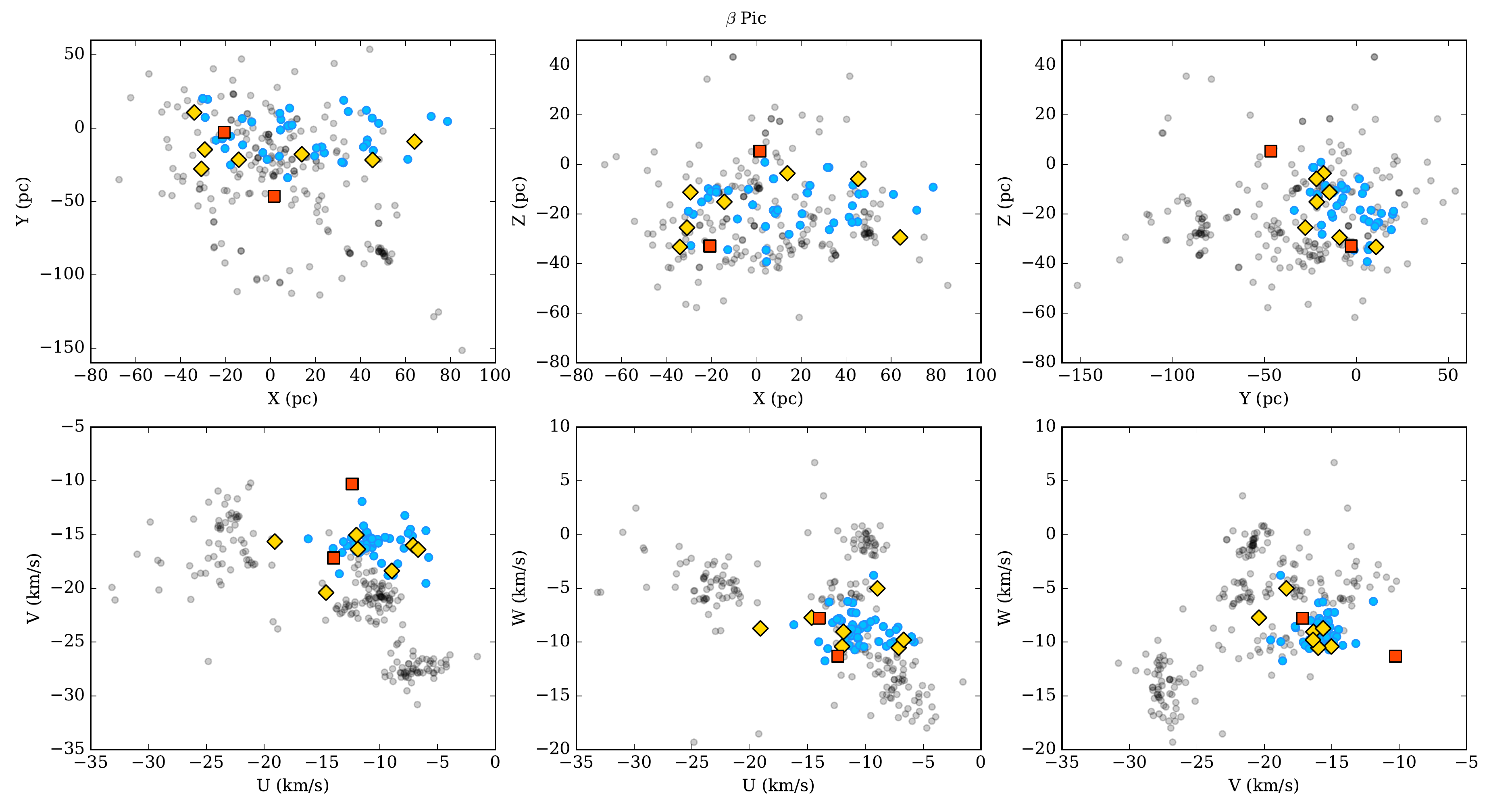}
\caption{A comparison of XYZUVW distributions of previously known $\beta$ Pic members from (blue circles) with newly confirmed members (yellow diamonds), and new discoveries (red squares).  Black symbols are the same as in Figure 5.}  
\end{figure*}

\begin{figure}
\plotone{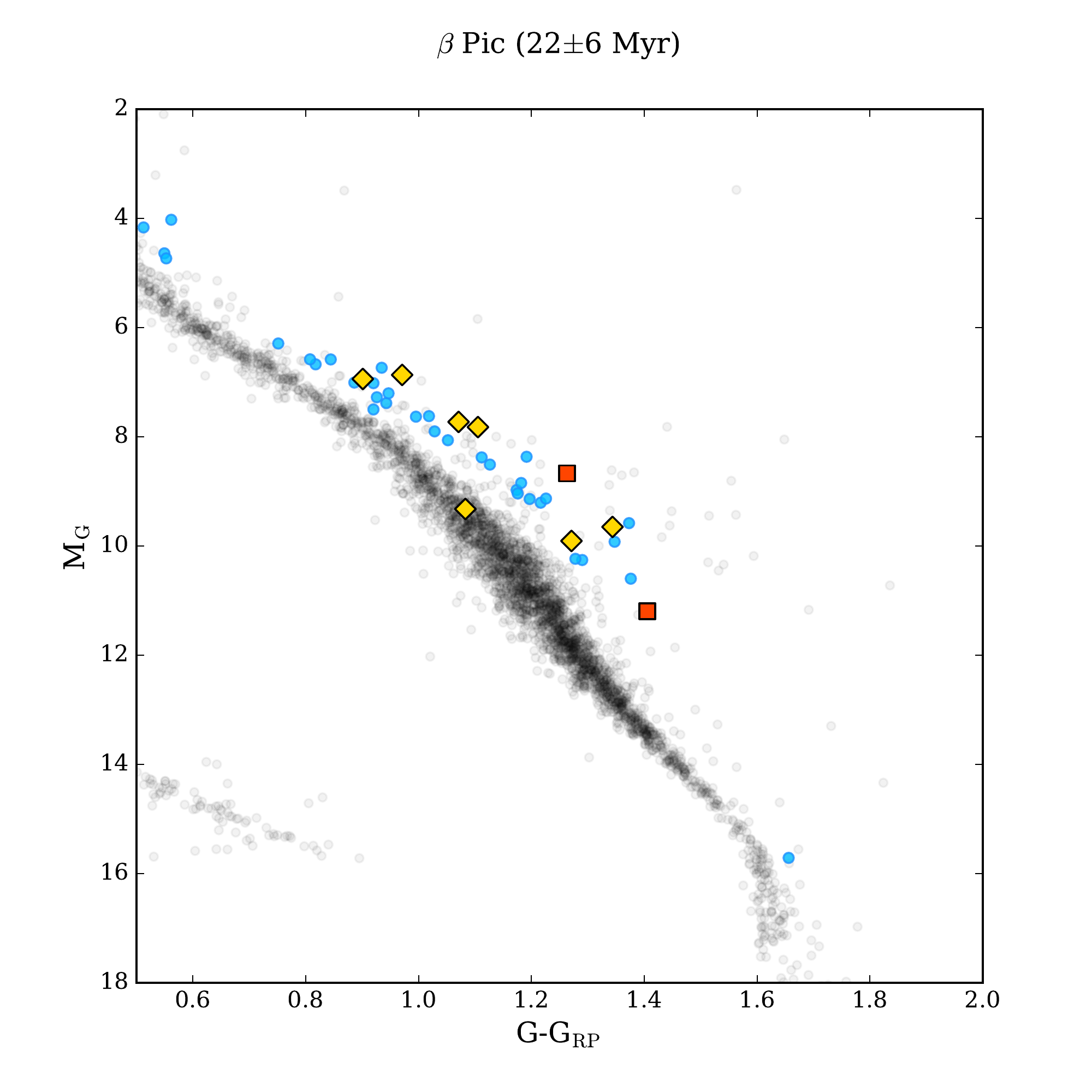}
\caption{CMD of previously known $\beta$ Pic members from \cite{gagne18a} (blue circles) with newly confirmed members (yellow diamonds), and new discoveries (red squares).  For reference, we show all objects from {\it Gaia} DR2 within 25 pc as background gray symbols.}
\end{figure}

\subsubsection{Carina}

We find 1 completely new Carina member, reassign membership to Carina for 2 objects previously suggested to belong to different groups, confirm Carina membership for 8 previously suggested candidates, and reject 1 Carina candidate.  Figure 11 shows the 3D XYZUVW distributions of previously known members from \cite{gagne18a} and newly confirmed Carina members.  Figure 12 shows a color-magnitude diagram of previously known and newly confirmed Carina members using {\it Gaia} DR2 parallaxes and photometry.   

2MASS 07065772$-$5353463 is suggested as a Columba member in \cite{malo13, malo14} without a parallax measurement.  Using our radial velocity, which is consistent with the value found in \cite{malo14}, and the {\it Gaia} DR2 parallax, we find much better agreement with the Carina Association.  We consider 2MASS 07065772$-$5353463 a bona fide Carina member.

2MASS 14284804$-$7430205 was presented as an Argus candidate in \cite{malo13}, but was found to not match any groups in \cite{malo14} once a radial velocity was measured.  Using our more precise radial velocity and the parallax from {\it Gaia} DR2, we find 2MASS 14284804$-$7430205 has kinematics consistent with Carina membership. The low probability found using BANYAN $\Sigma$ for this object (4.2\%) is due to discrepant predicted and actual distances (62.1$\pm$1.5 pc and 58.8$\pm$0.1 pc, respectively), a difference of $\sim$2.0$\sigma$.  

2MASS 06112997$-$7213388, 2MASS 06234024$-$7504327, and 2MASS 09032434$-$6348330 were suggested to be Carina members in \cite{malo13, malo14}, but without parallax measurements.  2MASS 08412528$-$5736021 and 2MASS 09315840$-$6209258 were assigned ambiguous membership in \cite{malo13, malo14}, with similar probabilities of belonging to Columba or Carina (and $\beta$ Pic for 2MASS 09315840$-$6209258).  We find good agreement with known Carina members when using our measured radial velocities and parallaxes from {\it Gaia} DR2.  We consider 2MASS 06112997$-$7213388, 2MASS 06234024$-$7504327, 2MASS 09032434$-$6348330, 2MASS 08412528$-$5736021, and 2MASS 09315840$-$6209258 bona fide Carina members.  Using our radial velocities and a {\it Gaia} DR2 parallaxes, we find that these objects match well with known Carina members.  All age information is consistent with Carina membership, so we therefore assign 2MASS 08412528$-$5736021 and 2MASS 09315840$-$6209258 bona fide Carina membership.  2MASS 08412528$-$5736021 was also found to have a co-moving companion in Section 4.2.

\cite{gagne18b} suggested 2MASS 08040534$-$6316396 and 2MASS 09180165$-$5452332 as possible Carina members using {\it Gaia} DR2 astrometry, but lacking radial velocities.  2MASS 08040534$-$6316396 was also suggested as a possible Carina member in \cite{bowler19}.  Our radial velocities, combined with their age information, confirms Carina membership for both objects. 

2MASS 08194309$-$7401232 is presented as a Carina candidate in \cite{gagne15} without a parallax or radial velocity measurement.  We confirm this object as a Carina member with our measured radial velocity and the parallax from {\it Gaia} DR2.

2MASS 08185942$-$7239561 was considered a Carina candidate in \cite{malo13}, but was subsequently rejected in \cite{malo14}, though without a parallax measurement.  We agree with this assessment using our radial velocity and the parallax from {\it Gaia} DR2.

\begin{figure*}
\plotone{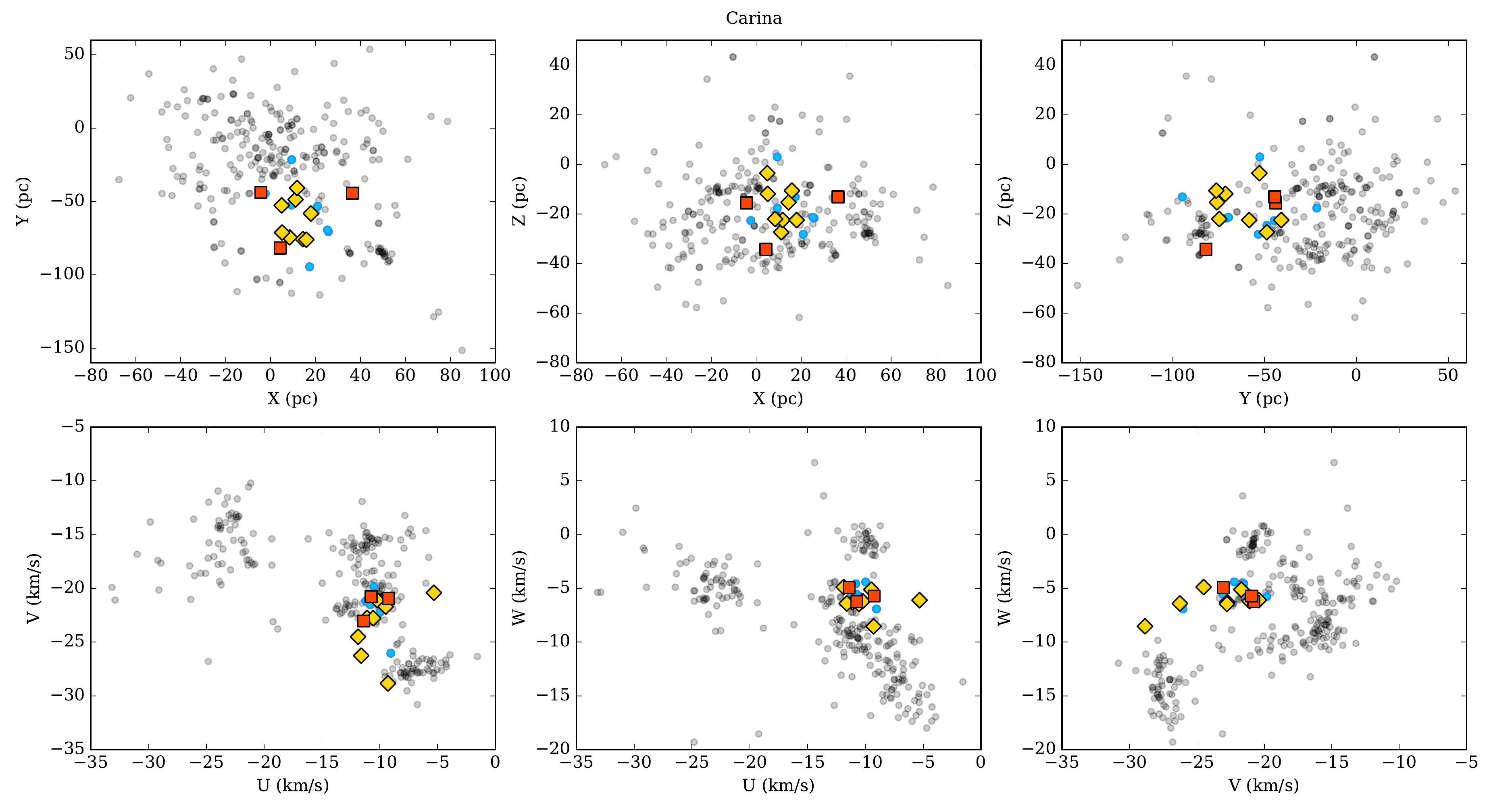}
\caption{A comparison of XYZUVW distributions of previously known Carina members from \cite{gagne18a} (blue circles) with newly confirmed members (yellow diamonds), and new discoveries (red squares).  Black symbols are the same as in Figure 5.}  
\end{figure*}

\begin{figure}
\plotone{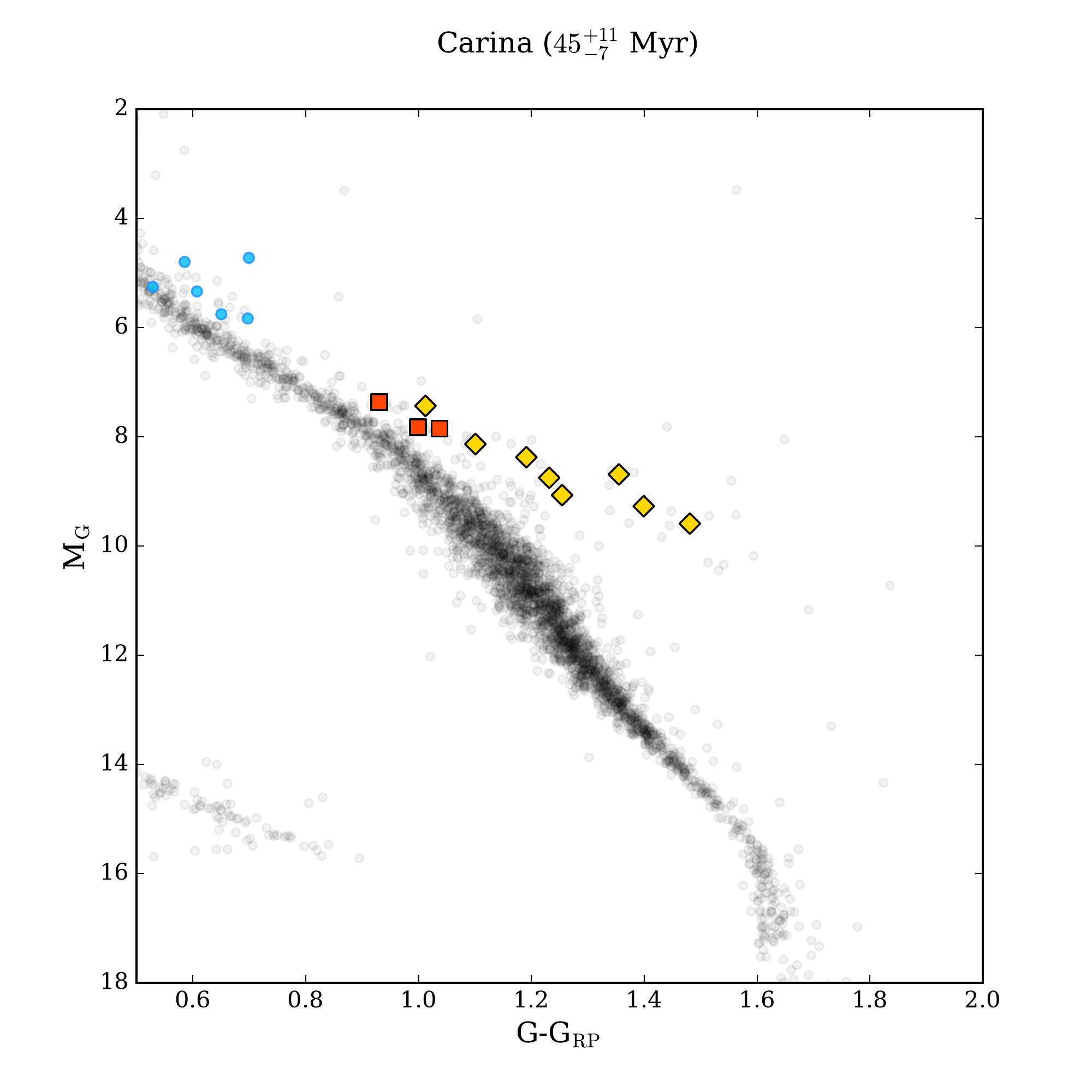}
\caption{CMD of previously known Carina members from \cite{gagne18a} (blue circles) with newly confirmed members (yellow diamonds), and new discoveries (red squares).  For reference, we show all objects from {\it Gaia} DR2 within 25 pc as background gray symbols. }  
\end{figure}

\subsubsection{Carina-Near}

In total, we have discovered one new Carina-Near member, confirming membership for one additional object, and reassigned membership to Carina-Near for 4 objects.  Figures 13 and 14 compare the 3D XYZUVW distributions and CMD positions of previously known members from \cite{gagne18a} and newly confirmed Carina-Near members.  Interestingly, 2MASS 06134171$-$2815173 and 2MASS 06380031$-$4056011 have the latest spectral types of known Carina-Near members (M3.5), with the exception of SIMP J013656.6+093347 \citep{gagne17b}.   

2MASS 06134171$-$2815173 is a new Carina-Near member presented in this work.  Its kinematics match well with other Carina-Near members, though BANYAN $\Sigma$ give a relatively low Carina-Near membership probability (9.4\%).  The small probability is due to a slight mismatch between the predicted distance (28.9$\pm$3.8 pc) and the measured distance (41.2$\pm$0.5 pc), a difference of $\sim$2.9$\sigma$.  However, all youth diagnostics are consistent with the age of Carina-Near ($\sim$200 Myr), and the thus conclude that 2MASS 06134171$-$2815173 is a new Carina-Near member.  Note that this star is part of a quadruple system found in Section 4.2.
  
2MASS 06380031$-$4056011 and 2MASS 19435432$-$0546363 were suggested to be potential Argus members in \cite{malo13} without parallax or radial velocity information.  Using parallaxes from {\it Gaia} DR2 and our measured radial velocities, we instead find excellent agreement with Carina-Near.  With age information consistent with Carina-Near membership, we conclude that 2MASS 06380031$-$4056011 and 2MASS 19435432$-$0546363 are bona fide Carina-Near members.  Note that 2MASS 06380031$-$4056011 was found to have a co-moving companion in Section 4.2.

\cite{ried14} classified 2MASS 20284361$-$1128307 as a potential Argus member using their measured parallax, but without a radial velocity.  \cite{bowler15} also considered 2MASS 20284361$-$1128307 an Argus candidate (referencing Shkonik et al.\ in prep).  We find that 2MASS 20284361$-$1128307 is a better match to Carina-Near, sharing both age and kinematic properties with known members, even though BANYAN $\Sigma$ does produce a small but not insignificant probability of belonging to Argus (26.4\%). 

2MASS 07170438$-$6311123 was presented as an ambiguous member in \cite{malo13}, possibly belonging to $\beta$ Pic, Columba, Carina, or AB Dor.  Our measured radial velocity combined with a {\it Gaia} DR2 parallax rule out membership to all of these groups.  We do find, however, a small probability of belonging to Carina-Near from BANYAN $\Sigma$.  This star is very likely young based on its X-ray luminosity and we include it as a new Carina-Near member. 

2MASS 08413264$-$6825403 was listed as a Carina-Near candidate in \cite{gagne18c}.  We confirm Carina-Near membership for this object using our measured radial velocity and age information.

2MASS 11462310$-$5238519 was suggested to be a possible Sco-Cen member in \cite{rod11} without a parallax or radial velocity. With parallax measurement from {\it Gaia} DR2 giving a distance within $\sim$60 pc, Sco-Cen membership is unlikely.  We instead find that 2MASS 11462310$-$5238519 is a better match to the Carina-Near Association, consistent with the lithium non-detection in its spectrum.  We consider 2MASS 11462310$-$5238519 a new Carina-Near member.  Note that this object also has a significant probability of belonging to the Argus association from BANYAN $\Sigma$ (39.2\%), but is a better match to Carina-Near (45.9\%).

\begin{figure*}
\plotone{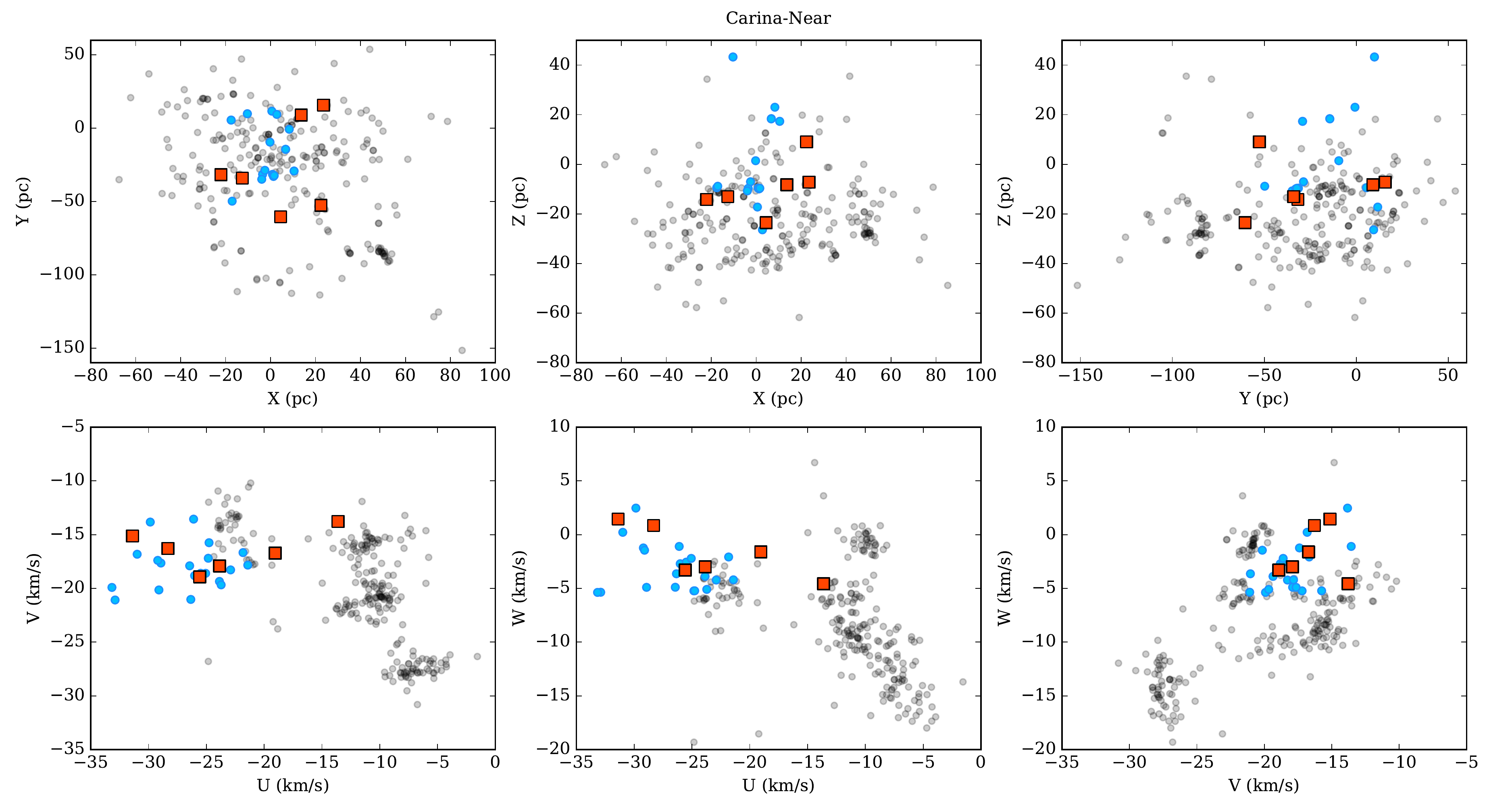}
\caption{A comparison of XYZUVW distributions of previously known Carina-Near members from \cite{gagne18a} (blue circles) with newly confirmed members (yellow diamonds), and new discoveries (red squares).  Black symbols are the same as in Figure 5.}  
\end{figure*}

\begin{figure}
\plotone{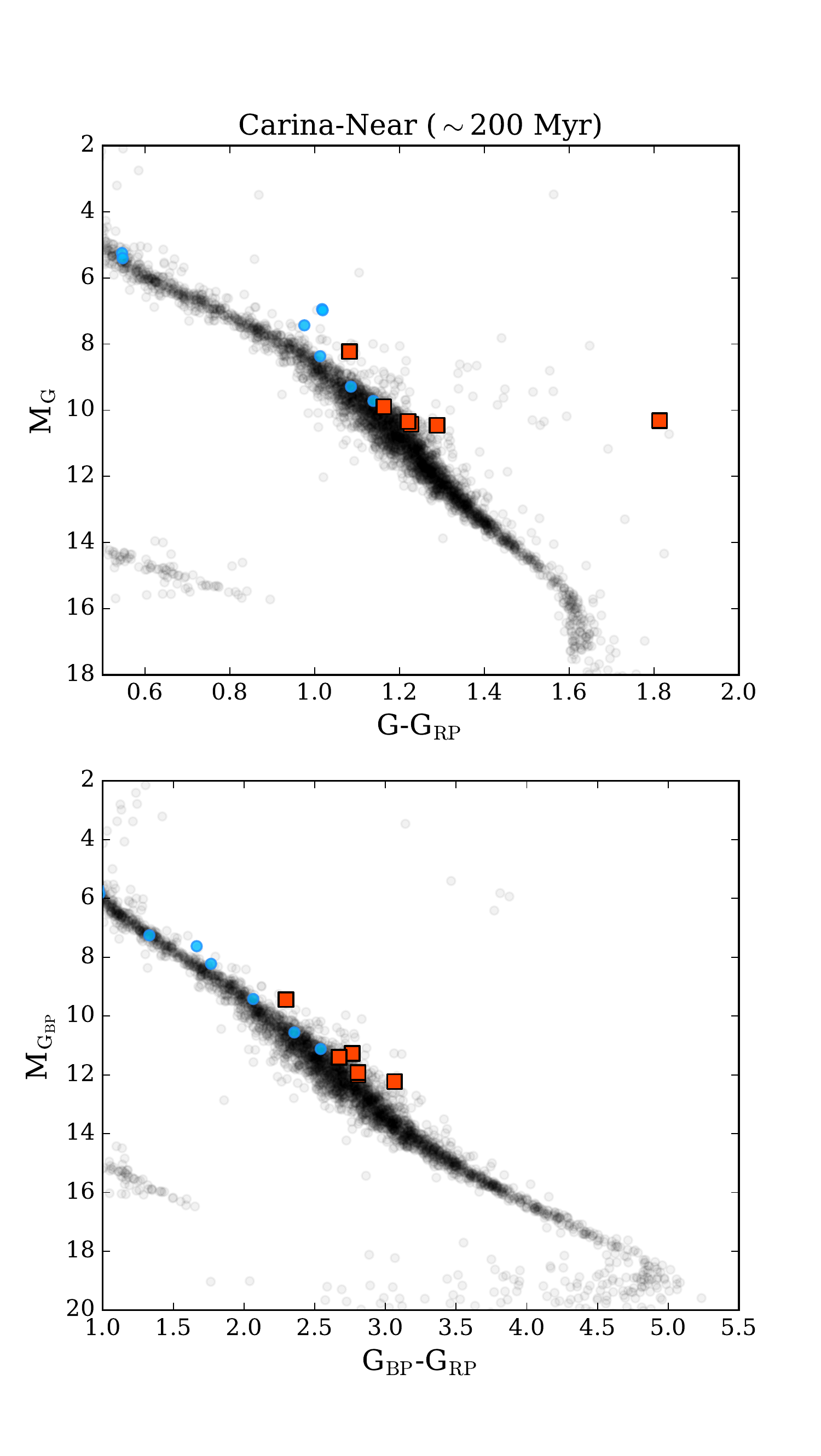}
\caption{CMDs of previously known Carina-Near members from \cite{gagne18a} (blue circles) with newly confirmed members (yellow diamonds), and new discoveries (red squares).  For reference, we show all objects from {\it Gaia} DR2 within 25 pc as background gray symbols. The outlying object in the upper panel is 2MASS 06134171$-$2815173, which likely has an inaccurate {\it Gaia}, as this object is not an outlier in the bottom panel using {\it Gaia} $G_{\rm RP}$ and $G_{\rm BP}$ magnitudes.  }  
\end{figure}

\subsubsection{Columba}
In total, we find one new Columba member, confirm 10 previously suggested Columba candidates as new bona fide members, while  5 previously suggested candidates are not found to be non-members.  Figure 15 shows the 3D XYZUVW distributions of bona fide members from \cite{gagne18a} and newly confirmed members.  Figure 16 shows a color-magnitude diagram of previously known members and newly confirmed Columba members using {\it Gaia} DR2 parallaxes and photometry.   

2MASS 03241504$-$5901125 was suggested to be a possible Horologium Association member in \cite{torres00}, but was subsequently rejected as a Tuc-Hor member in \cite{zuck01}.  This star was later proposed to be a Columba candidate in \cite{malo13,malo14}, but lacked a parallax measurement. Our measured radial velocity and the {\it Gaia} DR2 parallax show that the kinematics of this star are consistent with the Columba Association, though with a low probability from BANYAN $\Sigma$ (8.2\%).  The measured distance (86.6$\pm$1.2 pc) is 1.1$\sigma$ away from the optimum BANYAN $\Sigma$ distance (81.8$\pm$3.2 pc), while the measured radial velocity (18.3$\pm$1.2 km s$^{-1}$) is less than 1$\sigma$ away from the optimum radial velocity (16.6$\pm$0.6 km s$^{-1}$).  All age diagnostics confirm a young age for this object, and we posit that Columba membership is highly likely.

\cite{malo13} proposed 2MASS 03320347$-$5139550 and 2MASS 05195695$-$1124440 as possible Columba members, albeit without measured radial velocities or parallaxes.  The {\it Gaia} DR2 parallaxes for these objects, combined with our measured radial velocities confirm Columba kinematics.  We consider 2MASS 03320347$-$5139550 and 2MASS 05195695$-$1124440 bona fide Columba members, as age indicators for both objects are all consistent with Columba membership.  Note that 2MASS 03320347$-$5139550 and 2MASS 05195695$-$1124440 both have co-moving companions in Section 4.2.

2MASS 04091413$-$4008019, 2MASS 05100427$-$2340407, 2MASS 05100488$-$2340148, 2MASS 05241317$-$2104427, 2MASS 05395494$-$1307598, and 2MASS 05432676$-$3025129 were proposed as a possible Columba members in \cite{malo13, malo14}, though without parallax measurements.  Our measured radial velocities are consistent with those found in \cite{malo14}. Combined with the parallax measurements of these objects from {\it Gaia} DR2, we find space motions consistent with Columba membership, though 2MASS 05432676$-$3025129 returns a small Columba membership probability from BANYAN $\Sigma$ (1.3\%).  This small probability is because the predicted radial velocity (22.7$\pm$1.2 km s$^{-1}$) is slightly off from our measured radial velocity (28.2$\pm$1.4 km s$^{-1}$), a difference of 2.1$\sigma$.  Because all age information is also consistent with Columba membership, 2MASS 04091413$-$4008019, 2MASS 05100427$-$2340407, 2MASS 05100488$-$2340148, 2MASS 05241317$-$2104427, 2MASS 05395494$-$1307598, and 2MASS 05432676$-$3025129 are deemed to be bona fide Columba members.  

\cite{malo13, malo14} considered 2MASS 05425587$-$0718382 a Columba candidate, though without a parallax measurement.  While our radial velocity (38.81$\pm$5.58 km s$^{-1}$) is less precise than the one presented in \cite{malo14} (29.7$\pm$3.9 km s$^{-1}$), we find that neither provides a good match to other Columba members when combined with the parallax from {\it Gaia} DR2.  We therefore reject this object as a potential Columba member. 

2MASS 06262932$-$0739540 was deemed an ambiguous Columba candidate in \cite{malo13}, though lacked radial velocity and parallax measurements.  We find good agreement with other Columba members using our measured radial velocity and the parallax from {\it Gaia} DR2.  The well detected lithium for this object confirms its youth, and is consistent with Columba membership.  We consider 2MASS 06262932$-$0739540 a bona fide Columba member.

2MASS 04322548$-$3903153 was suggested as a Columba candidate member in both \cite{malo13} and \cite{rod13}, though neither study had a radial velocity or parallax for this object.  The {\it Gaia} DR2 parallax, combined with our radial velocity measurement, firmly rule out Columba membership.   

\cite{malo13} presented 2MASS 06511418$-$4037510 as a possible Columba member.  Our radial velocity measurement, combined with its {\it Gaia} DR2 parallax, rules out Columba membership.  2MASS 06511418$-$4037510 is likely a K-type giant based on its small parallax (0.7084$\pm$0.0234 mas).    

\begin{figure*}
\plotone{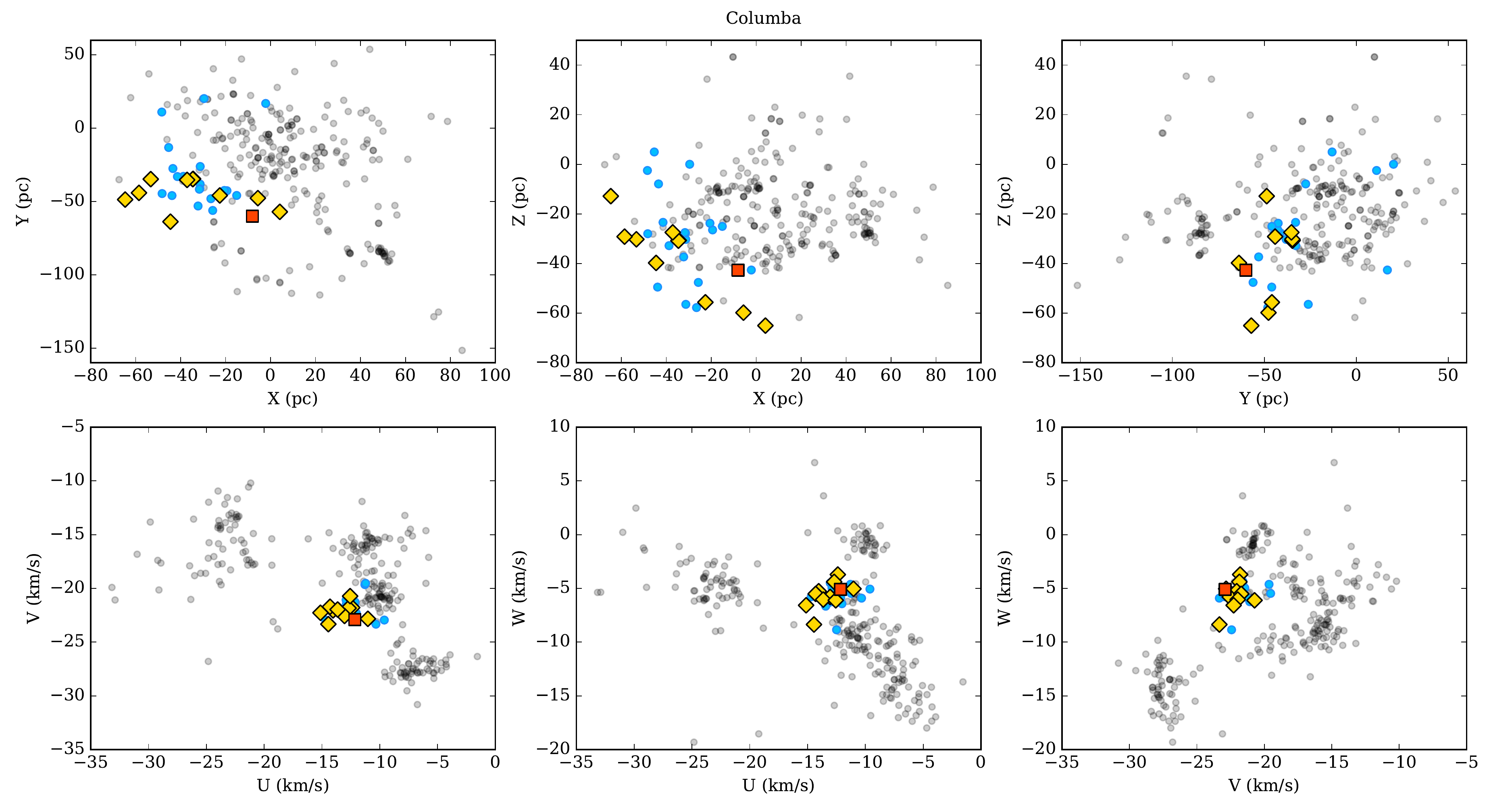}
\caption{A comparison of XYZUVW distributions of previously known Columba members from \cite{gagne18a} (blue circles) with newly confirmed members (yellow diamonds), and new discoveries (red squares).  Black symbols are the same as in Figure 5.}  
\end{figure*}

\begin{figure}
\plotone{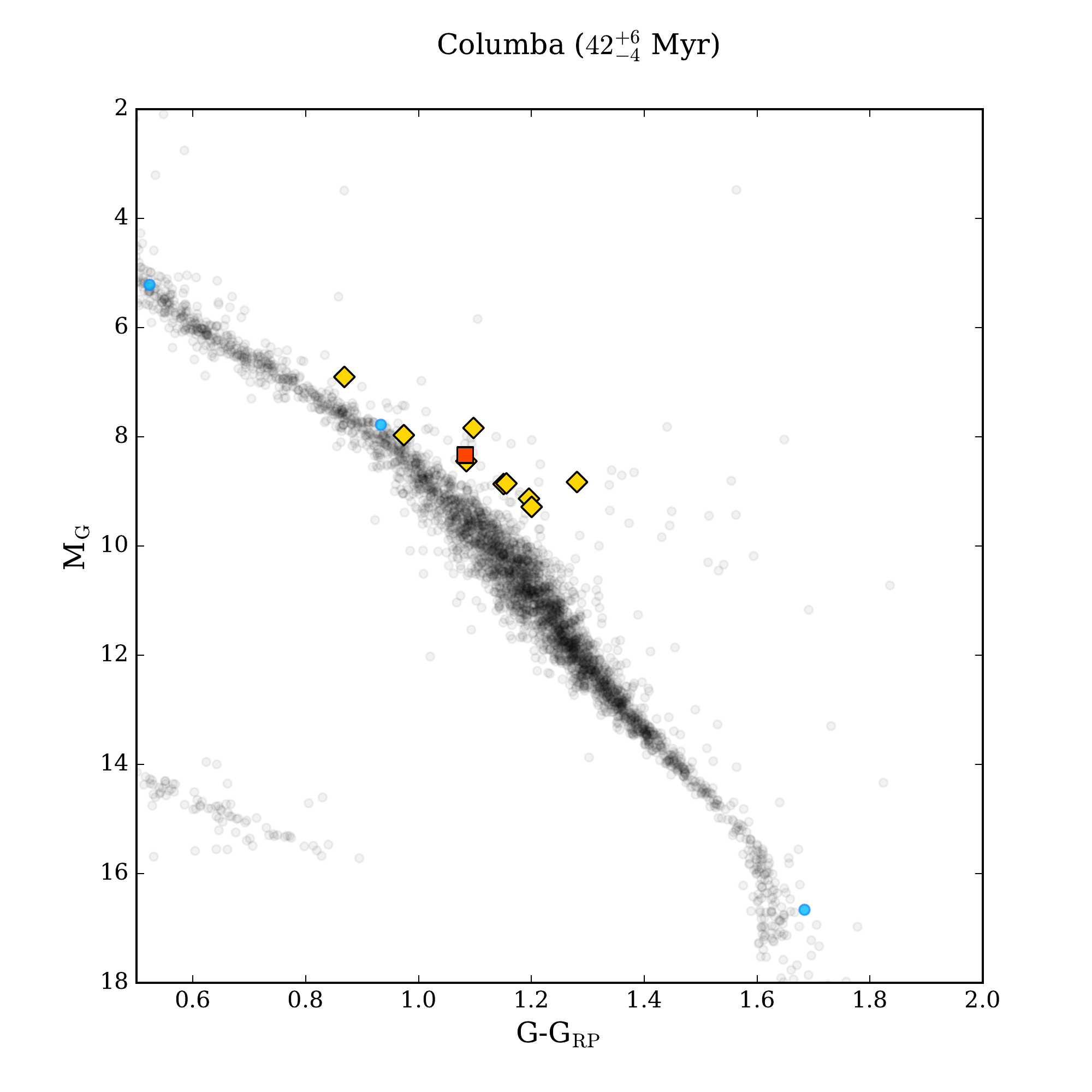}
\caption{CMD of previously known Columba members from \cite{gagne18a} (blue circles) with newly confirmed members (yellow diamonds), and new discoveries (red squares).  For reference, we show all objects from {\it Gaia} DR2 within 25 pc as background gray symbols. }  
\end{figure}

\subsubsection{$\epsilon$ Cha}
We confirm the membership of 4 $\epsilon$ Cha members using our measured radial velocities and parallaxes from {\it Gaia} DR2.  Figures 17 and 18 compare the 3D XYZUVW distributions and CMD positions of previously known members from \cite{gagne18a} and newly confirmed $\epsilon$ Cha members.    

2MASS 11493184$-$7851011 was proposed as an $\epsilon$ Cha member in \cite{torres08}.  It was later proposed to be a potential $\beta$ Pic member in \cite{malo13}, though the authors noted that they did not include $\epsilon$ Cha as a potential group and a parallax would clear up membership.  \cite{shk17} determine that 2MASS 11493184$-$7851011 is a bona fide $\beta$ Pic member.  \cite{zerjal19} also suggest 2MASS 11493184$-$7851011 as an $\epsilon$ Cha member. We find excellent agreement with $\epsilon$ Cha and consider this object to be a bona fide $\epsilon$ Cha member. 

2MASS 12194369$-$7403572 and 2MASS 12202177$-$7407393 were proposed to be $\epsilon$ Cha field members in \cite{torres08}.  \cite{murph13} used radial velocity measurements to reinforce membership.  Our measured radial velocities and {\it Gaia} DR2 parallaxes confirm $\epsilon$ Cha membership.  Note that the {\it Gaia} DR2 parallax uncertainty for 2MASS 12202177$-$7407393 is atypically large (0.6978 mas), and the reason for the relatively low probability from BANYAN $\Sigma$ is the discrepancy between its predicted distance (99.8$\pm$3.7 pc) and the measured distance (148.9$\pm$15.5 pc), a difference of 2.6$\sigma$.

2MASS 12210499$-$7116493 was suggested as an $\epsilon$ Cha member in \cite{kiss11} with a RAVE radial velocity, but lacking a parallax.  Our more precise radial velocity, a parallax from {\it Gaia} DR2, and a strong lithium detection, confirm $\epsilon$ Cha membership.

\begin{figure*}
\plotone{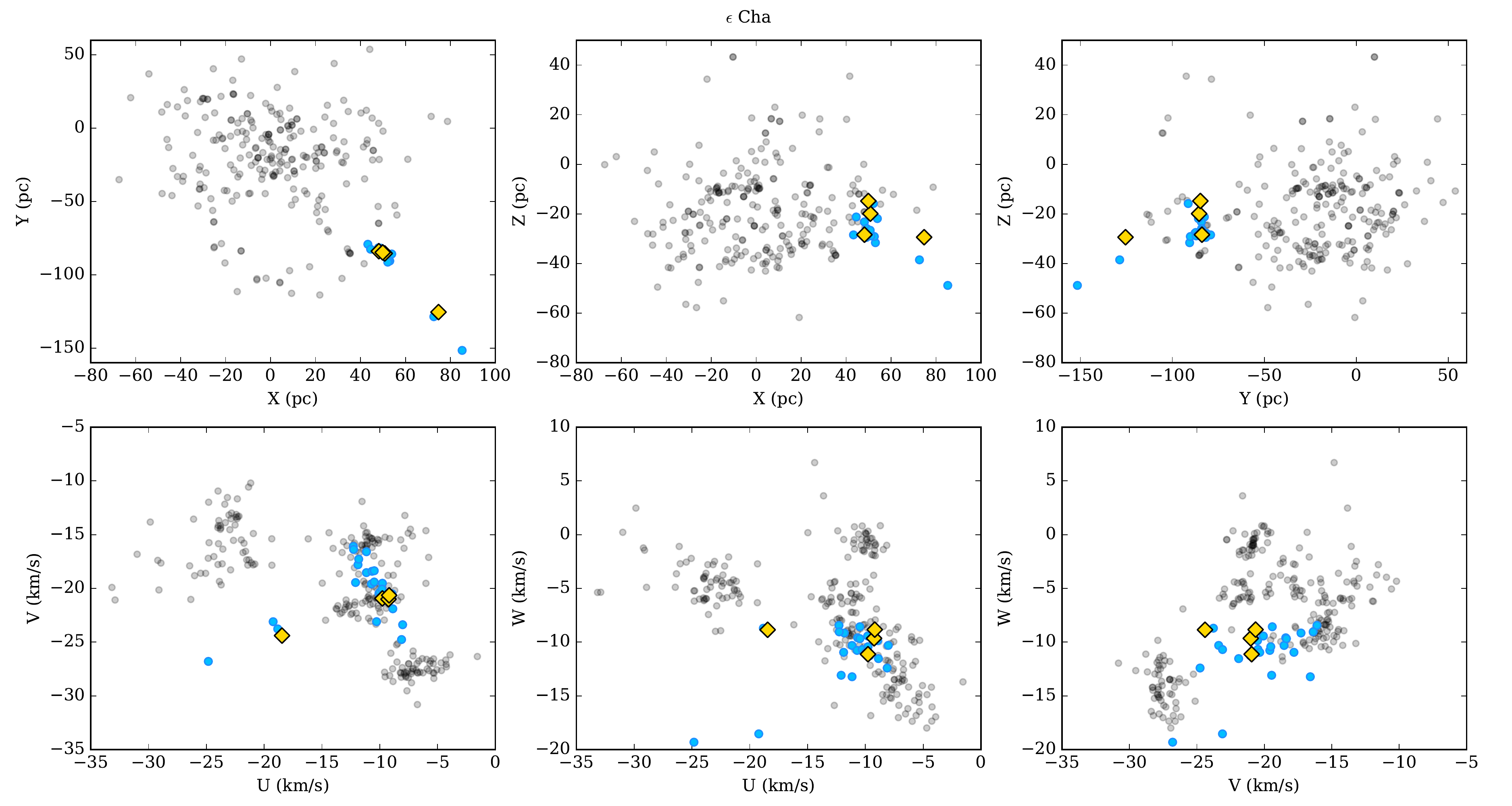}
\caption{A comparison of XYZUVW distributions of previously known $\epsilon$ Cha members from \cite{gagne18a} (blue circles) with newly confirmed members (yellow diamonds).  Black symbols are the same as in Figure 5.}  
\end{figure*}

\begin{figure}
\plotone{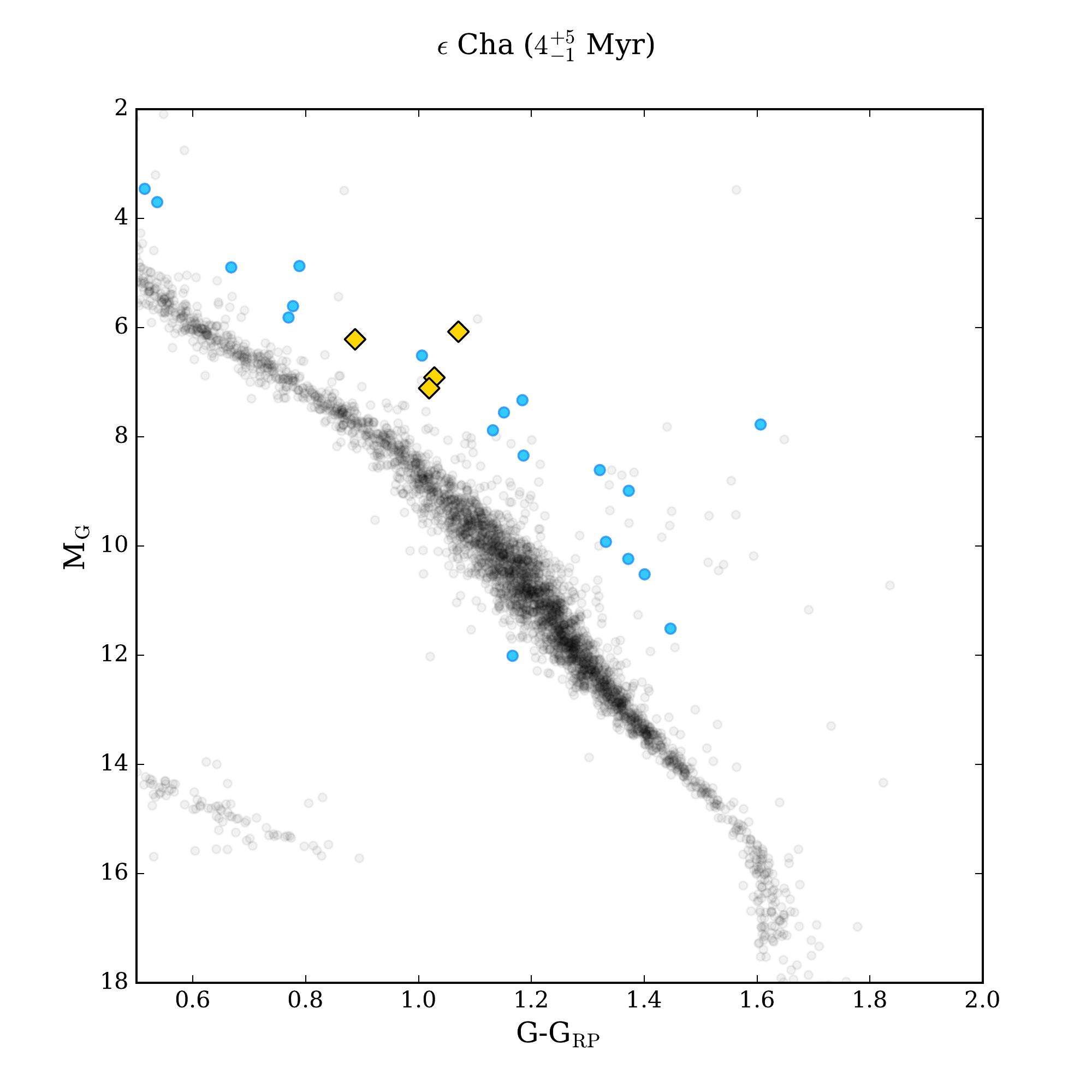}
\caption{CMD of previously known $\epsilon$ Cha members from \cite{gagne18a} (blue circles) with newly confirmed members (yellow diamonds).  For reference, we show all objects from {\it Gaia} DR2 within 25 pc as background gray symbols. The bona fide member below the main-sequence defined by the 25 pc sample is $\epsilon$ Cha 11, which is thought to be obscured by an circumstellar disk viewed edge-on \citep{luhman04}.}  
\end{figure}

\subsubsection{$\eta$ Cha}
Our measured radial velocities and the parallaxes from {\it Gaia} DR2 allow us to confirm the membership of 8 $\eta$ Cha members.  Figures 19 and 20 compare the 3D XYZUVW distributions and CMD positions of previously known members from \cite{gagne18a} and newly confirmed $\eta$ Cha members.    

2MASS 08361072$-$7908183 and 2MASS 08385150$-$7916136 were suggested as a members of $\eta$ Cha in \cite{song04} and \cite{lyo04}.  Our radial velocity measurements, plus the {\it Gaia} DR2 parallaxes confirm $\eta$ Cha membership.

2MASS 08413030$-$7853064 was suggested as an $\eta$ Cha member in \cite{lawson02}.  No radial velocity for this object has been presented previously, and our measurement confirms $\eta$ Cha membership.

2MASS 08413703$-$7903304, 2MASS 08422710$-$7857479, 2MASS 08423879$-$7854427, and 2MASS 08441637$-$7859080 were among the original $\eta$ Cha members proposed in \cite{mama99}.  Remarkably, we find no measurements of their radial velocities anywhere in the literature.  We provide the first radial velocities for these objects, all confirming $\eta$ Cha membership.  Note that 2MASS 08441637$-$7859080 currently lacks astrometry from Gaia.  The BANYAN $\Sigma$ percentage in Table 6 uses the statistical distance to this object from \cite{bell15} and the proper motion from \cite{zach17}.

2MASS 08440914$-$7833457 was suggested as an $\eta$ Cha member in \cite{song04}.  We confirm $\eta$ Cha membership with our measured radial velocity and the parallax from {\it Gaia} DR2.

\begin{figure*}
\plotone{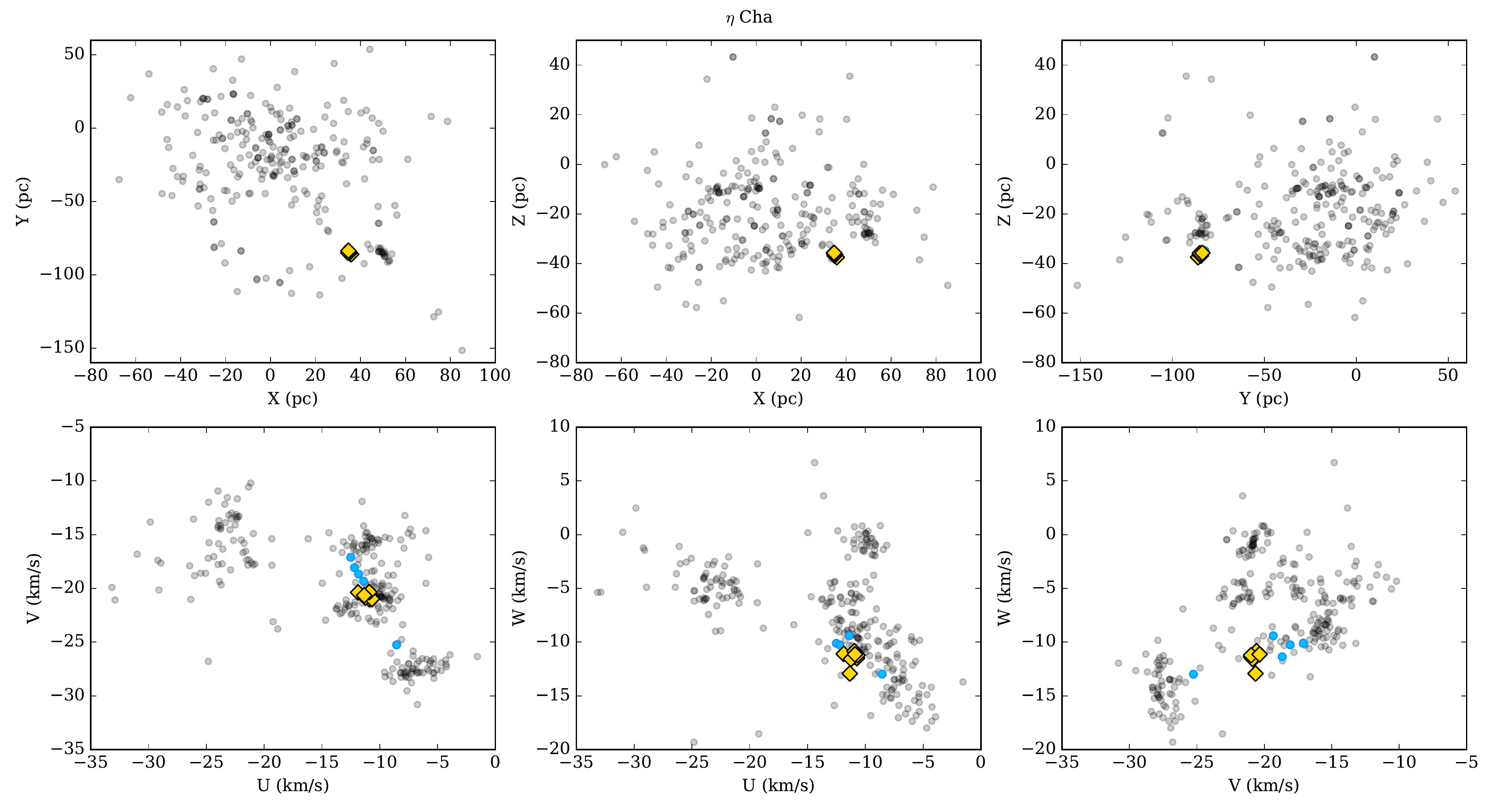}
\caption{A comparison of XYZUVW distributions of previously known $\eta$ Cha members from \cite{gagne18a} (blue circles) with newly confirmed members (yellow diamonds).  Black symbols are the same as in Figure 5.}  
\end{figure*}

\begin{figure}
\plotone{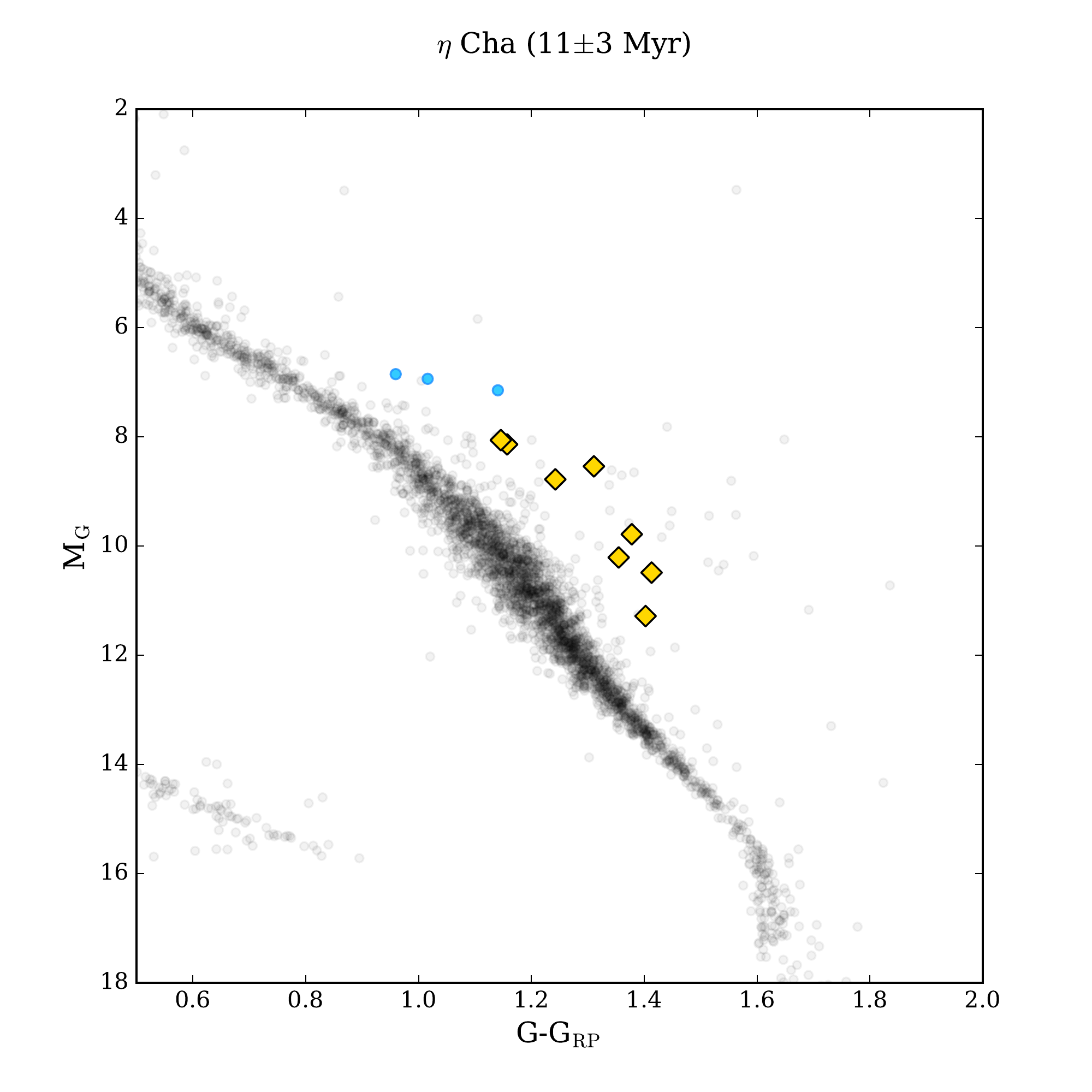}
\caption{CMD of previously known $\eta$ Cha members from \cite{gagne18a} (blue circles) with newly confirmed members (yellow diamonds).  For reference, we show all objects from {\it Gaia} DR2 within 25 pc as background gray symbols.}  
\end{figure}

\subsubsection{Octans}
2MASS 02411909$-$5725185 was suggested as a possible Octans member using {\it Gaia} DR2 kinematics (including a radial velocity measurement of 4.38$\pm$1.77 km s$^{-1}$) in \cite{gagne18b}.  Our more precise radial velocity (7.03$\pm$0.43 km s$^{-1}$) returns a 100\% probability of matching kinematics to known Octans members using BANYAN $\Sigma$.  All age information agrees with Octans membership.  Note that this star has a co-moving companion in Section 4.2.  We do not plot Octans figures, as we have only one confirmed member.  

\subsubsection{Tucana-Horologium}
We reassign membership for 3 objects as Tuc-Hor members, confirm Tuc-Hor membership for 4 previously suggested candidates, and find 2 previously suggested candidates to be non-members.  Figure 21 shows the 3D XYZUVW distributions of previously known members from \cite{gagne18a} and newly confirmed Tuc-Hor members.  Figure 22 shows a color-magnitude diagram of previously known and newly confirmed Tuc-Hor members using {\it Gaia} DR2 parallaxes and photometry.   

2MASS 00153670$-$2946003 (aka GJ 3017) was suggested as a Tuc-Hor candidate in \cite{gagne18b} using its {\it Gaia} DR2 parallax and a radial velocity of 0$\pm$5 km s$^{-1}$ from \cite{bard14} (this is likely a mistaken reference, with this measurement  being from \citealt{kunder17}).  Our more precise radial velocity for this object (0.61$\pm$1.31 km s$^{-1}$) is in agreement with the measurement presented in \cite{gagne18b} and puts 2MASS 00153670$-$2946003 squarely in the XYZUVW space as bona fide Tuc-Hor members.  A lithium non-detection is typical for M4 members of Tuc-Hor (\citealt{rod13}, \citealt{kraus14}), and the X-ray, UV, and H$\alpha$ activity levels of 2MASS 00153670$-$2946003 are all consistent with Tuc-Hor membership.  We conclude that 2MASS 00153670$-$2946003 is a bona fide member of Tuc-Hor. 

2MASS 03454058$-$7509121, 2MASS 19225071$-$6310581, and 2MASS 23204705$-$6723209 were suggested as a Tuc-Hor candidates in \cite{malo13,malo14} with a radial velocity measurements of 13.1$\pm$3.2 km s$^{-1}$, 6.4$\pm$1.5 km s$^{-1}$, and 6.6$\pm$0.3 km s$^{-1}$ respectively.  Our more precise radial velocities for 2MASS 03454058$-$7509121 and 2MASS 19225071$-$6310581 (13.81$\pm$0.65 km s$^{-1}$ and 0.84$\pm$1.00 km s$^{-1}$, respectively) and parallaxes from {\it Gaia} DR2 confirm their status as a Tuc-Hor members.  Our radial velocity for 2MASS 23204705$-$6723209 is less precise, though consistent with that measured in \cite{malo14} (6.13$\pm$1.26 km s$^{-1}$).  The {\it Gaia} DR2 parallax for 2MASS 23204705$-$6723209 confirms it as a Tuc-Hor member.  We also find a co-moving companion to 2MASS 23204705$-$6723209 in Section 4.2. 

2MASS 05142736$-$1514514 and 2MASS 05142878$-$1514546 were both assigned ambiguous Tuc-Hor/Columba membership in \cite{malo13}, and later suggested as Columba candidates in \cite{malo14}.  2MASS 06002304$-$4401217 was suggested as potential Columba member in \cite{malo13, malo14}.  We find that the kinematics of all three of these objects make them likely members of Tuc-Hor, which is reflected in their BANYAN $\Sigma$ probabilities. We consider all three objects Tuc-Hor members.  Note that we identify a third component of the 2MASS 05142736$-$1514514 and 2MASS 05142878$-$1514546 system, as well as a co-moving companion to 2MASS 06002304$-$4401217 in Section 4.2.

2MASS 02511150$-$4753077 and 2MASS 06434532$-$6424396 were suggested as a members of the Horologium association in \cite{torres00}.  We find that the kinematics of these objects match none of the groups evaluated here.

\begin{figure*}
\plotone{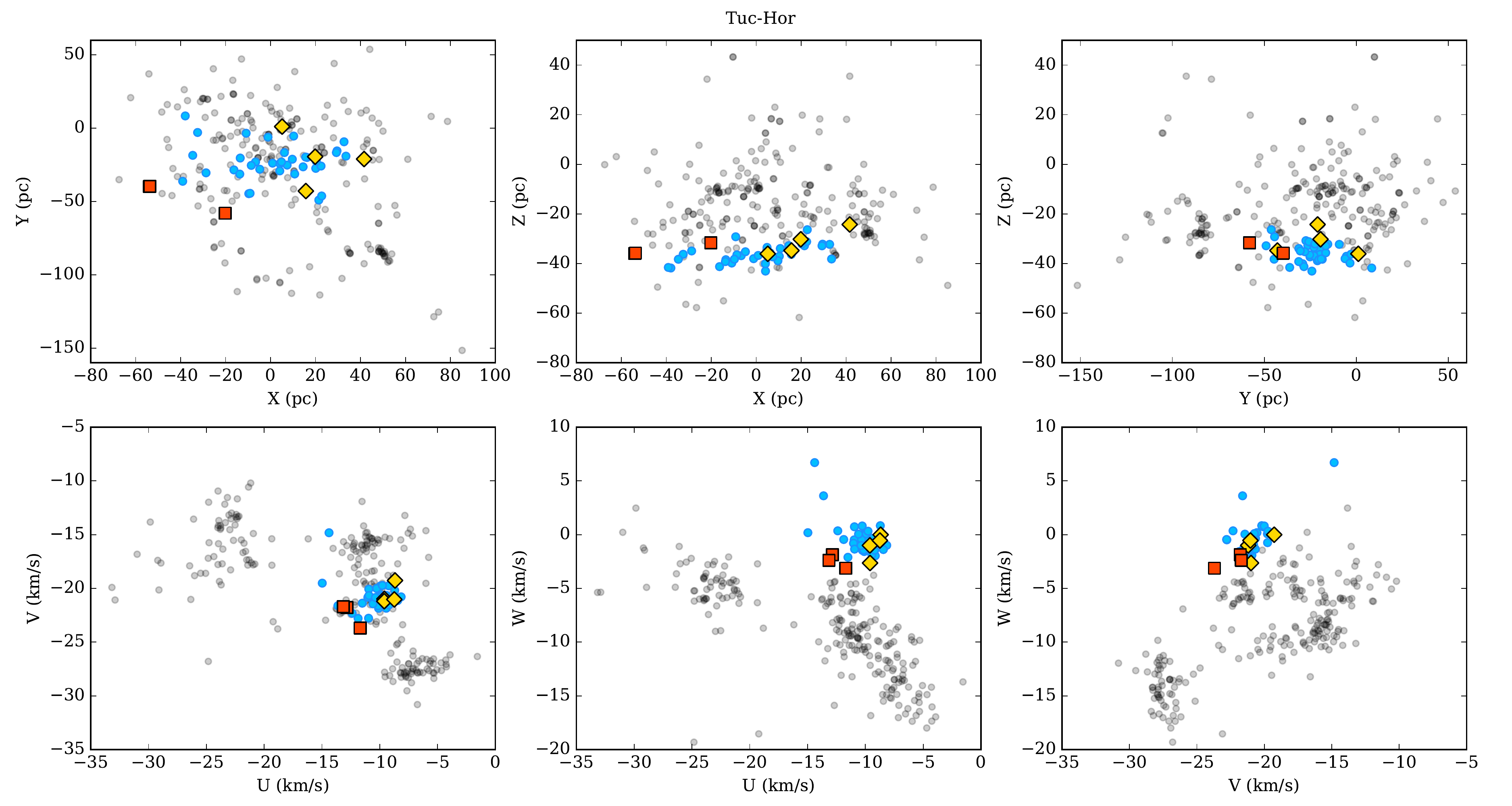}
\caption{A comparison of XYZUVW distributions of previously known Tuc-Hor members from \cite{gagne18a} (blue circles) with newly confirmed members (yellow diamonds), and new discoveries (red squares).  Black symbols are the same as in Figure 5.}  
\end{figure*}

\begin{figure}
\plotone{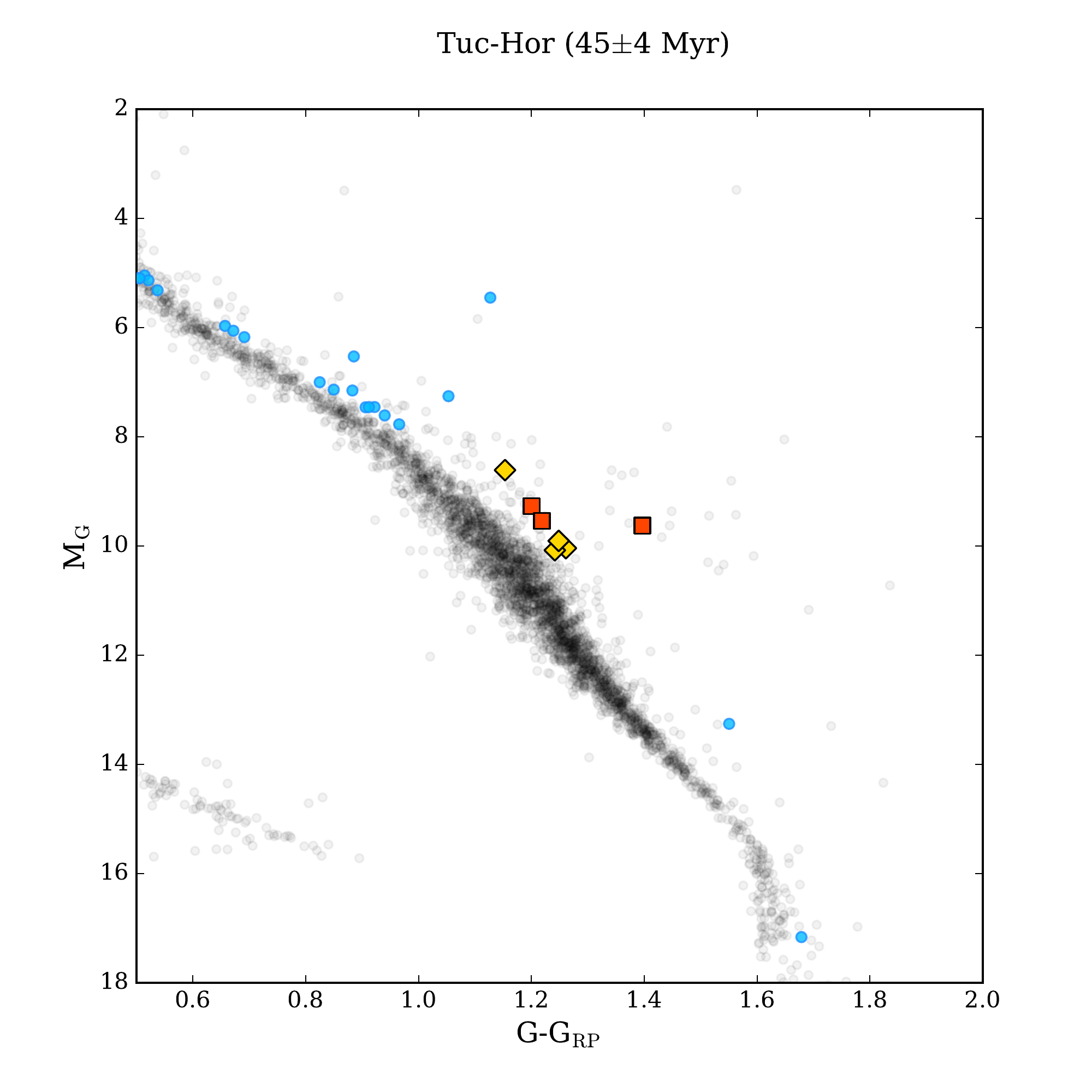}
\caption{CMD of previously known Tuc-Hor members from \cite{gagne18a} (blue circles) with newly confirmed members (yellow diamonds), and new discoveries (red squares).  For reference, we show all objects from {\it Gaia} DR2 within 25 pc as background gray symbols. }  
\end{figure}

\subsection{Low-Probability Members}
The BANYAN $\Sigma$ algorithm models each association with multivariate Gaussians in six dimensional XYZUVW space \citep{gagne18a}.  However, moving groups and associations are not necessarily Gaussian in shape \citep{larson81}.  This also means that groups are constrained by the current census of known members, which in many instances is still incomplete, occasionally leading to low probabilities for high likelihood members (e.g., \citealt{lee18}).  For this reason, we performed an additional check to see if there are any potential group members in our sample missed by the BANYAN framework.  We first found the central UVW coordinates of each group considered in Section 4.3 by finding the average position from previously known members from \citep{gagne18a} combined with newly confirmed members from this work.   We then calculate the dispersion in U, V, and W and search for objects within 3$\sigma$ of each velocity component.  We excluded objects that were previously determined to be a member of a group in Section 4.3 and objects with large astrometric uncertainties (e.g., objects without {\it Gaia} parallaxes).  Seven objects were found to match the kinematics of a known group.  We consider these objects possible members, and provide their information in Table 8.  We also include the $\sigma$ difference between the optimal distance and radial velocity values from BANYAN $\Sigma$ and those measured in this work defined as the difference between the optimal and measured values divided by the combined uncertainty. 

2MASS 04424932$-$1452268 and 2MASS 05240991$-$4223054 were rejected as AB Dor members in Section 4.3.1, but we find them to be potential fringe members of AB Dor here.  All age diagnostics are consistent with AB Dor membership for both objects.  2MASS 07343426$-$2401353, 2MASS 09423823$-$6229028, and 2MASS 13591045$-$1950034 were all rejected as Argus members in Section 4.3.2, however we find that the are potential edge members of Carina-Near.  Note that 2MASS 07343426$-$2401353 is also found to be a possible fringe member of Columba.  2MASS 07343426$-$2401353 was rejected as a potential $\beta$ Pic member in Section 4.3.3, and we find it to be a potential Carina-Near member here.  Age diagnostics are consistent with Carina-Near for all four objects.  Lastly, 2MASS 11091606$-$7352465 is found to be a potential member of AB Dor, $\beta$ Pic, or Carina.  This object was also found to have ambiguous moving group membership in \cite{malo14}, possibly belonging to $\beta$ Pic, Carina, or Argus.  The CMD position and age diagnostics for 2MASS 11091606$-$7352465 are more consistent with $\beta$ Pic and Carina than AB Dor.

\begin{longrotatetable}
\begin{deluxetable}{lrrrrrrrrrrrrr}
\tabletypesize{\footnotesize}
\tablecaption{Low-Probability Candidate Members}
\tablehead{
\colhead{2MASS} & \colhead{Spec.} & \colhead{Li EW\tablenotemark{a}} & \colhead{H$\alpha$ EW} & \colhead{log $L_{\rm X}$} & \colhead{$f_{\rm FUV}$/$f_{\rm J}$} & \colhead{$f_{\rm NUV}$/$f_{\rm J}$} & \colhead{XYZ} & \colhead{UVW}  & \colhead{$\sigma$$_{\rm dist}$} & \colhead{$\sigma$$_{\rm RV}$}  \\
\colhead{Name} & \colhead{Type} & \colhead{(\AA)} & \colhead{(\AA)} & \colhead{(erg s$^{-1}$)} & & &  \colhead{(pc)} & \colhead{(km s$^{-1}$)} & \colhead{}  }
\startdata
\colrule
\sidehead{AB Dor}
\colrule
04424932$-$1452268 & M4.0 & \dots & -6.00 & 29.00$\pm$0.10 & \dots & 1.75e-04 & (-24.41,-15.55,-20.33) & (-1.67,-30.52,-13.76) & 4.45 & 0.16 \\
05240991$-$4223054 & M0.5 & \dots & -2.41 & 29.80$\pm$0.09 & \dots & 1.84e-04 & (-31.89,-77.48,-55.42) & (-11.00,-31.02,-11.84) & 3.08 & 1.74\\
11091606$-$7352465 & M3.0 & \dots & -3.49 & 29.59$\pm$0.13 & 6.03e-05 & 2.25e-04 & (27.81,-57.30,-14.03) & (-7.11,-24.57,-7.39) & 1.80 & 1.60 \\
\colrule
\sidehead{$\beta$ Pic}
\colrule
11091606$-$7352465 & M3.0 & \dots & -3.49 & 29.59$\pm$0.13 & 6.03e-05 & 2.25e-04 & (27.81,-57.30,-14.03) & (-7.11,-24.57,-7.39) & 1.76 & 2.48 \\
\colrule
\sidehead{Carina}
\colrule
11091606$-$7352465 & M3.0 & \dots & -3.49 & 29.59$\pm$0.13 & 6.03e-05 & 2.25e-04 & (27.81,-57.30,-14.03) & (-7.11,-24.57,-7.39) & 4.34 & 1.68 \\
\colrule
\sidehead{Carina-Near}
\colrule
07343426$-$2401353 & M3.5 & \dots & -6.19 & 29.06$\pm$0.13 & \dots & \dots & (-21.54,-35.95,-1.41) & (-16.88,-18.69,-3.72) & 3.15 & 0.47\\
08224744$-$5726530 & M4.5 & \dots & -6.09 & 28.34$\pm$0.05 & \dots & \dots & (0.56,-12.64,-2.56) & (-36.48,-15.61,-5.83) & 8.84 & 2.99 \\
09423823$-$6229028 & M3.5 & \dots & -1.24 & 29.19$\pm$0.10 & \dots & \dots & (9.44,-41.05,-5.31) & (-30.35,-18.15,-8.55) & 8.80 & 2.22 \\
13591045$-$1950034 & M4.5 & \dots & -4.63 & 28.33$\pm$0.07 & 1.66e-05 & 6.18e-05 & (6.68,-4.86,6.99) & (-28.11,-16.99,-9.81) & 1.07 & 5.85 \\
\colrule
\sidehead{Columba}
\colrule
07343426$-$2401353 & M3.5 & \dots & -6.19 & 29.06$\pm$0.13 & \dots & \dots & (-21.54,-35.95,-1.41) & (-16.88,-18.69,-3.72) & 5.02 & 1.53 \\
\enddata
\end{deluxetable}
\end{longrotatetable}

\subsection{The Age of Carina}
The Carina association was originally part of a larger complex named GAYA2 in \cite{torres01} which included Carina, Tuc-Hor, and Columba, all of which were proposed to have similar ages ($\sim$30 Myr; \citealt{torres01}).  GAYA2 was eventually divided into Carina, Tuc-Hor, and Columba, with an updated list of Carina members in \cite{torres08}.  The age most often quoted for Carina, which we quote in this work, is 45$^{+11}_{-7}$ from \cite{bell15}, which is based on isochronal fitting to 12 group members, the lowest number of objects fit for a group in that work.  Furthermore, only 5 of the 12 objects examined had measured parallaxes, all with spectral types earlier than K2.  The remaining 7 objects (spectral types K7 to M4.5) only had statistical distances from either \cite{malo13} or \cite{malo14}.  

We note that the confirmed members of Carina found in this work are elevated above the main sequence more than expected in Figure 12.  For an empirical comparison, we plotted the CMD distribution of Carina members versus the presumed similar age Tuc-Hor association members found in Section 4.3.10 and the younger $\beta$ Pic group association in Figure 23.  While there is scatter among all groups, Carina members tend to rest higher above the main sequence than members of Tuc-Hor and more closely follow the sequence defined by $\beta$ Pic members, suggesting a younger age.         

\begin{figure*}
\plotone{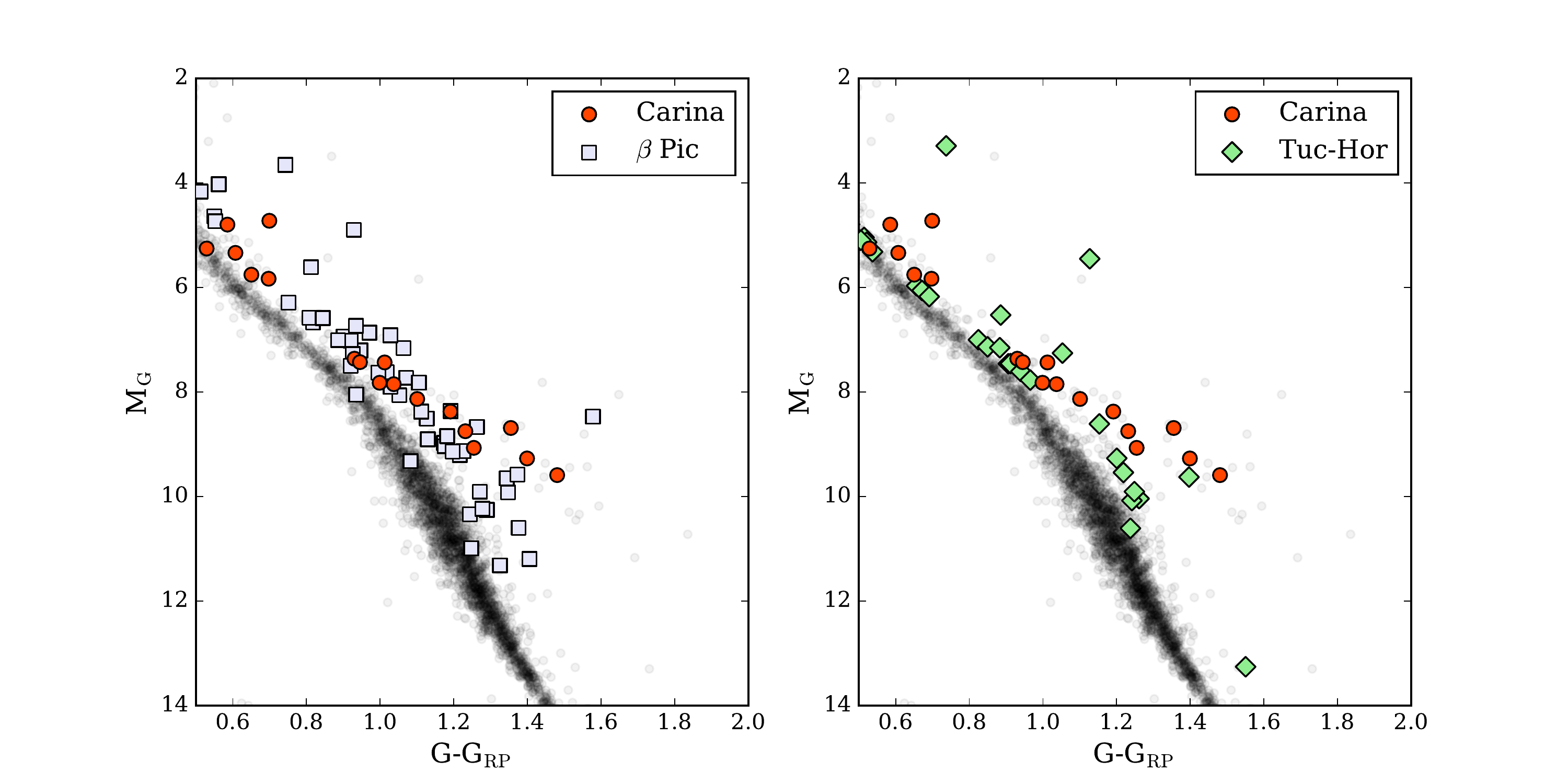}
\caption{CMD of previously known, new, and confirmed Carina members with members of the $\beta$ Pictoris moving group (left) and members of the Tucana-Horologium association (right).  For reference, we show all objects from {\it Gaia} DR2 within 25 pc as background gray symbols. }  
\end{figure*}

We also compare the lithium abundances of Carina members versus those from $\beta$ Pic and Tuc-Hor.  We took lithium abundances for $\beta$ Pic members from \cite{shk17}, and \cite{kraus14} for Tuc-Hor.  For Carina members, we use the compilation of lithium measurements from \cite{riedel17}.  A comparison of lithium abundances is shown in Figure 24.  While there is a dearth of known Carina members with spectral types between K2 and M0, the strong lithium detections for several members with spectral types $\geq$M0 and $<$M2 implies a younger age than Tuc-Hor, which is depleted for spectral types K7--M5.  The observed lithium abundances are more consistent with $\beta$ Pic, which is depleted for spectral types M1--M4.   Note that \cite{murph18} find an age for the disk-hosting Carina member WISE J0808$-$6443 \citep{silv16} of 33$^{+25}_{-15}$ Myr.  If a younger age is indeed warranted for Carina, the protoplanetary disk found around WISE J0808$-$6443 may pose less of a challenge to disk lifetime models.  A more thorough census of Carina membership could help to determine a more precise lithium depletion boundary age.  

\begin{figure}
\plotone{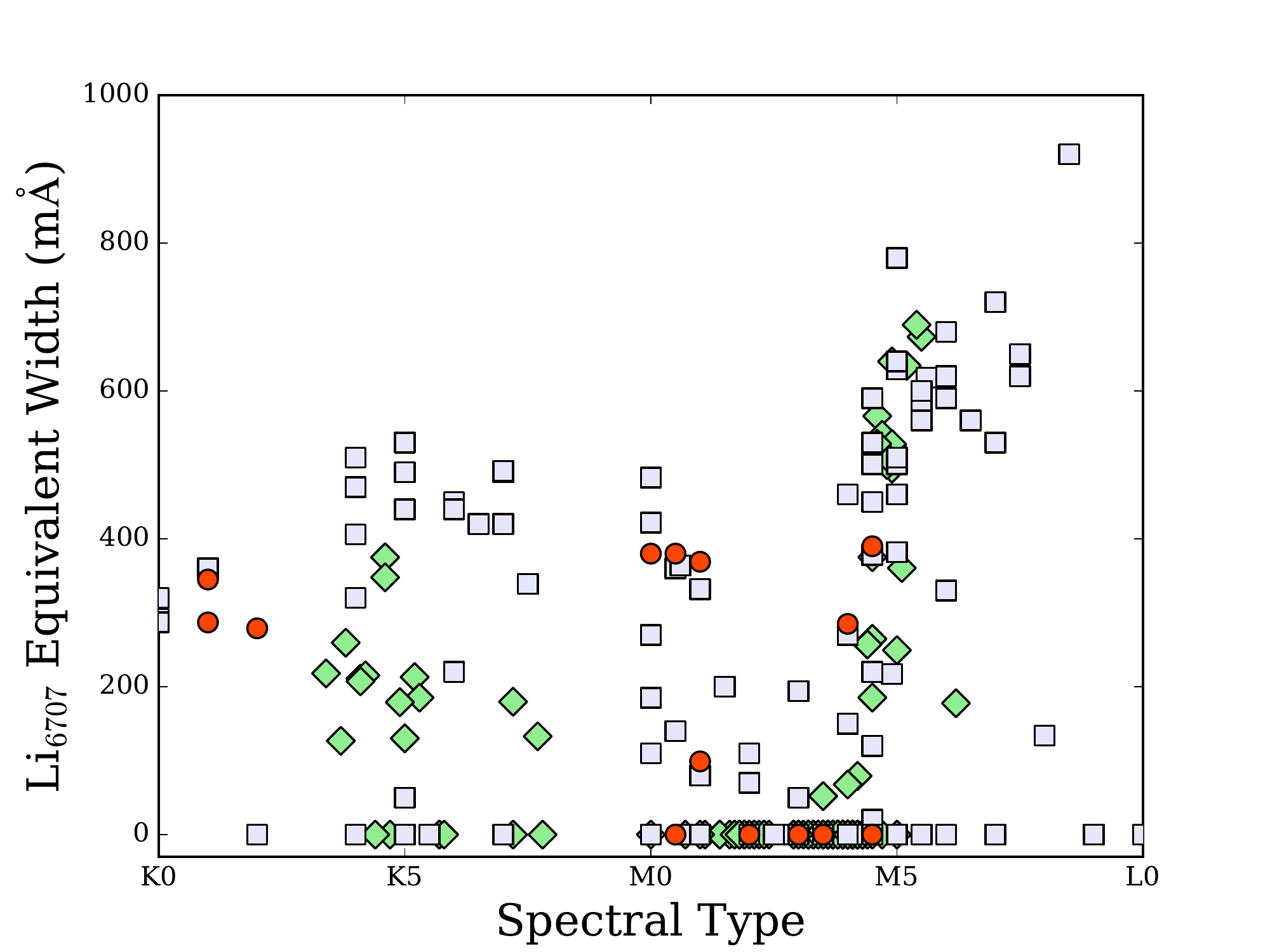}
\caption{Lithium $\lambda$6707 abundances of carina members (red circles) compared to those of the $\beta$ Pictoris moving group and the Tuc-Hor association.  }  
\end{figure}

\section{Conclusion}
 
Using high-resolution optical spectra, we measured radial velocities and H$\alpha$ and Li $\lambda$6707 equivalent widths for 336 candidate low-mass stars, with the aim of identifying new young stars in the solar neighborhood.  Combining our radial velocities with astrometry from {\it Gaia} DR2, we calculated 3D kinematics for each member, and evaluated their potential membership in known moving groups and associations.  We discovered 4 completely new moving group members, reassigned membership for 10 objects, and confirmed association membership for 62 additional stars.  We further rejected 44 suggested members, highlighting the importance of full kinematics when evaluating moving group membership.  We also investigated our measured H$\alpha$ equivalent widths for evidence of ongoing accretion, and find 5 likely accretors within our sample.  We used {\it Gaia} DR2 astrometry to search for co-moving companions to objects in our sample, identifying 69 co-moving systems.  In addition, we found that the age diagnostics for previously known and new Carina members confirmed in this work indicate an age nearer to that of $\beta$ Pic ($\sim$22 Myr), half the age of previous estimates.  Our catalog also contains numerous additional young stars that do not belong to any currently known moving groups or associations.  Such objects may help to locate and define new young groups in the solar neighborhood.  
 
\acknowledgments
A.S.\ and E.S.\ appreciate support from NASA/Habitable Worlds grant NNX16AB62G (PI E. Shkolnik).  This work has made use of data from the European Space Agency (ESA) mission {\it Gaia} (\url{https://www.cosmos.esa.int/gaia}), processed by the {\it Gaia} Data Processing and Analysis Consortium (DPAC, \url{https://www.cosmos.esa.int/web/gaia/dpac/consortium}). Funding for the DPAC has been provided by national institutions, in particular the institutions participating in the {\it Gaia} Multilateral Agreement.  This work is based on observations made with the NASA Galaxy Evolution Explorer. {\it GALEX} is operated for NASA by the California Institute of Technology under NASA contract NAS5-98034.  This publication makes use of data products from the Two Micron All Sky Survey, which is a joint project of the University of Massachusetts and the Infrared Processing and Analysis Center/California Institute of Technology, funded by the National Aeronautics and Space Administration and the National Science Foundation.  This research has made use of the SIMBAD database, operated at CDS, Strasbourg, France.

\end{document}